\documentclass[10pt,prd,fleqn,superscriptaddress,notitlepage,nofootinbib,preprintnumbers,nobalancelastpage]{revtex4-1}
\pdfoutput=1
\usepackage{amsmath,amssymb,graphicx,xspace}
\usepackage{todonotes}

\newcommand{\NNLOJET}{\textsc{NNLOjet}\xspace}
\newcommand{\PWG}{\texttt{POWHEG BOX}\xspace}
\newcommand{\PY}{\texttt{Pythia 8}\xspace}

\newcommand{\HW}{\texttt{Herwig 7}\xspace}
\newcommand{\HJets}{\texttt{HJets}\xspace}
\newcommand{\VBFNLO}{\texttt{VBFNLO}\xspace}
\newcommand{\Sherpa}{\textsc{Sherpa}\xspace}

\newcommand{\VBF}{VBF\xspace}
\newcommand{\VH}{VH\xspace}
\newcommand{\ttH}{ttH\xspace}

\begin{document}
\title{A comparative study of Higgs boson production from vector-boson fusion}
\preprint{FERMILAB-PUB-21-218-T, IPPP/20/101, MCNET-21-08, KA-TP-08-2021, OUTP-21-14P, ZU-TH 22/21, CERN-TH-2021-081}
\author{A.~Buckley}
\affiliation{School  of  Physics  and  Astronomy,  University  of  Glasgow,  Glasgow,  G12  8QQ,  UK}
\author{X.~Chen}
\affiliation{Physik-Institut, Universit{\"a}t Z{\"u}rich, CH-8057 Z{\"u}rich, Switzerland}
\affiliation{Institute for Theoretical Physics, Karlsruhe Institute of Technology, 76131 Karlsruhe, Germany}
\affiliation{Institute for Astroparticle Physics, Karlsruhe Institute of Technology, 76344 Eggenstein-Leopoldshafen, Germany}
\author{J.~Cruz-Martinez}
\affiliation{Dipartimento di Fisica, Universita degli Studi di Milano and INFN, Sezione di Milano}
\author{S.~Ferrario~Ravasio}
\affiliation{Centre for Theoretical Physics, Oxford University, Oxford, OX1 3PU, UK}
\affiliation{Institute for Particle Physics Phenomenology, Durham University, Durham, DH1 3LE, UK}
\author{T.~Gehrmann}
\affiliation{Physik-Institut, Universit{\"a}t Z{\"u}rich, CH-8057 Z{\"u}rich, Switzerland}
\author{E.W.N.~Glover}
\affiliation{Institute for Particle Physics Phenomenology, Durham University, Durham, DH1 3LE, UK}
\author{S.~H{\"o}che}
\affiliation{Fermi National Accelerator Laboratory, Batavia, IL, 60510, USA}
\author{A.~Huss}
\affiliation{Theoretical Physics Department, CERN, 1211 Geneva 23, Switzerland}
\author{J.~Huston}
\affiliation{Michigan State University, East Lansing, MI, 48824, USA}
\author{J.~M.~Lindert}
\affiliation{Department of Physics and Astronomy, University of Sussex, Brighton BN1 9QH, UK}
\author{S.~Pl{\"a}tzer}
\affiliation{Institute for Mathematics and Physics, University of Vienna, 1090 Wien, Austria}
\author{M.~Sch{\"o}nherr}
\affiliation{Institute for Particle Physics Phenomenology, Durham University, Durham, DH1      3LE, UK}

\begin{abstract}
  The data taken in Run~II at the Large Hadron Collider have started to probe Higgs boson production at high transverse momentum. Future data will provide a large sample of events with boosted Higgs boson topologies, allowing for a detailed understanding of electroweak Higgs boson plus two-jet production, and in particular the vector-boson fusion mode (\VBF).
  We perform a detailed comparison of precision calculations for Higgs boson production in this channel, with  particular emphasis on large Higgs boson transverse momenta, and on the jet radius dependence of the cross section. We study fixed-order predictions at next-to-leading order and next-to-next-to-leading order QCD, and compare the results to NLO plus parton shower~(NLOPS) matched calculations. The impact of the NNLO corrections on the central predictions is mild, with inclusive scale uncertainties of the order of a few percent, which can increase with the imposition of kinematic cuts. We find good agreement between the fixed-order and matched calculations in non-Sudakov regions, and the various NLOPS predictions also agree well in the Sudakov regime. We analyze backgrounds to \VBF Higgs boson production stemming from associated production, and from gluon-gluon fusion. At high Higgs boson transverse momenta, the $\Delta y_{jj}$ and/or $m_{jj}$ cuts typically used to enhance the \VBF signal over background lead to a reduced efficiency. We examine this effect as a function of the jet radius and using different definitions of the tagging jets. QCD radiative corrections increase for all Higgs production modes with increasing Higgs boson $p_T$, but the proportionately larger increase in the gluon fusion channel results in a decrease of the gluon-gluon fusion background to electroweak Higgs plus two jet production upon requiring exclusive two-jet topologies. We study this effect in detail and contrast in particular a central jet veto with a global jet multiplicity requirement.
\end{abstract}

\maketitle

\section{Introduction}
\label{sec:introduction}
\begin{figure}[t]
  \includegraphics[width=0.75\textwidth]{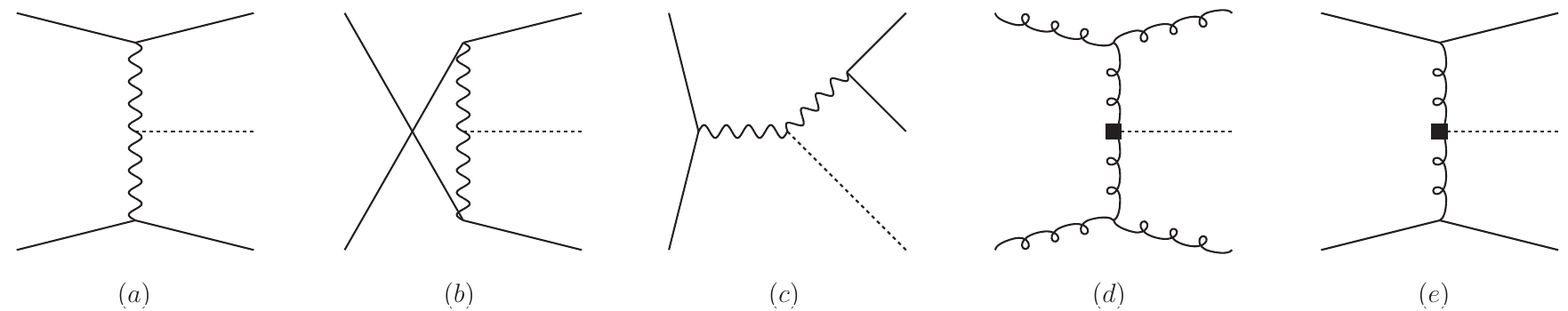}
  \caption{Leading order contributions to $pp\to h+2$~jets.
  Figures~(a) and~(b) show VBF-type diagrams, Fig.~(c) is of the Higgs-Strahlung type with $V\to jj$,
  and Figs.~(d) and~(e) are representative gluon-fusion-type diagrams (where the black square
  represents the effective vertex in the Higgs Effective Field Theory or the top-quark loop in full Standard Model).
  \label{fig:vbflo}}
\end{figure}

After the discovery of the Higgs boson in 2012~\cite{Aad:2012tfa,Chatrchyan:2012ufa}, the primary focus of Higgs physics has entered the era of precision phenomenology to measure its properties and in particular its kinematic behavior, couplings to Standard Model particles and to new physics. 
Vector-boson fusion (\VBF) is one of the crucial production channels for the Higgs boson at the LHC, where the Higgs boson is created by annihilation of virtual $W$ or $Z$ bosons, radiated off initial-state quarks in a $t$-channel scattering process. 
The experimental signature consists of a Higgs boson (or its decay products) and two forward jets allowing the determination of the Higgs boson couplings to electroweak (EW) gauge bosons. The \VBF cross section is currently known experimentally to the order of 20-25\%~\cite{ATLAS-CONF-2020-027,Sirunyan:2021ybb}.
This precision will greatly improve with the integrated luminosity expected at the high-luminosity LHC~\cite{HL-HE-LHC}. 
This large integrated luminosity will also allow for a more detailed investigation of Higgs boson production at high $p_T$, where possible new physics can enter~\cite{Greljo:2015sla,Araz:2020zyh}.

To interpret the upcoming precise measurements in various fiducial regions, theory predictions of the Higgs-plus-two-jet final states need to be performed with the highest available accuracy for all relevant production channels. 
Figure~\ref{fig:vbflo} exemplifies the corresponding leading order (LO) contribution in the Standard Model for gluon-gluon fusion~(ggF), vector-boson fusion (\VBF) and $Z$/$W$ associated production (\VH). 
\VBF production and \VH production (with the vector boson decaying hadronically) are indistinguishable in their final state and coupling order. 
Theoretical predictions usually refer to these as two different channels, mainly distinguished by phase space regions in which one or the other topology contributes dominantly to the final cross section~\cite{Ciccolini:2007jr,Ciccolini:2007ec,Campanario:2018ppz}. 
These two channels, together with their quantum interference, are the major contributors to the EW $Hjj$ production mode.

Phenomenologically, vector-boson fusion is characterised by topologies where both jets are separated by a large rapidity gap with no significant hadronic activity in the central region~\cite{Rainwater:1997dg,Rainwater:1998kj,Rainwater:1999sd,Plehn:1999nw,Eboli:2000ze}.
The structure function approach is based on such observation to capture the dominant contribution for Higgs boson production in associated with at least two forward jets. 
It includes higher order QCD corrections from initial state hadrons with a color-connected forward jet without any color flow towards the other initial and color-connected final states.  
This approach is exact at LO ~\cite{Cahn:1983ip,Kane:1984bb} and has been extended to next-to-leading order (NLO)~\cite{Han:1992hr,Figy:2003nv,Bolzoni:2010xr} and next-to-next-to-leading order (NNLO)~\cite{Cacciari:2015jma,Cruz-Martinez:2018rod} QCD accuracy with fully differential predictions.
Inclusive QCD corrections are known even with up to next-to-next-to-next-to leading order (N$^3$LO) corrections, and are typically of the order of a few percent, with little change in the kinematic distributions~\cite{Dreyer:2016oyx}. 
Parton shower matched predictions at NLO QCD~\cite{Nason:2009ai,Platzer:2011bc,Frixione:2013mta} have been compared to NNLO QCD calculations for key observables~\cite{Jager:2020hkz}.
Non-factorizable corrections beyond the structure function approach have also been estimated in Refs.~\cite{Liu:2019tuy,Dreyer:2020urf} and were found to not be significant.
The resulting remaining scale uncertainties are of the order of a few percent. 
Higher-order EW corrections to EW $Hjj$ production are known up to NLO~\cite{Ciccolini:2007jr,Ciccolini:2007ec,Denner:2014cla} incorporating both the VBF and VH modes, and their interference. 

In the full Standard Model theory, interference effects between $s$-channel (from \VH production) and $u$-channel topologies at leading order (LO) and $t$-channel color exchange at higher orders might conceivably lead to a partial closure of the central rapidity gap, if no additional fiducial cuts are applied.
A full next-to-leading order (NLO) calculation, including all QCD and EW effects shows that typical \VBF cuts very efficiently eliminate such contributions~\cite{Ciccolini:2007jr,Ciccolini:2007ec}.
Away from \VBF cuts, for a relatively small rapidity gap of the two leading final state jets, \VH production is the dominant channel of EW $Hjj$ production mode.
Testing the EW couplings of the Higgs boson becomes more challenging in those fiducial regions due to the large contribution from ggF channel.
With the same Higgs-plus-two-jet final states,  corrections from the ggF channel up to NLO QCD are calculated in the effective field theory of the infinite heavy top quark in~\cite{Campbell:2006xx,vanDeurzen:2013rv}.
Top quark mass effects were recently studied using an approximation to the full SM theory~\cite{NNLOJETSecDec}.
To estimate the contributions from the ggF channel within or away from \VBF cuts,  Higgs-plus-multi-jet merging effects are studied in~\cite{Andersen:2018tnm}.

The primary goal of this phenomenological study is to investigate the kinematic behavior of the Higgs boson production processes at high Higgs boson $p_T$, \emph{i.e.} in  a regime that has not been fully explored theoretically or experimentally to date, to examine the acceptance for  \VBF production measurements, and to propose new techniques to reduce backgrounds arising from other Higgs boson production modes.%
\footnote{The final state for $t\bar{t}H$ is more complex then for any of the other Higgs boson processes considered in this paper~\cite{ATLAS:2018mme,CMS:2018uxb}. We do not consider the $t\bar{t}H$ process in detail in this study, as we are primarily concerned with the backgrounds posed by gluon-gluon fusion, and by \VH.  } Simple jet counting can be used to improve the separation of  EW $Hjj$ production from gluon-gluon fusion, but a separation of \VBF from \VH becomes more difficult at high $p_T$.
In spite of this difficulty, it is still important to determine the coupling of the Higgs boson to vector bosons at high $p_T$, where new physics has the possibility of affecting both \VH and \VBF production~\cite{Araz:2020zyh}.

We will explore the robustness of the theoretical predictions in the high $p_T$ VBF region by comparing fixed-order results at NLO and NNLO with NLO predictions matched to parton showers (NLOPS), with a particular emphasis on the dependence of the cross sections on the jet radius. This comparison study is similar in spirit to the study performed in the context of the Les Houches 2017 workshop~\cite{Bellm:2019yyh}, which focused on gluon-gluon fusion Higgs boson production in the presence of an additional jet. The current study was initiated during the Les Houches 2019 workshop~\cite{Amoroso:2020lgh}, where the fixed order results for the Higgs transverse momentum distribution were presented. 

This paper is organized as follows. In Section~\ref{sec:setup} we introduce the theoretical inputs and computational tools used in this study, and in Section~\ref{sec:characteristics} we detail the general characteristics of Higgs boson production processes. Section~\ref{sec:transverse} describes the evolution of the Higgs boson kinematics as the Higgs boson transverse momentum increases, including the impact of the jet $R$-dependence, and Section~\ref{sec:discrimination} discusses the use of variables such as the dijet mass and the dijet rapidity separation for discrimination of the \VBF cross section from other Higgs boson production modes. Section~\ref{sec:topology} describes the impact of further event topology cuts on \VBF discrimination and Section~\ref{sec:conesize} examines the impact of the jet size on the Higgs + two or more jets cross sections at both fixed order and with NLOPS predictions. In Section~\ref{sec:tools} we compare the different computational approaches and estimate the uncertainties related to the NLOPS simulations used by experiments. We conclude in Section~\ref{sec:conclusions}.

\section{Theoretical inputs and computational tools}
\label{sec:setup}
To provide theoretical guidelines for experimental analysis and simplified template cross sections (STXS), we adopt the same jet definition as in ATLAS measurements~\cite{Buhrer:2019npn,ATLAS:2019jst} and study differential observables using fiducial bins suggested by STXS stage 1.1~\cite{Berger:2019wnu}. We study on-shell Higgs bosons of $m_{\text{H}}=125$~GeV, produced in the \VBF, \VH, and ggF channel with a center-of-mass energy of 13~TeV. The Higgs boson is required to be within the central region of detectors ($|y_{\text{H}}|<2.4$) in association with at least two accompanying jets. The jets are clustered using the {\tt FastJet} implementation of the anti-$k_T$ algorithm~\cite{Cacciari:2011ma}, and must satisfy the following cuts:
\begin{equation}
p_T^{\text{jet}} > 30\ \text{GeV}, \qquad  |y^{\text{jet}}|<4.4;
\end{equation}
We vary the jet radius $R$ from 0.3 to 1.0 with steps of 0.1 and study the impact on selected differential observables. The ranking of jets is by default according to the magnitude of transverse momentum. When specified, jets are alternatively also ranked by the absolute value of their rapidity. The EW parameters are defined in the G$_\mu$-scheme with G$_\mu=1.1663787\cdot 10^{-5}$ GeV$^{-2}$, and the gauge boson masses and widths set to:
\begin{align}
    m_{W}&=80.379 \text{~GeV}, & \Gamma_{W}&=2.085 \text{~GeV}. \\
    m_{Z}&=91.188 \text{~GeV}, & \Gamma_{Z}&=2.495 \text{~GeV}.
\end{align}
The value of the electroweak coupling constant in this scheme is $\alpha_{\rm EM}^{G_\mu}\approx 1/132.233$.

Theoretical uncertainties are estimated by varying QCD renormalization ($\mu_R$) and factorization ($\mu_F$) scales independently by a factor of two around the central scale of $\mu=H_T^{\text{parton}}/2$ defined uniformly for all processes at the parton level, with
\begin{equation}\label{eq:murf}
   H_T^{\text{parton}} = \sqrt{m^2_{\text{H}}+p_{T,\text{H}}^2} + \sum_{i \in \text{partons}} p_{T, i}\;.
\end{equation}
The sum over partons in the second term considers any final state QCD partons including from real radiation and from the $V\to jj$ decay in the \VH process.

We use 7-point scale variation, eliminating the two antipodal combinations $(\mu_R,\mu_F)=(\mu/2,2\mu)$ and $(\mu_R,\mu_F)=(2\mu,\mu/2)$. For the choice of parton distribution functions (PDFs) for initial states, we use the central group of PDF4LHC15\_30 NNLO PDFs~\cite{Butterworth:2015oua} for each calculation.
Predictions have been generated with \NNLOJET, \Sherpa, \PWG (matched to \PY and \HW) and \HW standalone. The detailed setup for each of these programs is listed below. Fixed-order results are nominally computed using \NNLOJET, and all other simulations have been cross-checked against \NNLOJET up to NLO precision.
NLOPS matched simulations presented here include parton-shower effects and simulations of beam remnant dynamics, but no hadronization or multi-parton interactions.
Analyses are performed with Rivet~\cite{Buckley:2010ar,Bierlich:2019rhm}.

In order to make the results comparable to the fixed-order prediction, in all NLOPS simulations we set the renormalization, factorization and resummation scale to $\mu=H_T^{\rm parton}/2$.
The running coupling used is chosen consistent with the PDF, and 
the CMW scheme~\cite{Catani:1990rr} is used in the shower algorithm (and in the POWHEG Sudakov form factor) to include the two-loop cusp anomalous dimension in the shower algorithms, \emph{i.e.} with $\alpha_s(M_Z)\approx0.126$, with the exception of the \PY shower, which uses by default one-loop running with $\alpha_s(M_Z)=0.137$.

In the gluon-gluon fusion production channel, the Higgs boson is created via a top-quark loop in association with two additional jets. Theoretical predictions for this configuration are typically computed at NLO in the Higgs effective field theory (HEFT). However, with kinematic cuts favoring large dijet invariant mass and rapidity separation, our analysis involves a wide range of energy scales, such that finite top quark mass effects can play an important role, and the HEFT results may be strongly modified. The full NLO QCD results for Higgs plus one-jet production in the Standard Model have become available only recently~\cite{Lindert:2018iug,Jones:2018hbb}, and a similar computation for Higgs-plus two jet production is not yet on the horizon. The best available technique to include top-quark mass effects in our predictions for gluon fusion would be a local reweighting strategy~\cite{Buschmann:2014sia,Greiner:2016awe,NNLOJETSecDec}. In this method, the ratio between Standard Model and effective theory results is computed at LO and applied to the HEFT prediction as a local multiplicative correction. For a wide range of differential observables, the results are in excellent agreement with a posteriori reweighting at the histogram level~\cite{Chen:2016zka,Becker:2020rjp}, which motivates us to choose the yet simpler histogram-based reweighting method to include top-quark mass effects in our calculations. For a given observable $\mathcal{O}$ from the gluon-gluon fusion production channel, the Standard Model and HEFT predictions of differential cross sections at the same fixed order is used to define the  reweighting coefficient:
\begin{equation}
    R_{\text{FO}}(\mathcal{O})=\frac{d \sigma^{\text{SM}}_{\text{FO}}}{d\mathcal{O}}\bigg/\frac{d \sigma^{\text{HEFT}}_{\text{FO}}}{d\mathcal{O}}.
\end{equation}
The approximated Standard Model (approx SM) predictions including higher order corrections (noted as FO') from HEFT and the heavy quark mass effect from the Standard Model are defined at histogram level as
\begin{equation}
\frac{d \sigma^{\text{approx SM}}_{\text{FO'}}}{d\mathcal{O}} = R_{\text{FO}}(\mathcal{O})\frac{d \sigma^{\text{HEFT}}_{\text{FO'}}}{d\mathcal{O}}.
\label{eq:approxSMdef}
\end{equation}
The state-of-the-art accuracy of the reweighting factor is at NLO for Higgs transverse momentum distribution, $R_{\text{NLO}}(p_{T,H})$. A recent study of finite top mass effect for a wide range of observables for Higgs-plus one and two jet production proves the reliability of multiplicative reweighting~\cite{NNLOJETSecDec}. In the following discussion, if not specifically stated, our approximated Standard Model predictions include corrections from $R_{\text{LO}}(\mathcal{O})$.

\subsection{\NNLOJET}
\label{sec:setup_nnlojet}
Parton level fixed-order predictions for the \VBF channel are calculated using the \protect\NNLOJET\ package, including QCD corrections up to NNLO~\cite{Cruz-Martinez:2018rod} in the structure function approximation~\cite{Han:1992hr}, i.e.\ omitting non-factorizable corrections at NLO and at NNLO. 
The antenna subtraction formalism~\cite{GehrmannDeRidder:2005cm,GehrmannDeRidder:2005aw,Daleo:2006xa,Daleo:2009yj,GehrmannDeRidder:2012ja,Currie:2013vh} is used to regulate IR divergences at each stage of the fixed-order calculations and to provide fully differential results.
The \NNLOJET\ results agree with an independent calculation~\cite{Cacciari:2015jma} of \VBF at NNLO QCD, which employed the same approximations. A recent study~\cite{Liu:2019tuy} using the eikonal approximation estimates the non-factorizable corrections to be less than 2\% (with respect to LO) for differential observables.
\protect\NNLOJET also provides predictions for the gluon-gluon fusion production channel and is used to compute the histogram-based reweighting factors needed to include approximate top-quark mass effects in the NLO matched results.

\subsection{POWHEG BOX}
The \PWG{}~\cite{Frixione:2007vw,Alioli:2010xd} is a general framework for implementing NLO calculations that can be matched to parton showers using the POWHEG method~\cite{Nason:2004rx}.
Higgs production in \VBF is implemented in the {\tt VBF\_H} code~\cite{Nason:2009ai}.

In this study we also consider the Higgsstrahlung background process, which is simulated using the {\tt ZH} and {\tt WH} generators~\cite{Luisoni:2013cuh}\footnote{A more recent version of the {\tt HW} and {\tt HZ} generators (with leptonic $W/Z$ decay), which also includes EW corrections, is implemented in the {\tt POWHEG BOX RES} framework~\cite{Granata:2017iod}.}, that have been modified to perform the hadronic decay of the $Z$ and of the $W$ boson.
To get the correct matrix elements for the leading order hadronic $V=Z,W$ decay, we include the correct color factor $N_c$ in the squared amplitudes and we modify the vector and axial couplings in the decay if $V$ is the $Z$ boson.
Due to the fact that $V$ is a colorless object,
there is no interference between NLO QCD corrections in production, which are already implemented, and those in decay, so once the integral over the radiation phase space is performed, the additional $\mathcal{O}(\alpha)$ contribution  to the cross section, differential in the kinematics of the underlying Born configuration $\Phi_b$, is
\begin{equation}
d\sigma_{\rm NLO dec}(\Phi_b) = \frac{\alpha_s}{\pi} d\sigma_{\rm LO}(\Phi_b).
\label{eq:NLOdec}
\end{equation}
Thus, we include the term in Eq.~\eqref{eq:NLOdec} in the event weight (\emph{i.e.} in the $\tilde{B}$ function in POWHEG language) so that we reproduce the full inclusive (respect to additional QCD radiation) NLO calculation.
Since we do not modify the matrix elements for the real corrections, only initial-state radiation from the production process is generated.
However, once we match our generator to the \HW{} angular-ordered parton shower (or to the \PY{} shower), that contains matrix element corrections~(MEC) in the $V$ decay, also the hardest emission in decay is generated using the exact real matrix element, rather than the approximate shower kernel.
Thus, to have an ``effective" \PWG{} NLOPS generator for \VH with hadronic decay, it is sufficient to include the contribution in Eq.~\eqref{eq:NLOdec} in the weight of the event, which must be showered with a parton shower that implements matrix element corrections: in this way, the event weight is identical to the NLO cross section differential in the undelying Born kinematics and the hardest emission from the production process and from the $V$ decay is described using the exact matrix element.

Events produced with \PWG{} are then matched to the \PY default shower~\cite{Sjostrand:2014zea} and to the \HW angular-ordered parton shower~\cite{Bahr:2008pv,Gieseke:2003rz,Bellm:2019zci}.\footnote{We use version \PY version 8.2 and \HW version 7.2.}
As discussed in Ref.~\cite{Jager:2020hkz}, the default \PY global recoil for initial state radiation is not adequate for vector-boson fusion processes, and thus we employ the local dipole recoil~\cite{Cabouat:2017rzi}.

We rely only on the {\tt bornzerodamp}\footnote{If the real matrix element differs by more than a factor 5 compared to its collinear approximation, the contribution is considered non-singular.} mechanism to separate the real contribution into singular and non-singular parts, and we use {\tt hdamp}$=\infty$.
 The \VH{} process is quite sensitive to this choice, thus we also generated a sample with {\tt hdamp}$=m_{\text H}$, \emph{i.e.} considering a real emission as singular with probability ${\tt hdamp}^2/(p_T^2+{\tt hdamp}^2)$ . We use the running of the strong coupling provided by LHAPDF~\cite{Buckley:2014ana} and when computing the renormalization and factorization scales for the real correction we use the 4-body kinematics of the corresponding underlying Born configuration.
 The infrared cutoff for the first emission transverse momentum has been set to 1~GeV, which corresponds to the maximum between the \PY{} and \HW{} shower cutoffs.

\subsection{Herwig 7}
The \HW{} event generator\footnote{We use version 7.2.}
\cite{Bahr:2008pv,Bellm:2019zci} is using the \texttt{Matchbox} module \cite{Platzer:2011bc},
to automatically assemble fixed-order and parton shower matched
calculations. It can match both to the angular ordered \cite{Gieseke:2003rz} and
dipole shower algorithms \cite{Platzer:2009jq}, using input from
plugins providing matrix elements. The \VBFNLO{} program
\cite{Arnold:2008rz,Baglio:2014uba} serves as one such module,
providing NLO-QCD corrections to the $Hjj$ and $Hjjj$ production
processes in the \VBF approximation. The \HJets{} library
\cite{Campanario:2013fsa} is an alternative module. It provides matrix
elements and NLO-QCD corrections for the full EW $Hjj$
production process without resorting to the \VBF approximation.

For the \HW{} standalone samples we adopt the subtractive matching paradigm;
a hard veto scale cuts off the shower evolution at high transverse momenta. Its
central value should reflect the momentum transfer in the
hard process of interest, such that the shower evolution will not produce
jets with significantly harder transverse momenta. A smearing is
applied to the cutoff function. In this study, we choose the
so-called ``resummation'' profile studied in more detail in
\cite{Bellm:2016rhh,Cormier:2018tog}. Shower uncertainties are
evaluated by varying the hard veto scale. This should reflect the
bulk of the uncertainty both in the soft region and in regions which
will be improved through the NLO matching.

\subsection{\Sherpa}
We use a pre-release version of the Sherpa Monte
Carlo event generator~\cite{Gleisberg:2003xi,Gleisberg:2008ta,Bothmann:2019yzt},
based on version 3.0.0. The NLO matching is performed in the S-MC@NLO
approach~\cite{Hoeche:2011fd,Hoeche:2012ft}, and NLO computations are performed with
the help of Amegic~\cite{Krauss:2001iv,Gleisberg:2007md}, Comix~\cite{Gleisberg:2008fv}
and OpenLoops~\cite{Cascioli:2011va,Buccioni:2019sur}, as well as matrix elements from MCFM~\cite{Campbell:2019dru}.
For the \VBF and \VH simulations we have implemented dedicated loop matrix elements
in the structure function approximation within Sherpa,
such as to speed up event generation for this study.
We use a modified version of a parton shower algorithm~\cite{Schumann:2007mg}
based on Catani-Seymour dipole subtraction~\cite{Catani:1996vz,Catani:2002hc},
as well as the Dire parton shower~\cite{Hoche:2015sya}.

\section{General characteristics of gluon-gluon fusion,  \VBF and \VH production}
\label{sec:characteristics}
Higgs boson production in association with two jets can proceed through three major channels: gluon-gluon fusion~(ggF), vector-boson fusion (\VBF) and
$Z$/$W$ associated production (\VH). 
Unlike gluon-gluon fusion, \VBF and \VH production result in two jets in the final state at the Born level, where both jets are quark jets.
While the gluon-gluon fusion channel dominates the inclusive Higgs production cross section, imposing a two-jet requirement reduces its impact substantially.
The \VBF and \VH contributions are finally made measurable through suitable kinematical cuts.
A good discrimination between \VBF and the other production channels can then typically be achieved by the requirement that the tagging jets 
be widely separated in rapidity and/or have a large dijet invariant mass. This kinematic configuration is preferred by the $t$-channel electroweak
topology of the VBF process, while both gluon-gluon fusion and \VH, due to their $s$-channel nature, typically generate jets that have a small separation
in phase space. At higher perturbative orders the situation is largely unchanged, because color exchange effects between the two quark lines
in \VBF are highly suppressed. An additional discrimination against gluon-gluon fusion can be accomplished by means of a veto on additional jets. 

As there can be more than two jets in the final state, the two highest $p_T$ jets are typically chosen to tag the \VBF process. Using the two jets with the largest rapidity separation may, in principle, result in a more powerful discrimination of \VBF from the other Higgs boson processes, but this selection has not typically been used due to the difficulties in dealing with relatively low $p_T$ jets measured at forward rapidities, especially in the presence of large pileup. However, both ATLAS~\cite{CERN-LHCC-2020-007} and CMS~\cite{CMS:2667167} will have forward timing detectors in place for the HL-LHC, which should allow for a better handling of forward jets and make such measurements possible in the future.

As both \VBF and \VH production are $q\bar{q}^\prime$ initiated and can yield the same final state, there can be quantum-interference between the two. This interference, however, is suppressed by $1/N_c$ at leading order, with $N_c$ the number of colors. It is further kinematically suppressed, because the $t$-channel propagators in \VBF type diagrams and the $s$-channel propagators in \VH type diagrams cannot simultaneously go on-shell. Thus, electroweak $Hjj$ production is expected to be simply the incoherent sum of the two individual processes.

\begin{figure}[t!]
\centering
\includegraphics[width=.45\textwidth]{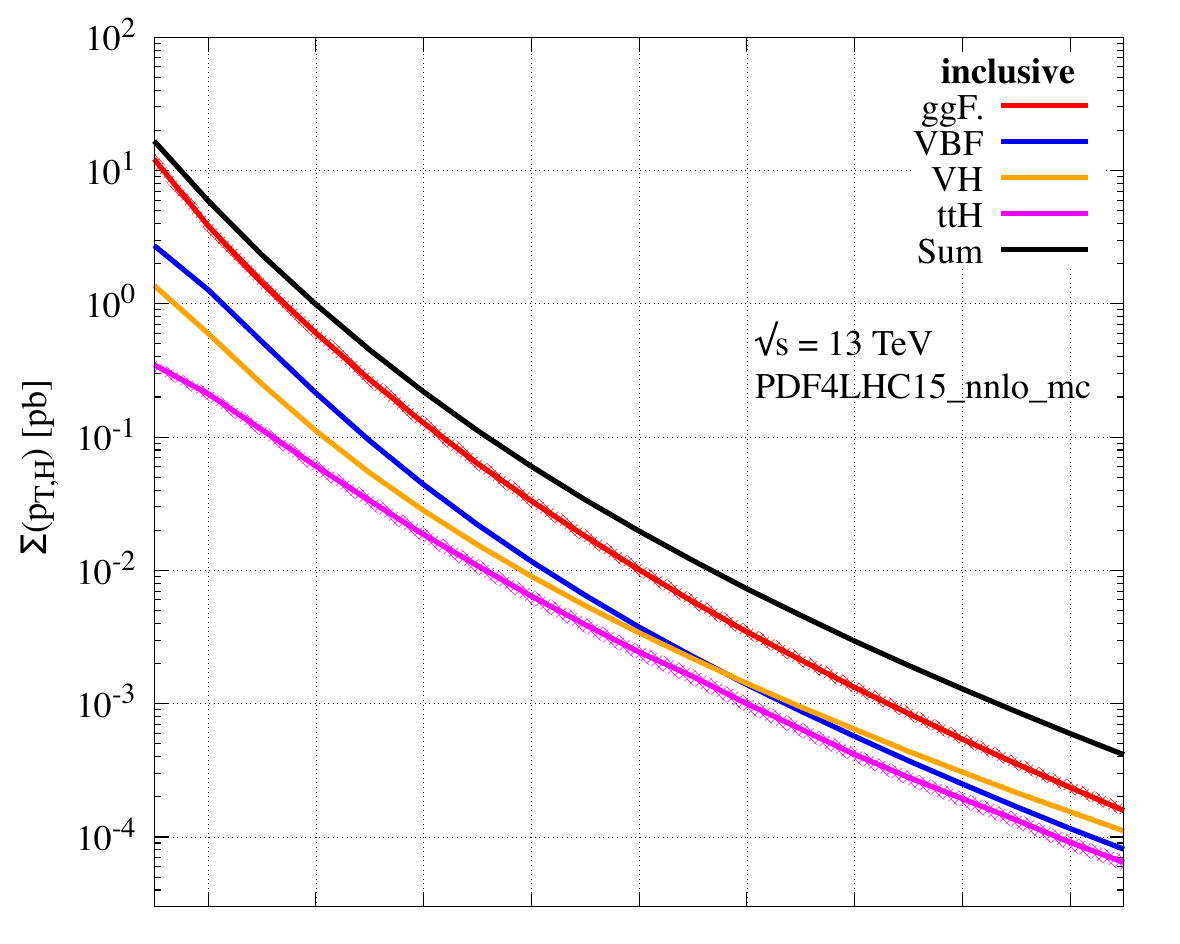}
\includegraphics[width=.45\textwidth]{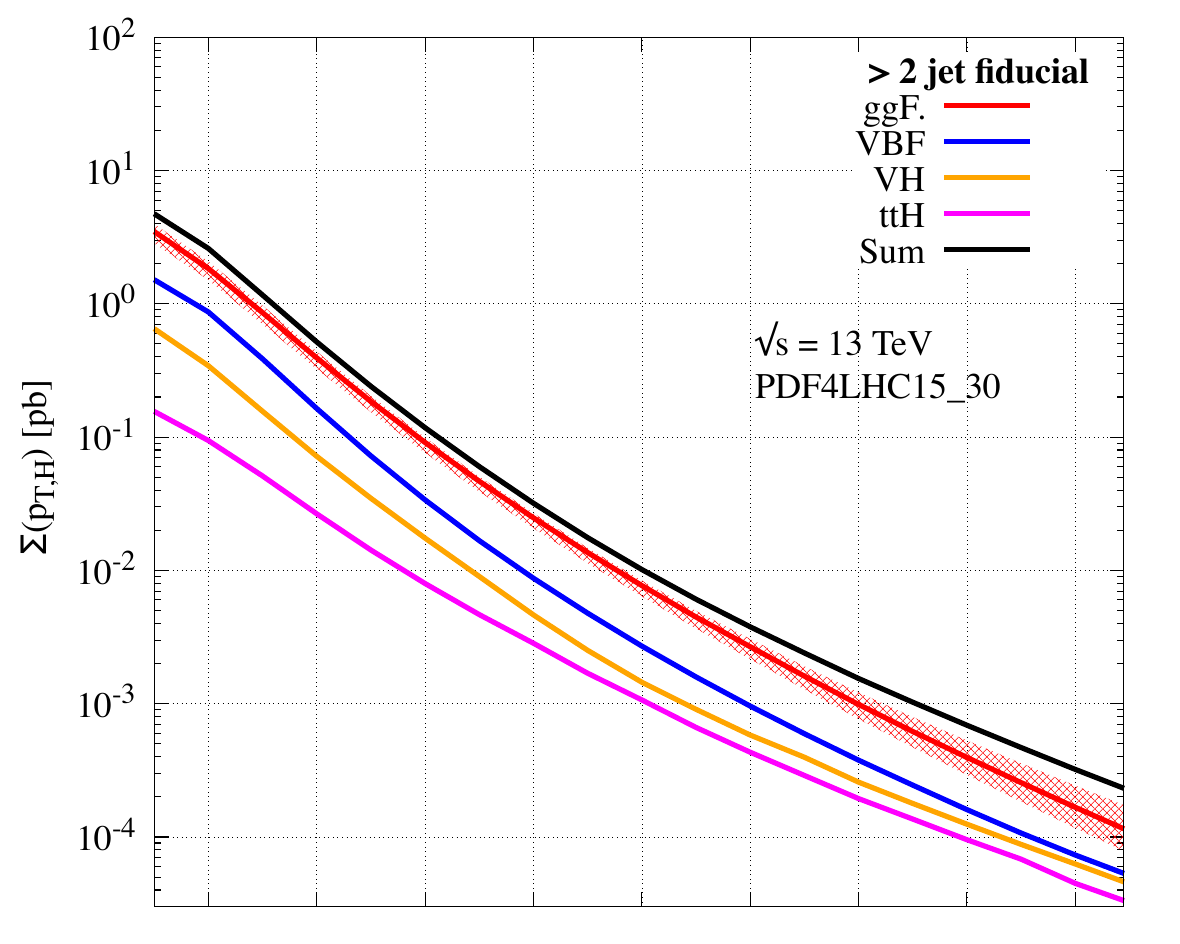}
\\\vspace{-1.1em}
\includegraphics[width=.45\textwidth]{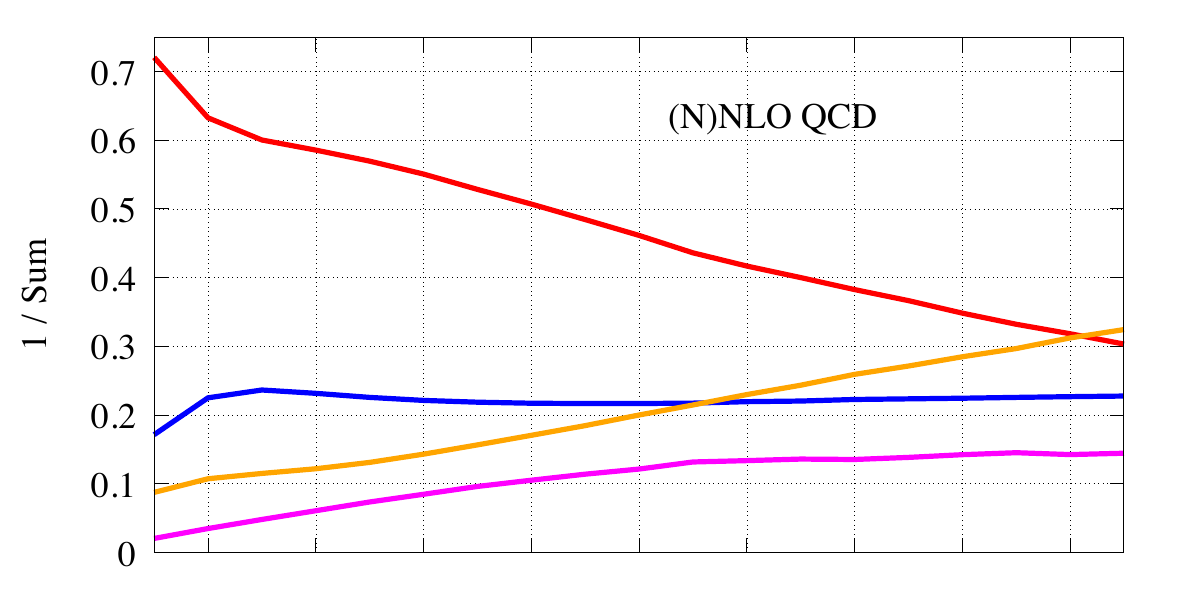}
\includegraphics[width=.45\textwidth]{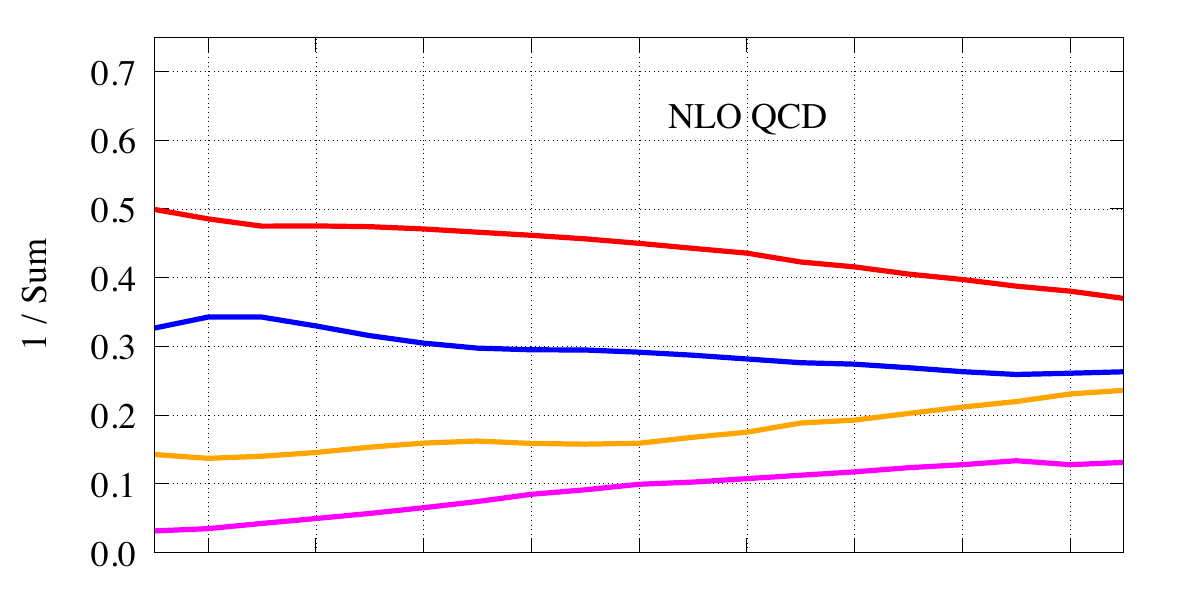}
\\\vspace{-1.1em}
\includegraphics[width=.45\textwidth]{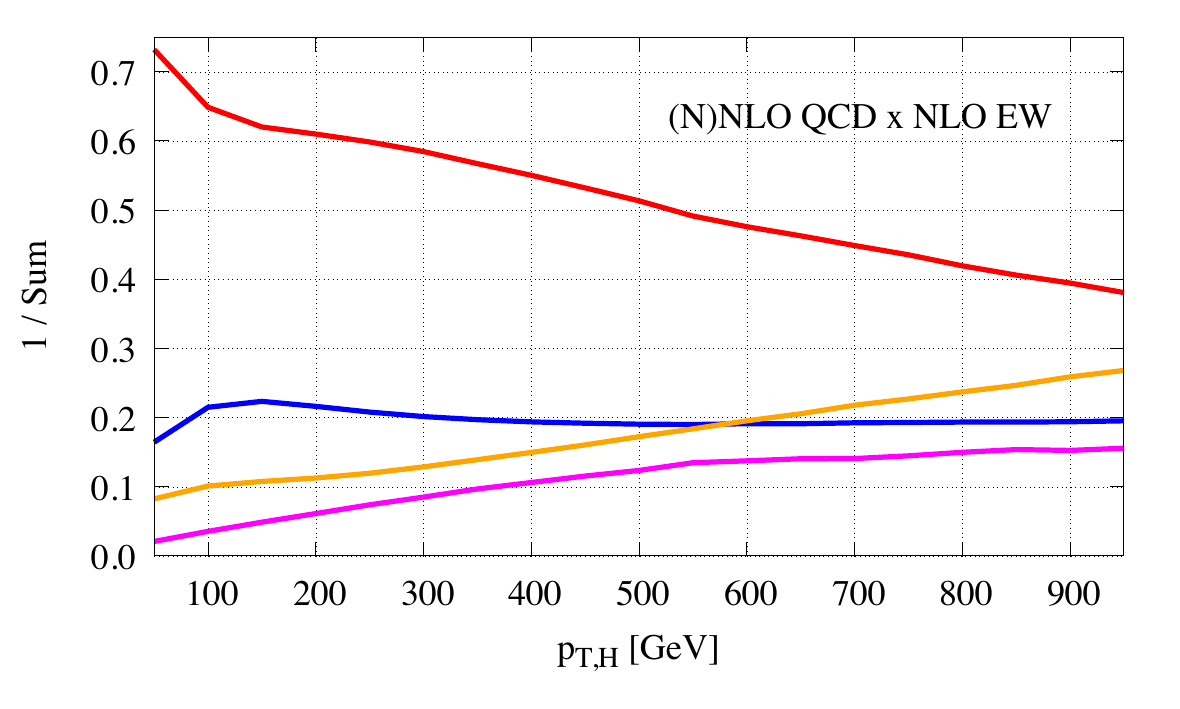}
\includegraphics[width=.45\textwidth]{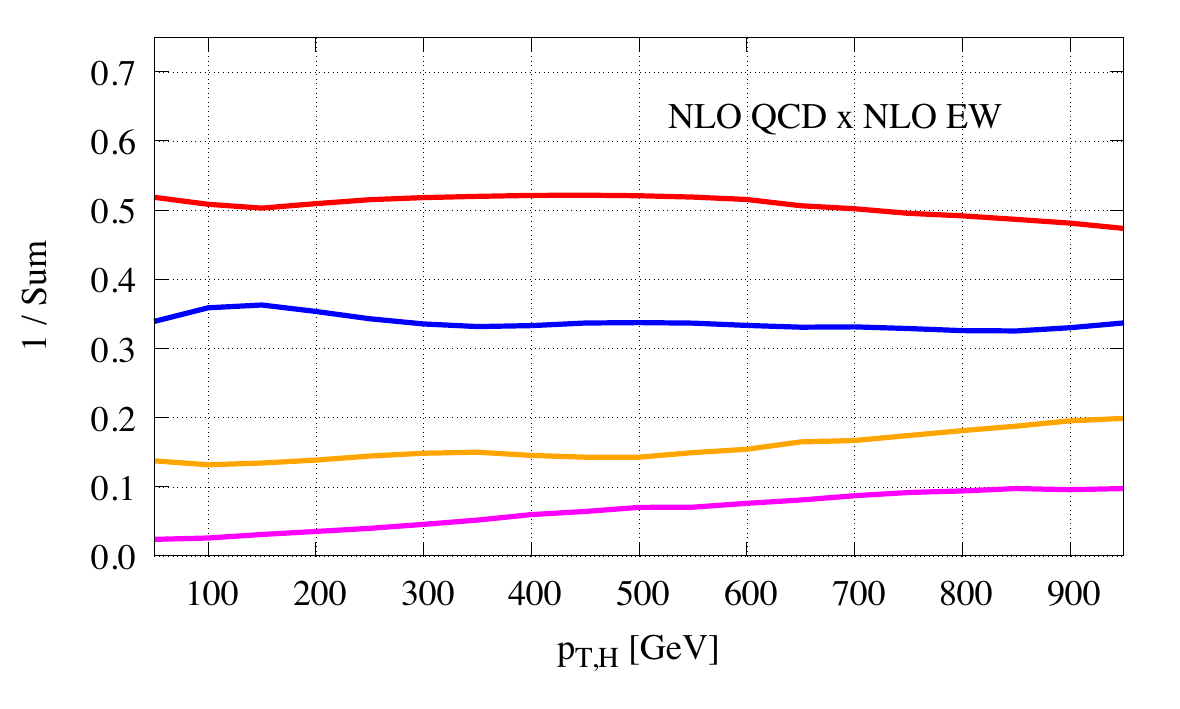}
\caption{Cumulative Higgs transverse momentum distributions for the ggF, \VBF, \VH and $t\bar{t}$H modes. The relative yield with respect to the sum of all shown modes is plotted in the lower ratio plots. On the left, the fully inclusive spectrum is shown with gluon-gluon fusion at NNLO QCD (combined with NLO mass effects), \VBF at NNLO QCD $\times$ NLO EW, \VH at NLO QCD $\times$ NLO EW and $t\bar{t}$H at NLO QCD $\times$ NLO EW, as provided in~\cite{Becker:2020rjp}. The upper ratio plot shows the relative yield including only higher-order QCD corrections, while the lower ratio plot combines QCD and EW corrections. 
On the right the
distribution is shown in the two-jet fiducial phase-space and using the setup considered in our analysis (at least two anti-$k_T$ jets with $p_T^{\text{jet}} > 30\ \text{GeV},  |y^{\text{jet}}|<4.4$). Here ggF(+2 jets) is shown at NLO~QCD (combined with LO mass effects), and \VBF, $V(\to jj)H$ and $t\bar{t}H$ (all-hadronic) at NLO QCD $\times$ NLO EW. See text for details.
}
\label{fig:rel_fraction}
\end{figure}
Figure~\ref{fig:rel_fraction} (left) shows in the lower panels the relative fraction in the cumulant inclusive spectrum
$\Sigma(p_T) \equiv \int_{p_T}^\infty \mathrm{d}p_{T,H} (\mathrm{d}\sigma / \mathrm{d}p_{T,H})$
of the different Higgs boson processes as a function of the Higgs boson transverse momentum. The upper ratio plot includes all processes including only higher-order QCD corrections, while in the lower panel also NLO EW corrections included. The top panel shows the absolute cumulant.
Besides gluon-gluon fusion~(ggF), \VBF and \VH production,  we also show Higgs boson production in association with a top-quark pair (\ttH) for reference. The setup for all these predictions follows~\cite{Becker:2020rjp} and fully inclusive on-shell processes are considered.
 The ggF prediction combines NNLO QCD~\cite{Chen:2014gva,Chen:2016zka} with NLO mass effects~\cite{Lindert:2018iug,Jones:2018hbb}, the \VBF prediction is at NNLO QCD~\cite{Cruz-Martinez:2018rod} combined with NLO EW, the \VH prediction is at NLO QCD combined with NLO EW, as is the \ttH prediction.
The NNLO predictions are computed with \NNLOJET (see above), while NLO mass effects to the ggF process are obtained from Ref.~\cite{Jones:2018hbb} as pT dependent multiplicative correction factors (see Ref.~\cite{Becker:2020rjp} for details). The NLO QCD corrections to \VH are computed with POWHEG-BOX~\cite{Luisoni:2013cuh}, while the NLO QCD corrections to \ttH and all the considered NLO EW corrections are computed with Sherpa+OpenLoops~\cite{Gleisberg:2008ta,Kallweit:2014xda,Schonherr:2017qcj,Buccioni:2019sur} (setup as defined in Ref.~\cite{Becker:2020rjp}). 
For the ggF process no EW corrections are considered, as these are not known at finite $p_T$. In any case, they are expected to be small (at the percent level), as there are no soft-collinear EW mass singular enhanced contributions. 
For the other processes the EW corrections are combined multiplicatively with the QCD predictions. This is motivated by the fact that the EW corrections at large $p_T$ are dominated by EW Sudakov logarithms, which factorize with respect to the QCD corrections. 
The gluon-gluon fusion process dominates for the transverse momentum range displayed, while the fraction of \VBF is roughly flat. The impact of \VH increases substantially for large $p_T$, due both to the $s$-channel nature of the process, and the enhancement from valence quark PDFs. \ttH yields the smallest contribution, however its relative contribution also increases at large $p_T$.
The EW corrections decrease substantially the relative yield of the \VH and \VBF processes at finite $p_T$.

In Fig.~\ref{fig:rel_fraction} (right), a requirement is made that the event contain at least two jets, using the fiducial phase-space considered in our analysis (at least two anti-$k_T$ jets with $p_T^{\text{jet}} > 30\ \text{GeV},  |y^{\text{jet}}|<4.4$), aiming to reduce the fraction from gluon-gluon fusion.
Here we consider ggF(+2 jets) at NLO~QCD (combined with LO mass effects), and \VBF, $V(\to jj)H$ and $t\bar{t}H$ (all-hadronic) at NLO QCD $\times$ NLO EW.
The setup for the ggF(+2 jets), \VBF and $V(\to jj)H$ NLO QCD predictions are detailed in Section~\ref{sec:setup}, while the $t\bar{t}H$ (all-hadronic) NLO QCD prediction is obtained with Sherpa, considering LO on-shell hadronic top decays. The  relative EW corrections are taken identical to the inclusive case, as their impact is driven by the kinematics of the hard scattering process. 
The gluon-gluon fusion relative yield remains rather flat across the whole range, as does the \VBF{} relative fraction.
Associated VH production, on the other hand,  increases with Higgs boson $p_T$,
however the increase is significantly milder than in the inclusive case.
  In fact, at very large $p_T$ the two jets from the vector boson decay in \VH are highly boosted and might regularly be merged into a single jet depending on the jet cone size. The two-jet requirement is then only fulfilled due to higher-order radiation. The \ttH process contributes up to 10\% at very large $p_T$. However, its impact is further suppressed via additional VBF-type selections (in particular requiring large dijet invariant masses) and/or b-jet

\begin{table}[t]
    \centering
    \bgroup
    \def\arraystretch{1.15}
    \setlength\tabcolsep{2.5mm}
    \begin{tabular}{l|rrrrrrrrrr}
        $\sigma$ [fb] & \multicolumn{2}{c}{$p_{T,h}\ge0$~GeV} & \multicolumn{2}{c}{$p_{T,h}\ge200$~GeV} & \multicolumn{2}{c}{$p_{T,h}\ge300$~GeV} & \multicolumn{2}{c}{$p_{T,h}\ge400$~GeV} & \multicolumn{2}{c}{$p_{T,h}\ge500$~GeV} \\
        & 2j incl. & 2j excl. & 2j incl. & 2j excl. & 2j incl. & 2j excl. & 2j incl. & 2j excl. & 2j incl. & 2j excl.\\\hline
        EW H+2j & 2565 & 2087 & 259 & 179 & 58.8 & 37.9 & 16.0 & 9.78 & 5.34 & 3.15 \\
        VBF+VH & 2555 & 2069 & 258 & 177 & 58.7 & 37.4 & 16.1 & 9.70 & 5.38 & 3.14 \\
        VBF & 1859 & 1586 & 183 & 134 & 39.9 & 27.2 & 11.0 & 7.14 & 3.61 & 2.27 \\
        VH & 696 & 483 & 74.8 & 42.7 & 18.8 & 10.2 & 5.09 & 2.56 & 1.77 & 0.87 \\
        ggF approx SM & 3219 & 2227 & 305 & 134 & 70.6 & 26.2 & 19.0 & 6.45 & 5.96 & 1.85
    \end{tabular}
    \egroup
    \caption{Expected cross sections at NLO QCD for \VBF, \VH and gluon-gluon fusion production of a Higgs boson with two or more jets, for different Higgs boson transverse momentum cuts and at $R=0.4$. }
    \label{tab:incl_njets_fraction}
\end{table}

In Table I, we list the cross sections for gluon-gluon fusion, \VBF, \VH, their incoherent sum and the full electroweak Higgs plus two-jet production process. The latter is labeled EW~H+2j in the following, and is obtained by computing all contributions to Higgs plus two jet final states at $\mathcal{O}(\text{EW})=3$ and at NLO QCD. It is well matched by the incoherent sum of \VBF and \VH, indicating the anticipated smallness of interference terms.
Table~\ref{tab:incl_njets_fraction} lists five different Higgs boson transverse momentum cuts and both inclusive and exclusive two-jet cross sections. An exclusive two-jet requirement significantly reduces the contribution from gluon-gluon fusion at high $p_T$, while maintaining essentially the same relative fraction of \VH and \VBF. A veto on the production of extra jets reduces the \VBF (\VH) cross section available for analysis, and also results in a reduced effective precision of the related theoretical predictions, necessitating the need for resummation of logarithms associated with the veto scale. Such logarithms can be described only approximately by parton showers, such that a large theoretical uncertainty remains in principle. However, we emphasize that fully automated tools are available~\cite{Gerwick:2014gya,Baberuxki:2019ifp,Baron:2020xoi,Caletti:2021oor}, which allow to compute NLL resummed predictions if the jet multiplicity requirement is global. We will discuss this point in more detail below.

\begin{figure}[p]
  \centering
  \begin{minipage}{.45\textwidth}
    \includegraphics[width=\textwidth]{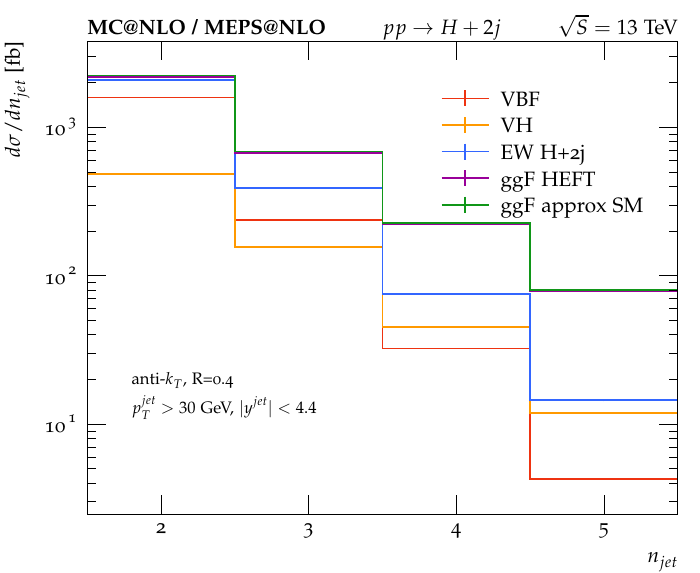}
    \includegraphics[width=\textwidth]{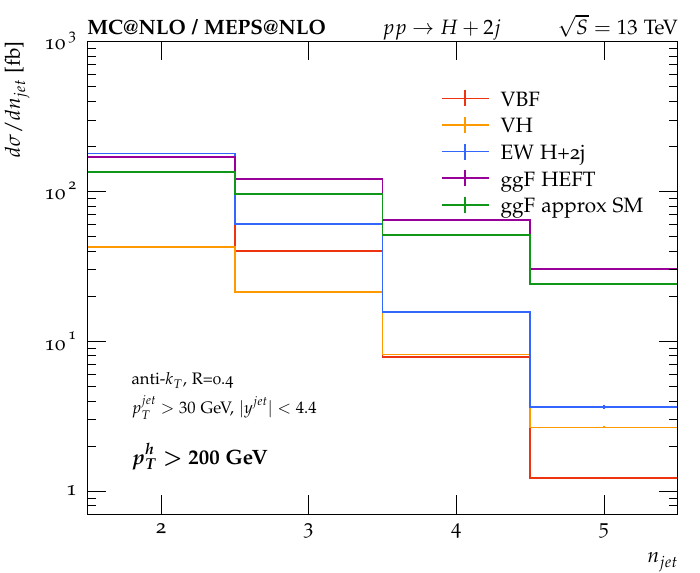}
    \includegraphics[width=\textwidth]{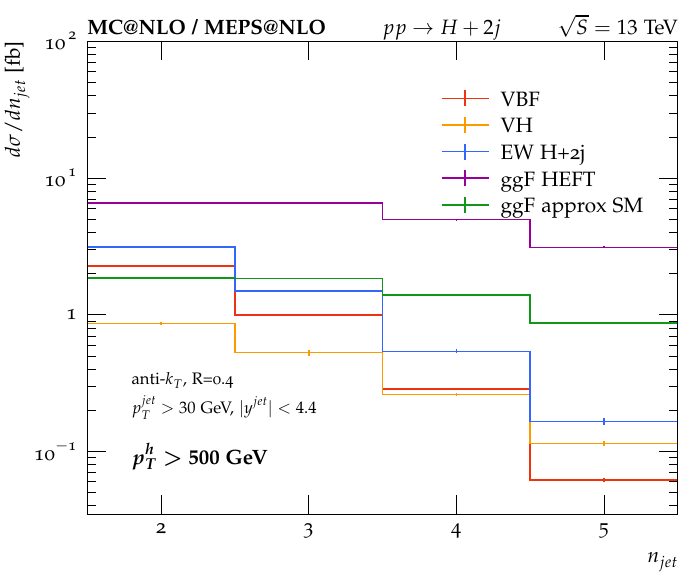}
  \end{minipage}\hfill
  \begin{minipage}{.45\textwidth}
    \includegraphics[width=\textwidth]{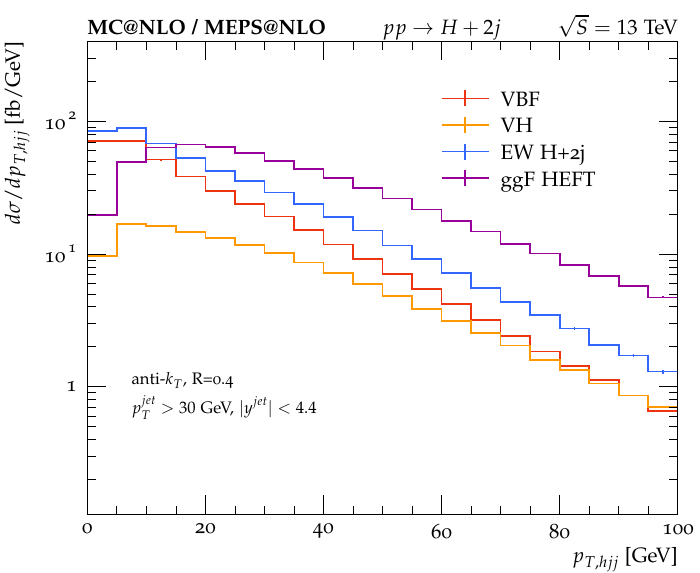}
    \includegraphics[width=\textwidth]{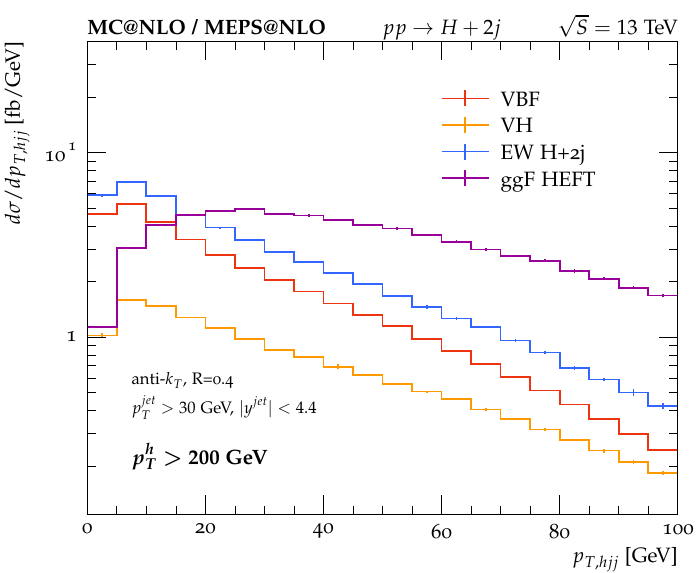}
    \includegraphics[width=\textwidth]{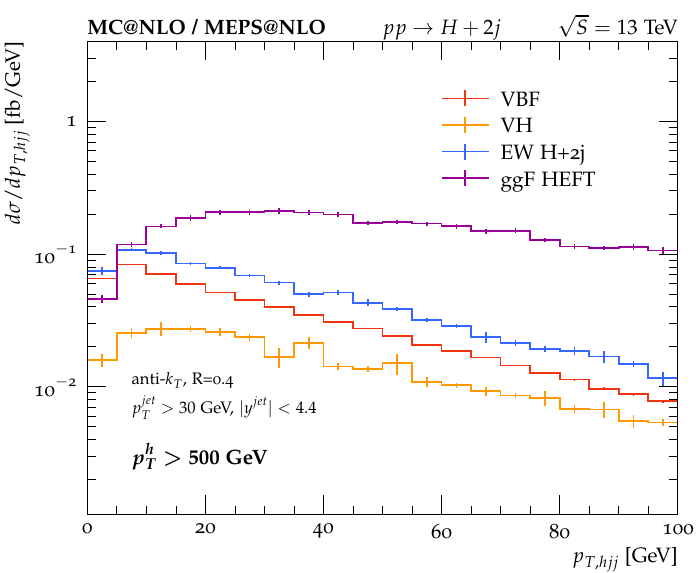}
  \end{minipage}
\caption{Distribution of the exclusive number of jets (left) and the transverse momentum of the Higgs-dijet pair (right).
We compare predictions from S-MC@NLO for VBF, VH and EW H+2j production to the irreducible background
from gluon fusion, computed in HEFT. See the main text for details.}
\label{fig:incl_njets}
\end{figure}
Figure~\ref{fig:incl_njets} (left) shows the exclusive jet multiplicity distribution for \VBF, \VH and $Hjj$ production (the sum of \VBF and \VH plus their interference). All distributions are computed using the S-MC@NLO technique inside the \Sherpa{} framework.
Predictions for the \VBF and \VH channels have been cross checked against {\tt Powheg+Herwig7} simulations, while the inclusive EW production mode
has been validated against \HW{} standalone predictions.
They are compared to the irreducible background from gluon-gluon fusion, which is computed using the MEPS@NLO merging method~\cite{Gehrmann:2012yg,Hoeche:2012yf} with up to one jet at NLO and up to four jets at LO accuracy. The gluon-gluon fusion contributions are shown in both the HEFT and approximate Standard Model, using the reweighting technique described in Sec.~\ref{sec:setup}.  We have verified that this prediction remains stable when adding an additional jet at NLO accuracy; hence we choose the computationally simpler setup in order to reduce the MC uncertainties. At least two jets are required at the analysis level in all simulations to allow for a comparison on equal footing.
The ratio of \VH to \VBF increases with jet multiplicity due to the larger phase space for initial-state radiation in \VH.

\begin{figure}[p]
  \centering
  \includegraphics[width=.6\textwidth]{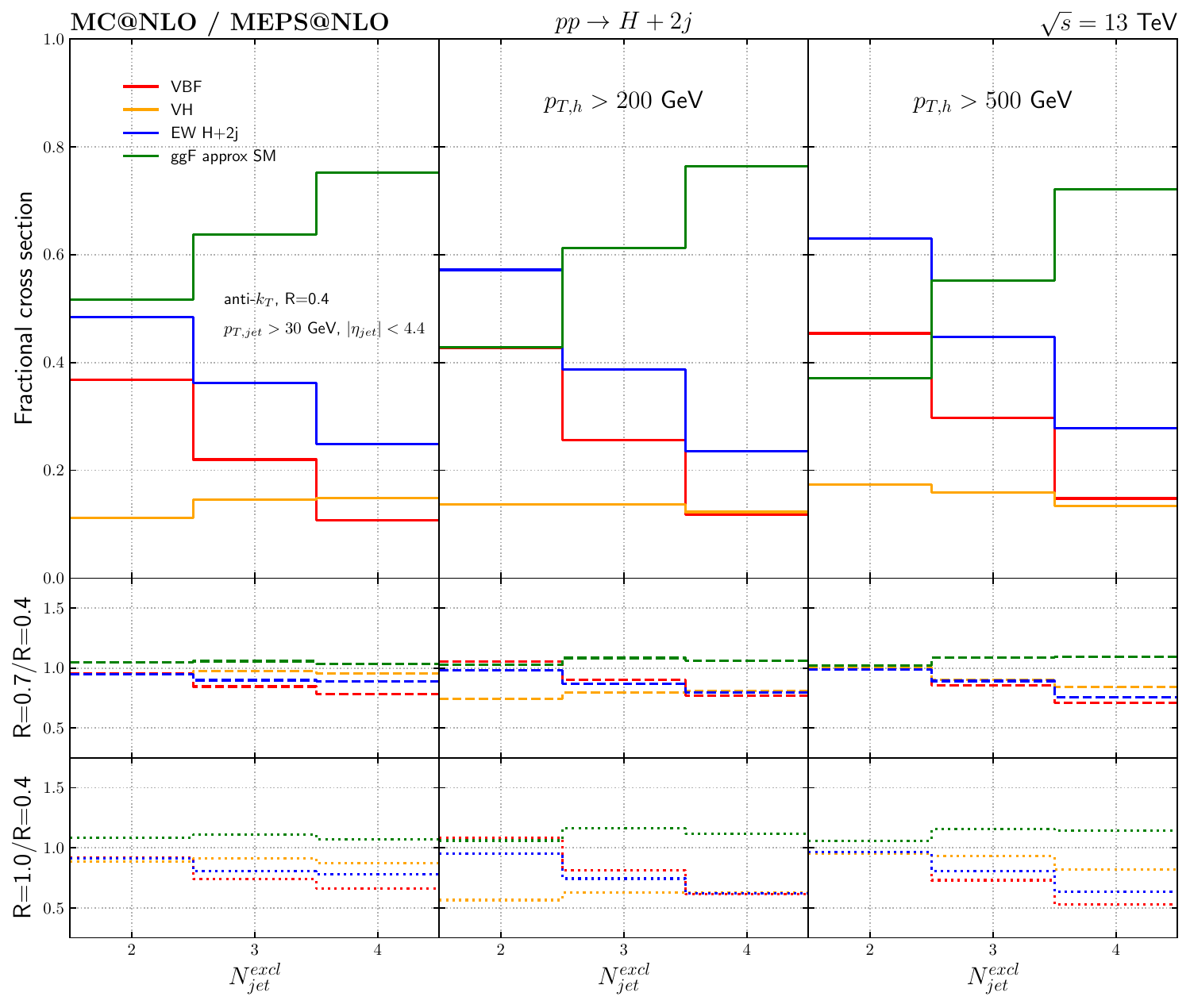}
\caption{Fractional cross sections of the various production channels as a function of the exclusive jet multiplicity. }
\label{fig:incl_njets_fraction}
\end{figure}

The right panels of Fig.~\ref{fig:incl_njets} show the transverse momentum distribution of the Higgs plus dijet pair. This observable is infrared unsafe and can therefore not be described reliably by fixed-order calculations. We observe a significant difference between the \VBF production mode and the \VH and gluon fusion channels, with the latter producing harder radiation due to the $s$-channel color exchange. In comparison to \VH, the gluon fusion process shows again more QCD activity, which can be ascribed to the larger color charges in the $gg$ initial state. Note that the gluon-fusion results are computed in the effective theory. For an infrared unsafe observable we cannot apply a histogram-based reweighting factor, but on physics grounds we expect the following behavior for the full Standard Model results: At small Higgs boson transverse momentum the shape of the distribution should be affected, and the differential cross section at high $p_{T, \text{H}jj}$ will be reduced. At large Higgs boson transverse momentum the cross section will be affected but the shape of the distribution should remain mostly the same. 
The larger QCD activity in the gluon fusion process can also be seen in Fig.~\ref{fig:incl_njets_fraction}, which shows the fractional cross section for each exclusive jet multiplicity: each bin
has been obtained dividing the corresponding bin
in the left panel of Fig.~\ref{fig:incl_njets} by the sum of the gluon-fusion (ggF approx SM) and the electroweak H+2 jets production cross section.
We further observe that for larger Higgs boson transverse momentum cuts, the sensitivity on the jet cone size $R$ increases for all three sub-processes as shown in the lower panels of Fig.~\ref{fig:incl_njets}.

The cross-section predictions from gluon-gluon fusion are dominant (relative to $Hjj$) for high jet multiplicities and reduce slightly as the Higgs boson $p_T$ is increased.
This is also seen in Table~\ref{tab:incl_njets_fraction} where the cross sections for \VBF, \VH (and their coherent sum), and gluon-gluon fusion are shown for five different Higgs boson transverse momentum cuts and for two or more jets in the final state, or exactly two jets in the final state.
For the highest Higgs boson $p_T$ cut, the gluon-gluon fusion background to $Hjj$ becomes smaller than the electroweak $Hjj$ production in the exclusive two-jet final state, while the efficiency of the electroweak $Hjj$ event selection is about 40\%.

\begin{figure}[p]
\centering
  \begin{minipage}{.295\textwidth}
    \includegraphics[width=\textwidth]{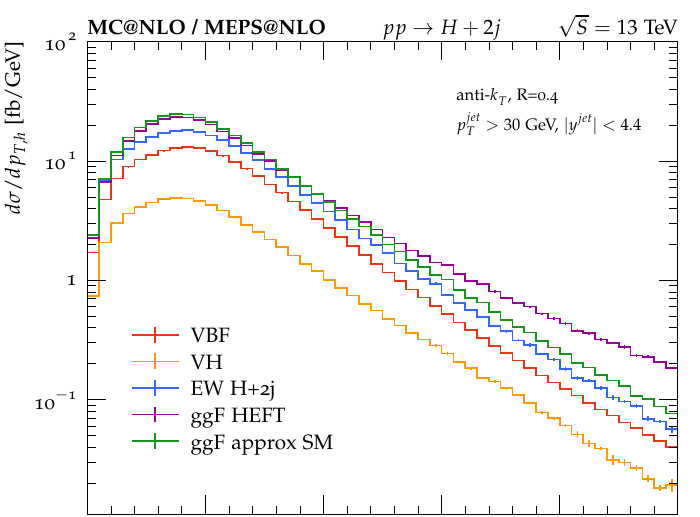}
    \includegraphics[width=\textwidth]{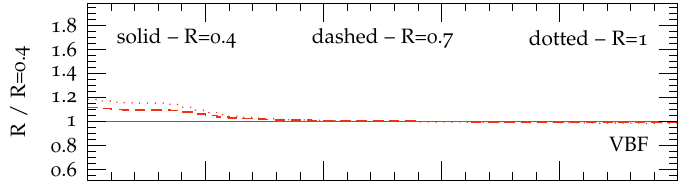}
    \includegraphics[width=\textwidth]{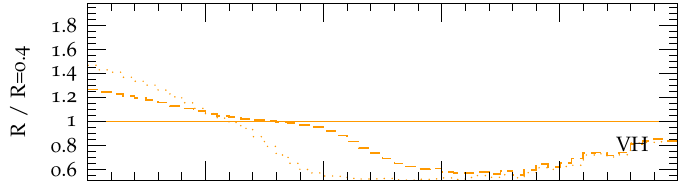}
    \includegraphics[width=\textwidth]{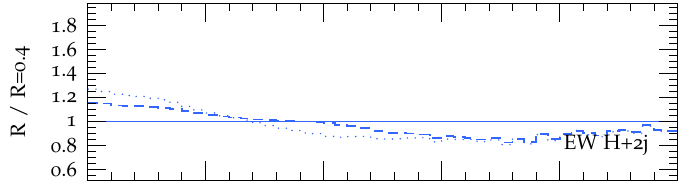}
    \includegraphics[width=\textwidth]{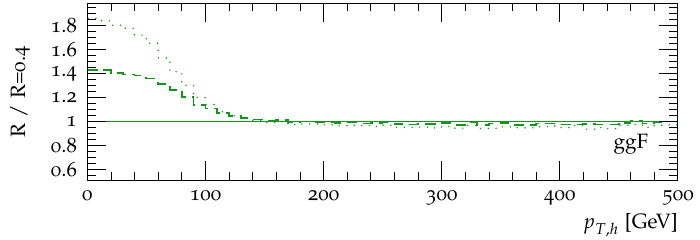}
  \end{minipage}\hskip 1cm
  \begin{minipage}{.295\textwidth}
    \includegraphics[width=\textwidth]{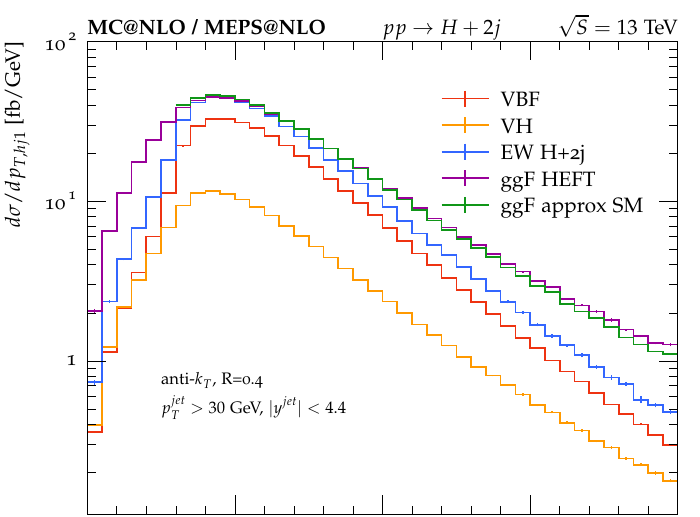}
    \includegraphics[width=\textwidth]{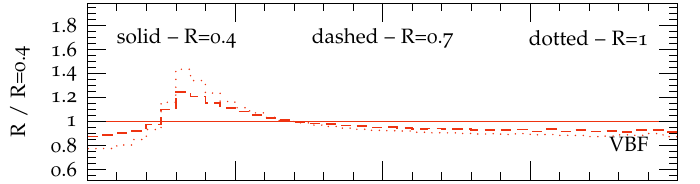}
    \includegraphics[width=\textwidth]{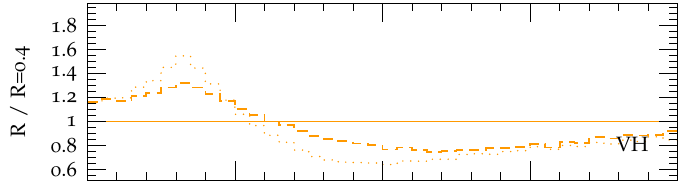}
    \includegraphics[width=\textwidth]{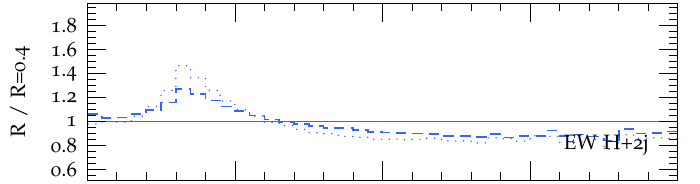}
    \includegraphics[width=\textwidth]{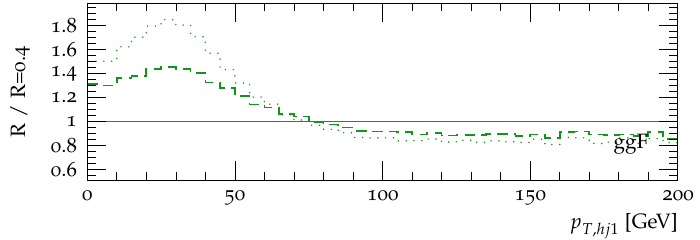}
  \end{minipage}
\caption{Higgs boson transverse momentum (left) and Higgs plus leading jet transverse momentum (right) distribution.
See Fig.~\ref{fig:incl_njets} and the main text for details.}
\label{fig:incl_pth}
\end{figure}

\section{Transverse momentum distributions and their evolution with Higgs boson $p_T$}
\label{sec:transverse}

Figure~\ref{fig:incl_pth} (left) shows the transverse momentum distribution of the Higgs boson for gluon-gluon fusion (both in the effective field theory and with top mass corrections), \VBF, \VH, and EW $Hjj$ production. The lower panels display the ratio of the predictions varying jet radius over the reference radius, $R=0.4$. A minimum of two jets is required. All processes shown have a peak at a transverse momentum in the range from 80-100~GeV. The gluon-gluon fusion predictions  grow with respect to \VBF/$Hjj$ as the Higgs boson transverse momentum increases. The ratio of $Hjj$ to \VBF is roughly constant as a function of $p_T$. For \VBF and gluon-gluon fusion, the cross section increases with increasing $R$ for Higgs transverse momenta less than 150~GeV. In this region, a larger value of $R$ makes it easier for the second jet to have a transverse momentum larger than the 30~GeV threshold. Gluon-gluon fusion has a larger $R$ dependence at low $p_T$ than \VBF.
\VH has a somewhat more complex behavior. The cross section increases for larger $R$ at low $p_T$, as also observed for the other two processes, but decreases for larger $R$ at high $p_T$. The high $p_T$ behavior  results from the loss of one of the jets from the $V$ decay,
due to the two quark originating from the $V$ decay merging at high $p_T$ with larger jet radius. This situation then requires a radiated jet to be above the transverse momentum threshold to fulfill the requirement of a second jet in the event.

The Higgs boson plus jet transverse momentum distribution is shown in the right panel of Fig.~\ref{fig:incl_pth}.
The gluon-gluon fusion process dominates for transverse momentum values below 30~GeV and above approximately 70~GeV,
and is approximately as large as $Hjj$ production in the intervening region.

\begin{figure}[p]
\centering
  \begin{minipage}{.2925\textwidth}
    \includegraphics[width=\textwidth]{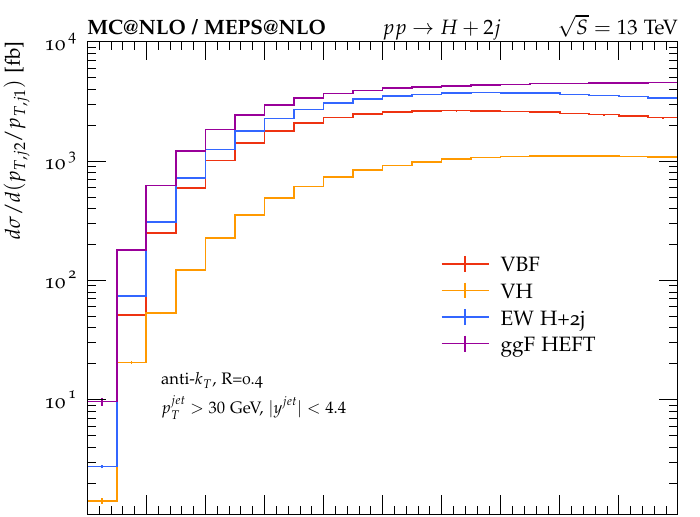}
    \includegraphics[width=\textwidth]{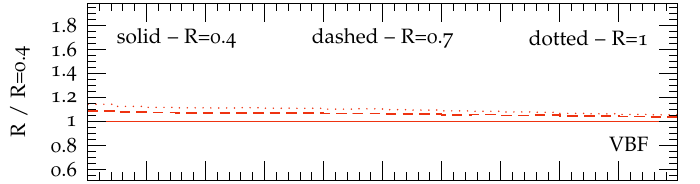}
    \includegraphics[width=\textwidth]{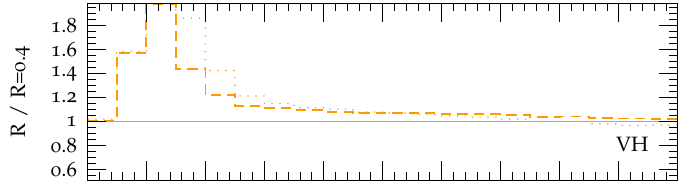}
    \includegraphics[width=\textwidth]{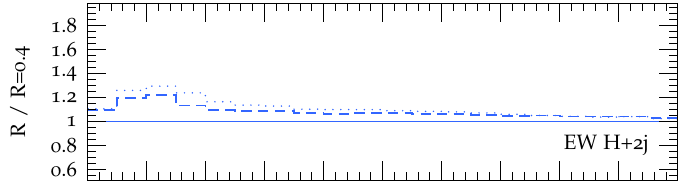}
    \includegraphics[width=\textwidth]{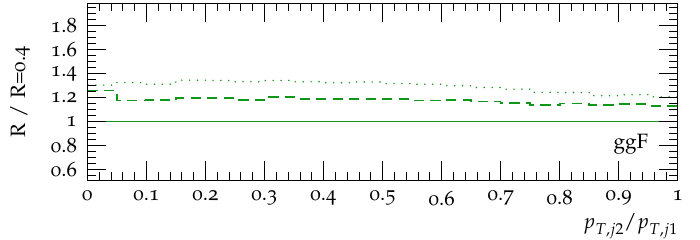}
  \end{minipage}\hfill
  \begin{minipage}{.2925\textwidth}
    \includegraphics[width=\textwidth]{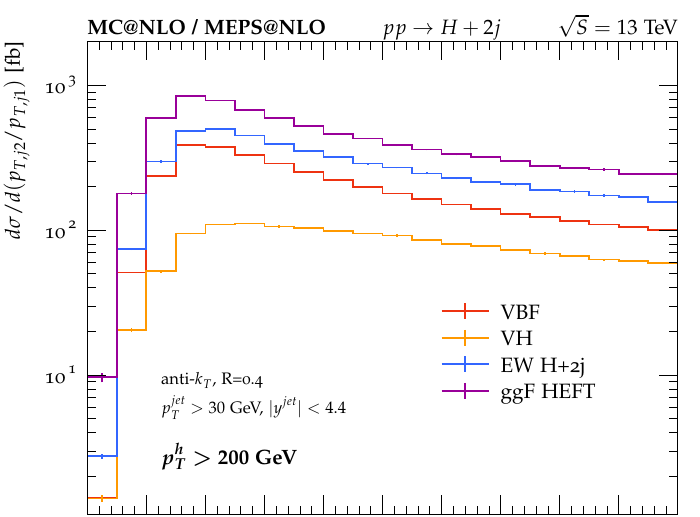}
    \includegraphics[width=\textwidth]{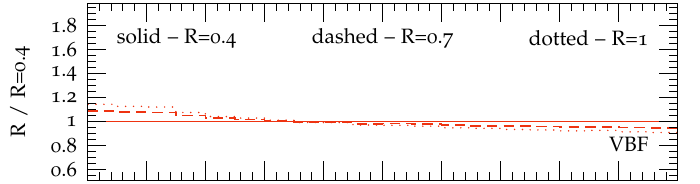}
    \includegraphics[width=\textwidth]{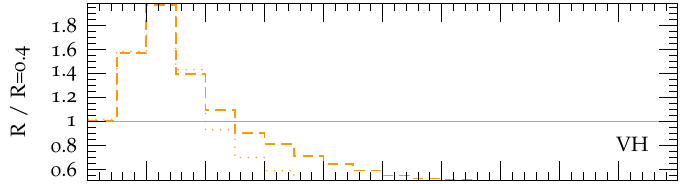}
    \includegraphics[width=\textwidth]{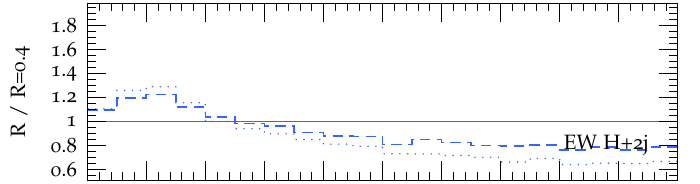}
    \includegraphics[width=\textwidth]{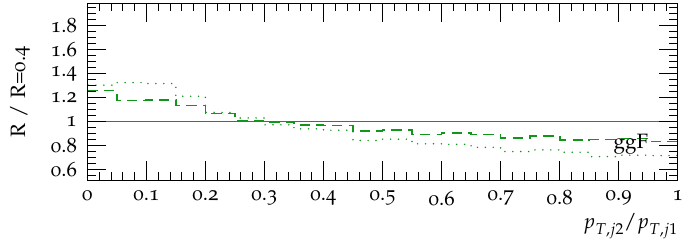}
  \end{minipage}\hfill
  \begin{minipage}{.2925\textwidth}
    \includegraphics[width=\textwidth]{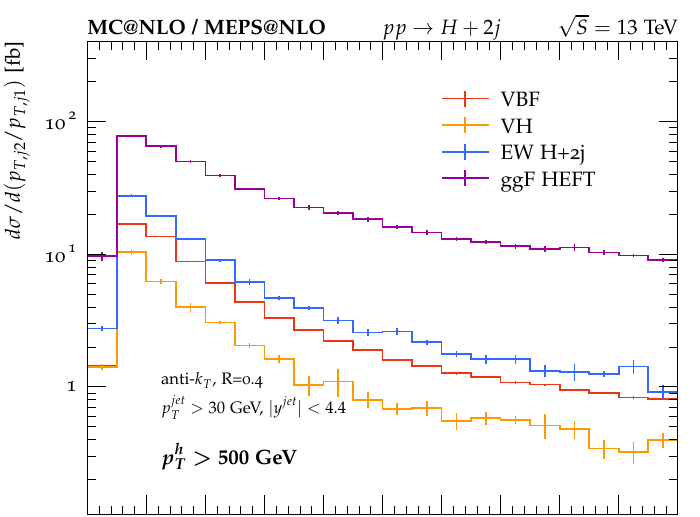}
    \includegraphics[width=\textwidth]{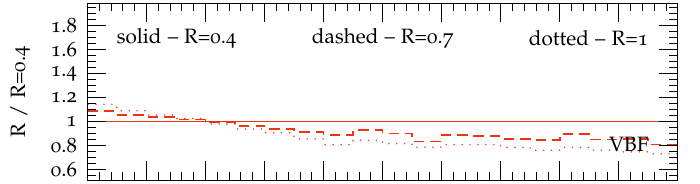}
    \includegraphics[width=\textwidth]{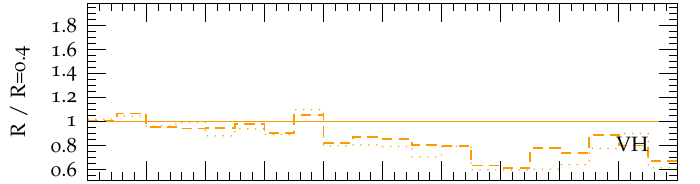}
    \includegraphics[width=\textwidth]{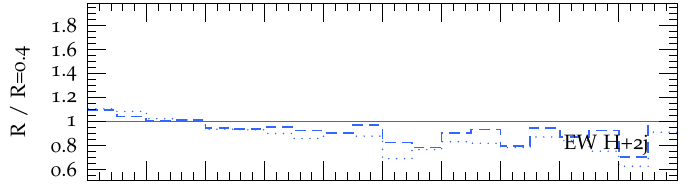}
    \includegraphics[width=\textwidth]{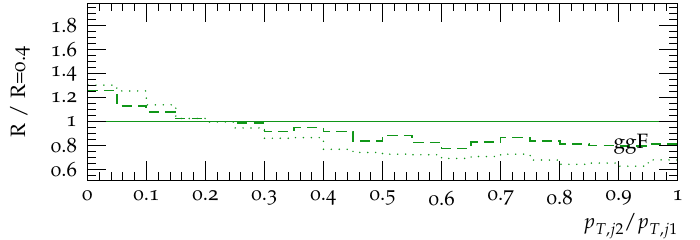}
  \end{minipage}\hfill
\caption{Ratio between the transverse momentum of the second leading jet and the leading jet. The left panels show inclusive predictions,
while the middle and right panels show results for a minimum Higgs transverse momentum of 200 and 500~GeV.
See Fig.~\ref{fig:incl_njets} and the main text for details.}
\label{fig:incl_pt2_pt1}
\end{figure}
\begin{figure}[p]
\centering
  \begin{minipage}{.2925\textwidth}
    \includegraphics[width=\textwidth]{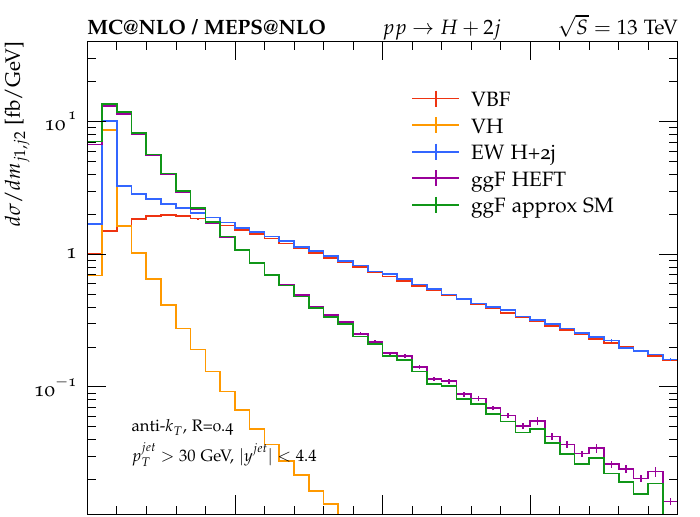}
    \includegraphics[width=\textwidth]{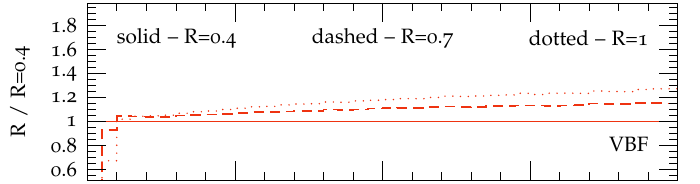}
    \includegraphics[width=\textwidth]{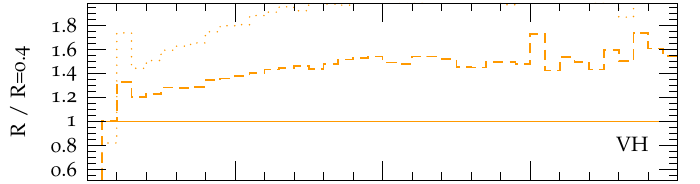}
    \includegraphics[width=\textwidth]{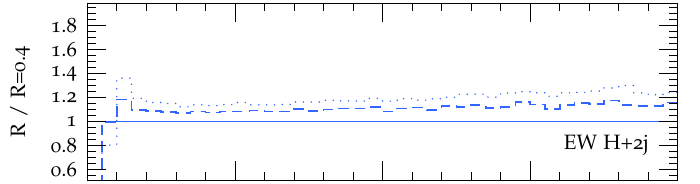}
    \includegraphics[width=\textwidth]{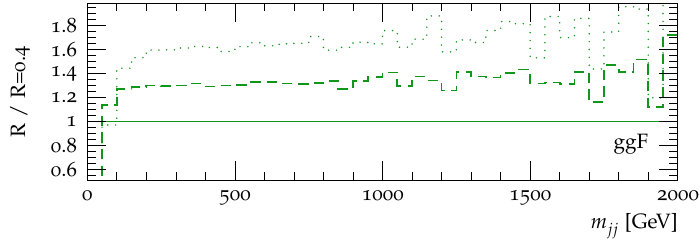}
  \end{minipage}\hfill
  \begin{minipage}{.2925\textwidth}
    \includegraphics[width=\textwidth]{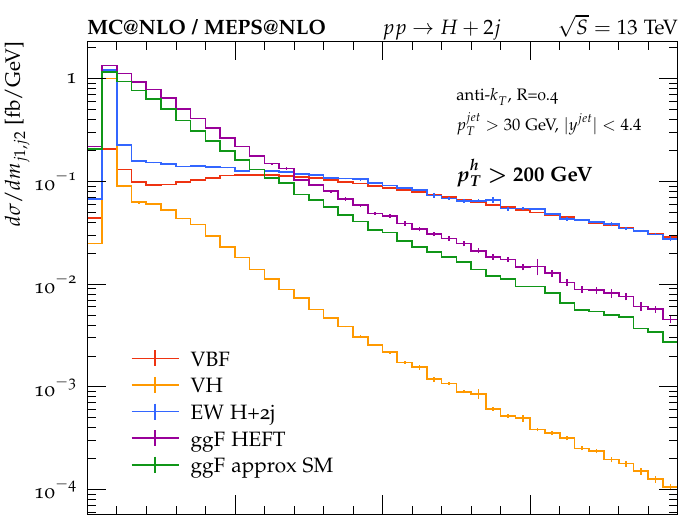}
    \includegraphics[width=\textwidth]{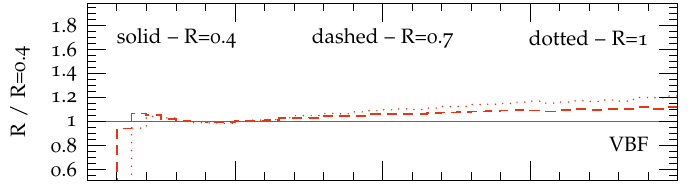}
    \includegraphics[width=\textwidth]{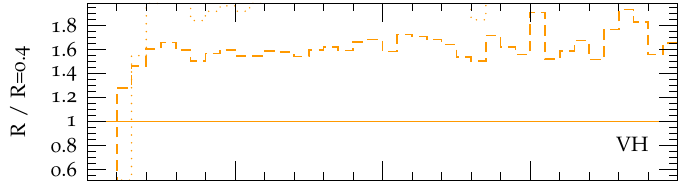}
    \includegraphics[width=\textwidth]{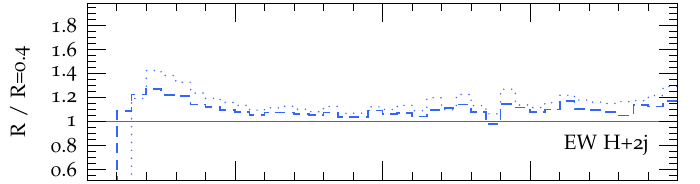}
    \includegraphics[width=\textwidth]{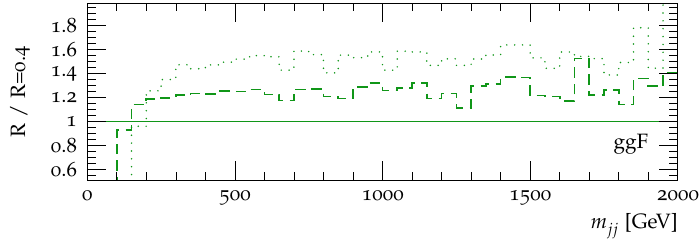}
  \end{minipage}\hfill
  \begin{minipage}{.2925\textwidth}
    \includegraphics[width=\textwidth]{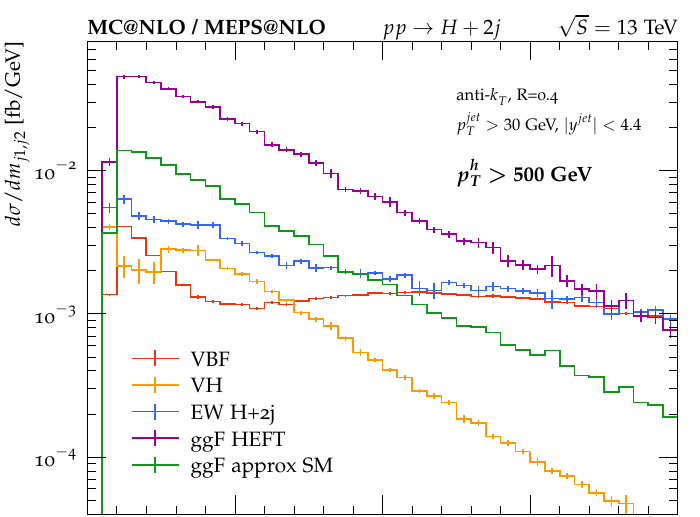}
    \includegraphics[width=\textwidth]{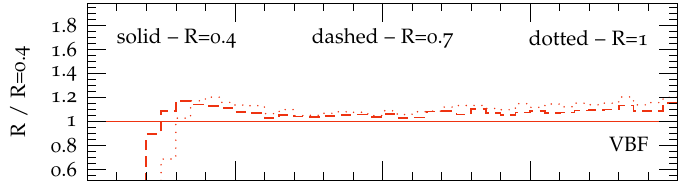}
    \includegraphics[width=\textwidth]{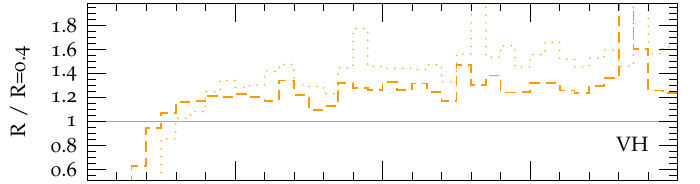}
    \includegraphics[width=\textwidth]{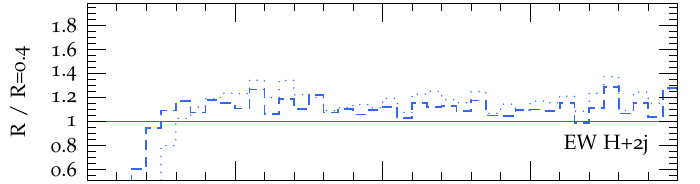}
    \includegraphics[width=\textwidth]{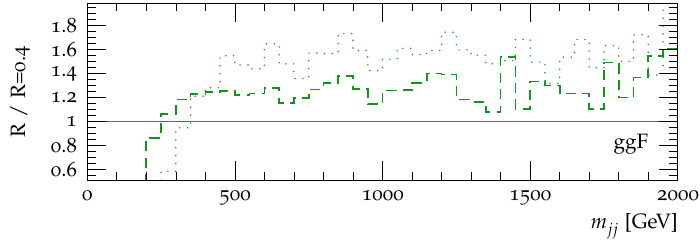}
  \end{minipage}\hfill
\caption{Dijet invariant mass distribution. The left panels show inclusive predictions,
while the middle and right panels show results for a minimum Higgs transverse momentum of 200 and 500~GeV.
See Fig.~\ref{fig:incl_njets} and the main text for details.}
\label{fig:incl_m_jj12}
\end{figure}
The transverse momentum imbalance between the two leading jets is shown in Fig.~\ref{fig:incl_pt2_pt1} for three different Higgs boson $p_T$ cuts, for all of the Higgs boson processes under consideration.
For the inclusive cuts, the two tagging jets have similar transverse momenta and the Higgs boson recoils against the dijet system. As the transverse momentum of the Higgs boson increases, the difference between the transverse momentum of the leading jet and of the second leading jet increases, again for all processes. The Higgs boson is essentially recoiling against the leading jet, with the second leading jet becoming soft. Note that the ratio between the transverse momentum of the second leading jet over the leading jet one falls more rapidly for \VH + \VBF than for gluon-gluon fusion for a Higgs boson transverse momentum larger than 500~GeV.

As the Higgs boson $p_T$ increases, the probability of producing additional jets increases, primarily because the amount of radiation from the leading jet increases.
Any additional jets can thus become comparable in transverse momentum to the second quark tagging jet, often replacing it as the jet with the second leading transverse momentum. If radiated from the leading jet, this additional jet will be much closer in phase space to the leading jet than to the other quark jet from the \VBF process, and thus will affect the \VBF tagging efficiency. The $R$-dependence in Fig.~\ref{fig:incl_pt2_pt1} is especially prominent in the jet ratio range less than 0.2, but also sizable for gluon-gluon fusion in the jet ratio range of 0.1.

\section{Dijet mass, $\Delta y_{jj}$ and further VBF discrimination}
\label{sec:discrimination}

\begin{figure}[p]
  \centering
  \includegraphics[width=.675\textwidth]{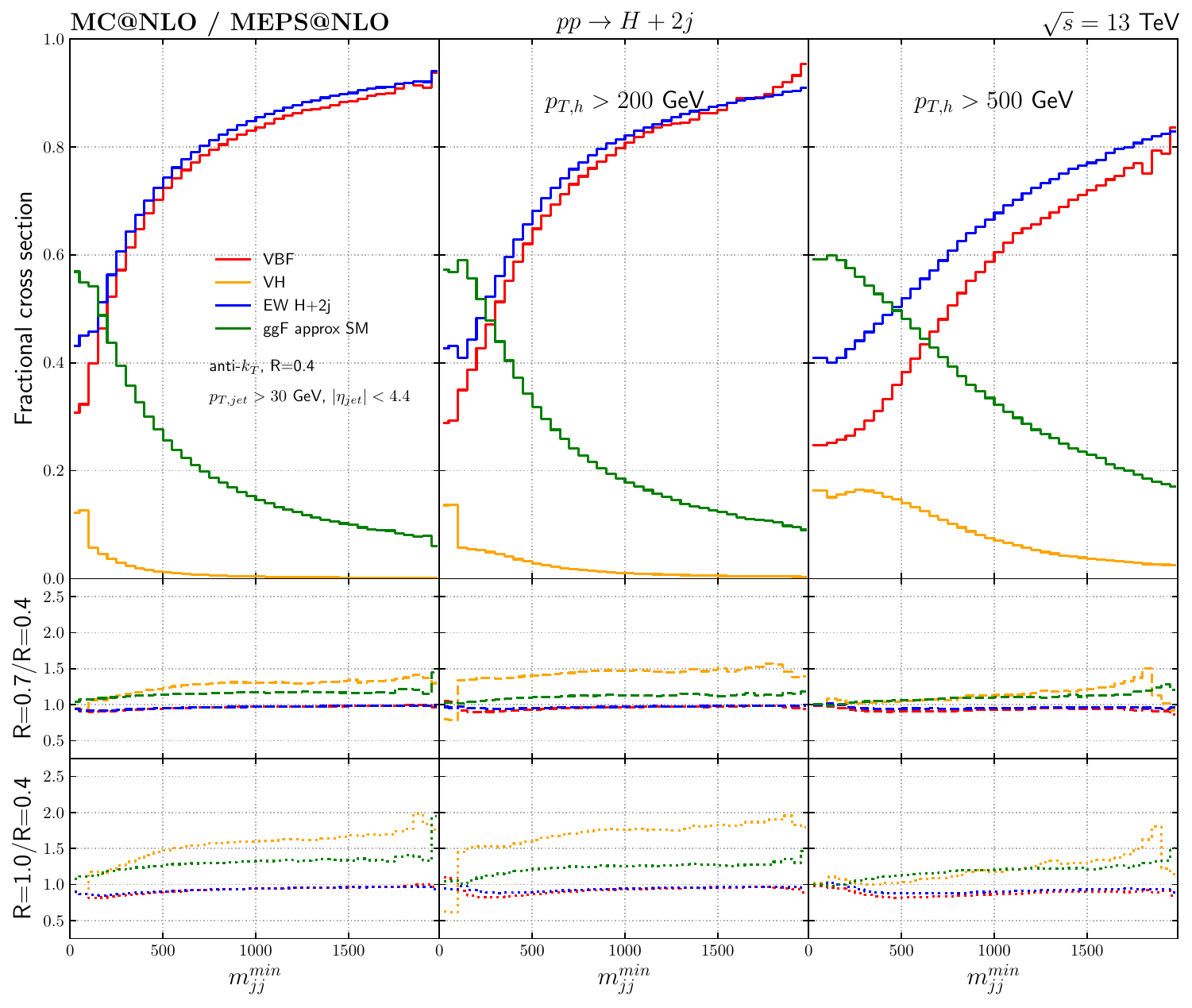}
  \caption{Fractional cross sections of the various production channels
    as a function of the minimum dijet mass (using the two leading jets) cross section.}
\label{fig:incl_m_jj12_fraction}
\end{figure}
The dijet mass distribution from the two leading jets in the event is shown in Fig.~\ref{fig:incl_m_jj12} for the three different Higgs boson $p_T$ requirements.
All predictions are relatively featureless except for \VH and $Hjj$ for which the $W/Z$ resonance region is clearly visible. Gluon-gluon fusion production is the largest process at low dijet mass.  For the inclusive case, \VBF saturates the $Hjj$ prediction for masses above 400~GeV and becomes larger than gluon-gluon fusion at about the same mass. The $R$ dependence of the cross sections for  both \VBF and $Hjj$
grows with increasing dijet mass, albeit with a relatively small slope. An $R$ dependence for the dijet mass cross section is also observed for gluon-gluon fusion, but with a magnitude that is relatively constant over the dijet mass range. At high Higgs boson $p_T$,  similar to the inclusive case, there is still a large enhancement of the cross section in the resonance region for the 200~GeV cut. In addition,  the $Hjj$ cross section is larger than the \VBF cross section out to a higher dijet mass. There is barely any resonance enhancement with the 500~GeV cut, but there is a considerable enhancement of the $Hjj$ cross section to that of the \VBF cross section for higher  dijet masses, up to  1~TeV. This is due to the merging of the two jets from the $V$ boson decay, with the dijet mass being created with an additional radiated jet. The $R$-dependence for \VH and gluon-gluon fusion is very large, decreasing as the Higgs boson $p_T$ increases. The $R$-dependence for \VBF is relatively small. The information in Fig.~\ref{fig:incl_m_jj12} is summarized in Fig.~\ref{fig:incl_m_jj12_fraction}, where the fractional cross sections of the production channels as a function of the minimal dijet mass (using the two leading jets) are shown for the three separate Higgs boson transverse momentum cuts.
For the inclusive case, \VBF becomes larger than gluon-gluon fusion for $m_{jj} \ge 200$~GeV. For a Higgs boson transverse momentum greater than 500~GeV, this does not happen until $m_{jj} \ge 700$~GeV, reducing the impact of an $m_{jj}$ cut to enhance \VBF production.

\begin{figure}[p]
  \centering
  \begin{minipage}{.2925\textwidth}
    \includegraphics[width=\textwidth]{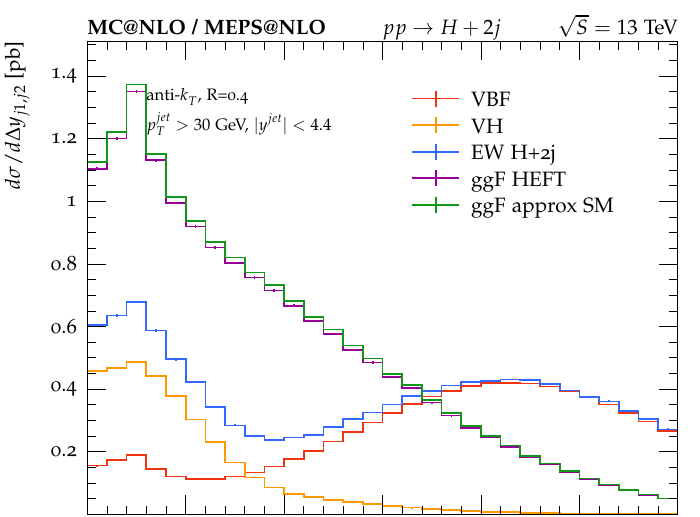}
    \includegraphics[width=\textwidth]{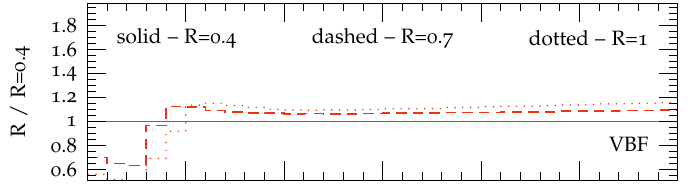}
    \includegraphics[width=\textwidth]{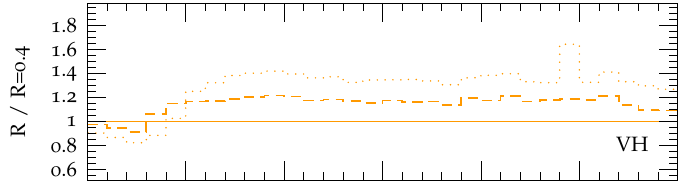}
    \includegraphics[width=\textwidth]{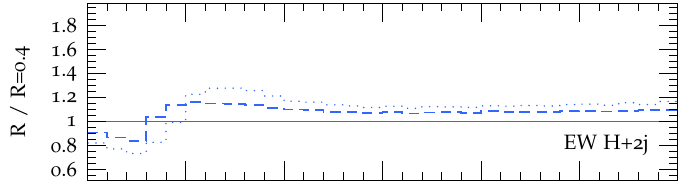}
    \includegraphics[width=\textwidth]{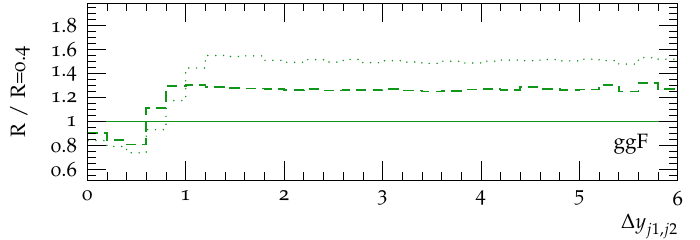}
  \end{minipage}\hfill
  \begin{minipage}{.2925\textwidth}
    \includegraphics[width=\textwidth]{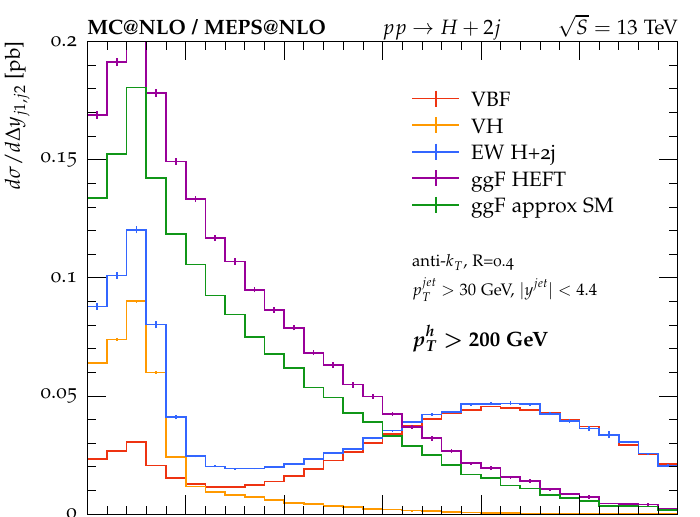}
    \includegraphics[width=\textwidth]{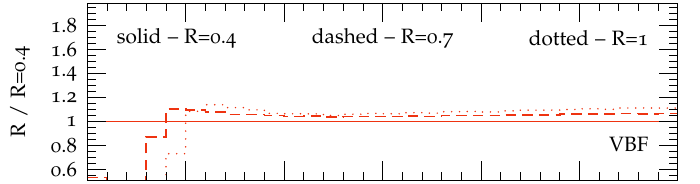}
    \includegraphics[width=\textwidth]{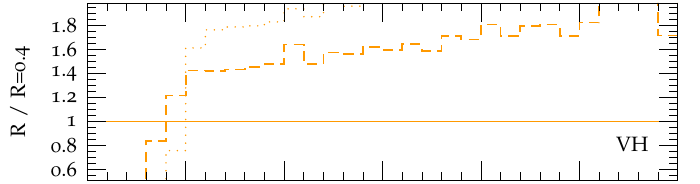}
    \includegraphics[width=\textwidth]{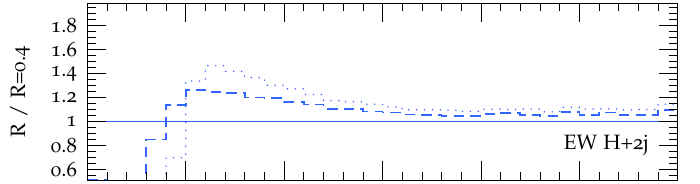}
    \includegraphics[width=\textwidth]{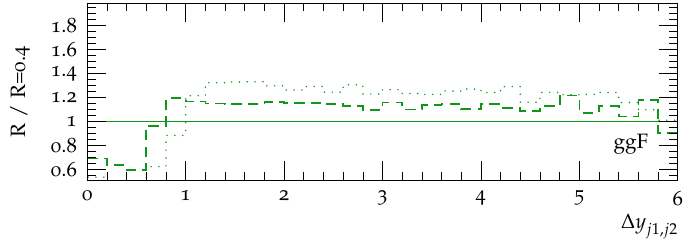}
  \end{minipage}\hfill
  \begin{minipage}{.2925\textwidth}
    \includegraphics[width=\textwidth]{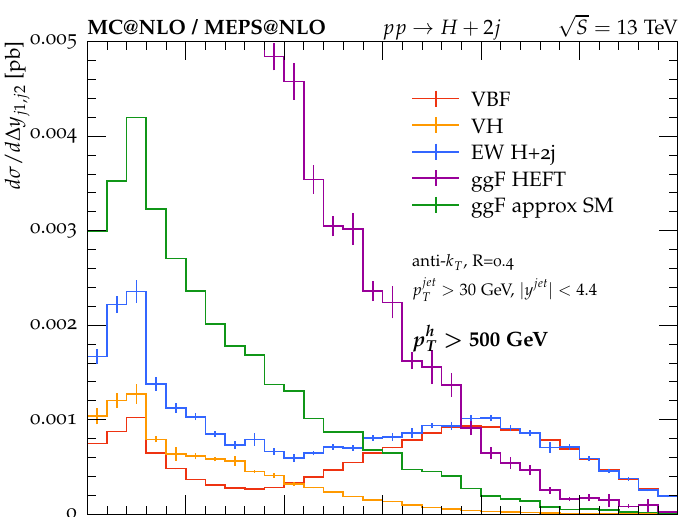}
    \includegraphics[width=\textwidth]{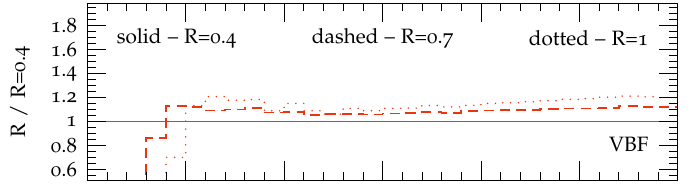}
    \includegraphics[width=\textwidth]{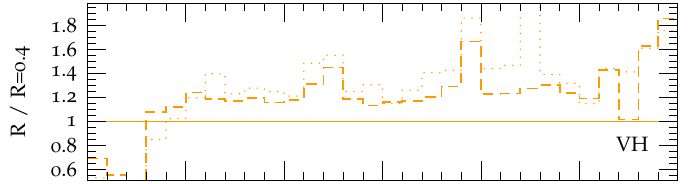}
    \includegraphics[width=\textwidth]{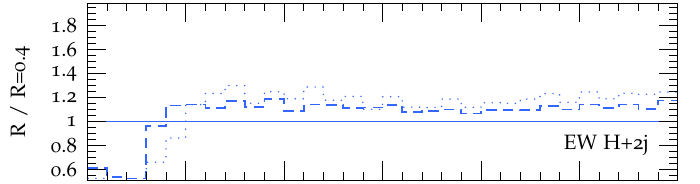}
    \includegraphics[width=\textwidth]{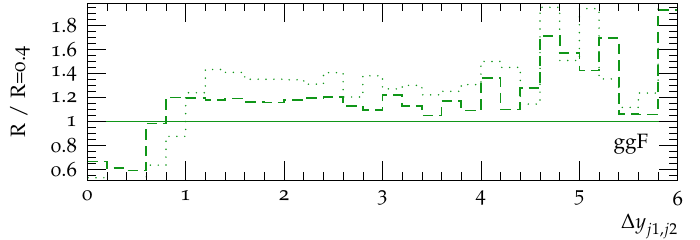}
  \end{minipage}
\caption{$\Delta y_{jj}$ distribution, using the two leading jets.
The left panels show inclusive predictions, while the middle and right panels
show results for a minimum Higgs transverse momentum of 200 and 500~GeV.
See Fig.~\ref{fig:incl_njets} and the main text for details.}
\label{fig:incl_delta_y_jj12}
\end{figure}
\begin{figure}[t]
  \centering
  \includegraphics[width=.65\textwidth]{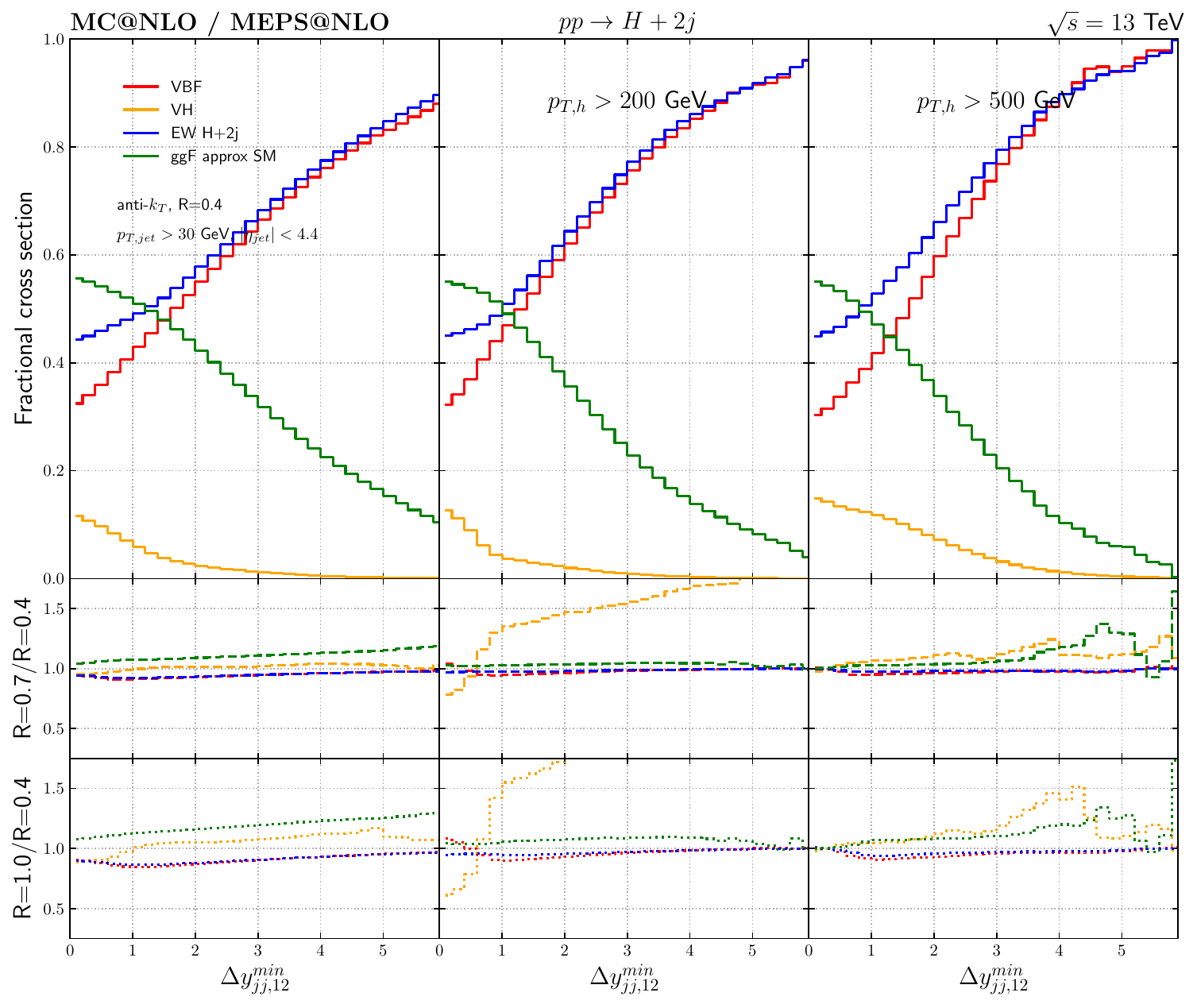}
  \includegraphics[width=.65\textwidth]{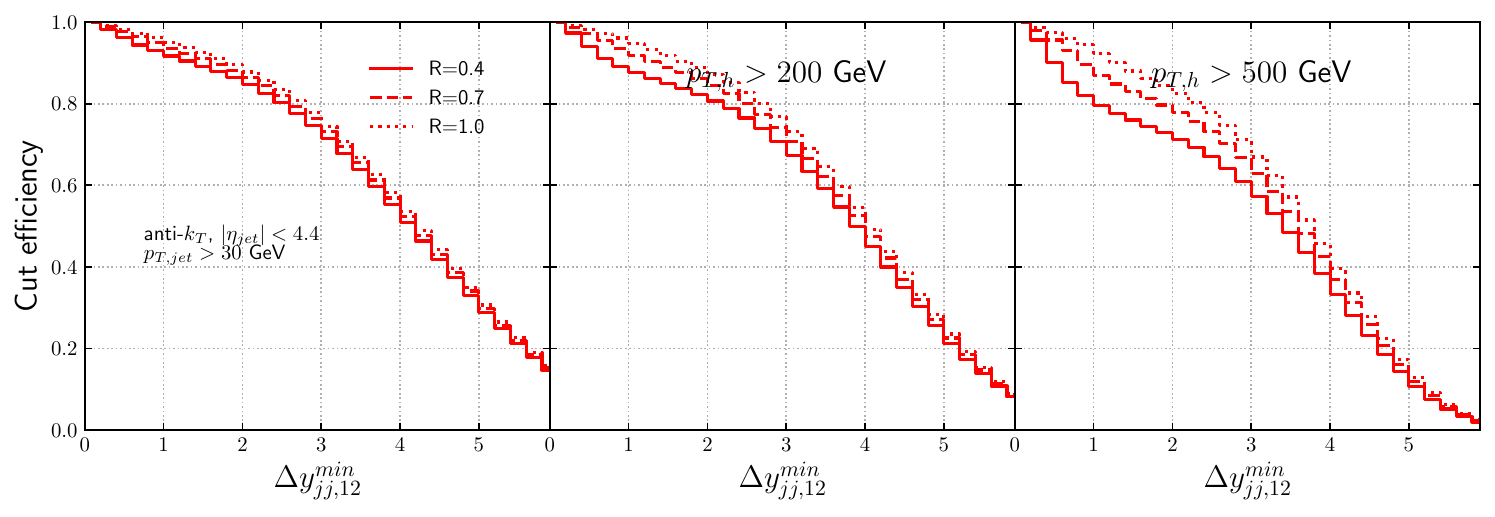}
\caption{Fractional cross sections of the various production channels as a function of the minimal dijet rapidity separation using the two leading jets ($\Delta y_{jj,12}$), for different Higgs boson $p_T$ cuts.
In the bottom panel we show the acceptance probability for the \VBF{} production as a function of $\Delta y_{jj,12}^{\min}$ for different jet radii.
}
\label{fig:incl_delta_y_jj12_fraction}
\end{figure}
The $\Delta y_{jj}$ distribution of the two leading jets is shown in Fig.~\ref{fig:incl_delta_y_jj12} for all processes.  Again, gluon-gluon fusion dominates for low $\Delta y_{jj}$, and $Hjj$ is larger than \VBF, with the cross-over point being in the region $\Delta y_{jj} \approx 3\ldots 3.5$.
One noticeable feature is the presence of a bump at low values of $\Delta y_{jj}$, for all processes,  larger at 500~GeV than at 200~GeV. For the highest Higgs boson $p_T$ cut for \VBF, the bump is as large as the usual peak at $\Delta y_{jj}$ around 4. This bump is due to hard gluon radiation off of the leading jet, sufficiently separated from the parent jet to be reconstructed as a separate jet. Let us consider the Born contribution in Fig.~\ref{fig:vbflo}, so that the two final state partons correspond to the two leading jets. In the factorised approach, we can imagine to add radiative corrections to the two quark lines separately, preserving the momentum flowing in the t-channel propagators.
As the {\it original} second tagging jet tends to be soft for large Higgs boson $p_T$, the {\it original} leading jet can fragment into two resolved jets, both with a $p_T$ higher that the {\it original} second jet, so that $\Delta y_{jj}$ computed using the two highest transverse momenta jets is smaller. The application of any significant $\Delta y_{jj}$ cut for \VBF identification then leads to the loss of these events. The relative fraction of EW $Hjj$ production to gluon-gluon fusion stays relatively constant  at the location of this $\Delta y_{jj}$ bump as the Higgs boson $p_T$ increases, as the \VH component of the bump is reduced as the \VBF contribution increases.

We now introduce the fractional cross section, \emph{i.e.} the relative contribution to the cross section for the several Higgs boson production processes, as a function of the lower cut on the rapidity separation between the two leading jets (\emph{i.e.} the ones with largest $p_T$) $\Delta y^{\min}_{jj}$ . Results are shown in the top panel of Fig.~\ref{fig:incl_delta_y_jj12_fraction}  for the three different Higgs boson $p_T$ cuts. \VBF becomes the dominant process in the $\Delta y^{\min}_{jj}$ range greater than 1.5 for all $p_T$ cuts. However, at high Higgs boson $p_T$, the \VH process increases in importance. The $R$ dependence of each contribution is shown in the middle panel. \VH greatly increases with larger $R$ for the Higgs boson $p_T$ range greater than 200~GeV, but not for Higgs boson $p_T$ greater than 500~GeV.
The fractional cross section for gluon-gluon fusion with a minimum $\Delta y_{jj}$ cut of 3 decreases as the Higgs boson $p_T$ increases.

The acceptance for the \VBF process as a function of $\Delta y^{\min}_{jj}$ is shown in the bottom panel of Fig.~\ref{fig:incl_delta_y_jj12_fraction}. There is essentially no $R$-dependence for inclusive production. The $R$-dependence becomes larger with higher Higgs boson transverse momentum. The impact of the low $\Delta y_{jj}$ bump on the \VBF acceptance decreases as the jet radius increases, as expected from the semi-collinear nature of the hard gluon radiation.
The acceptance for the \VBF cross section decreases as the Higgs boson $p_T$ increases, as for boosted topologies the two leading jets tend to be close in rapidity.

\begin{figure}[t]
  \centering
  \includegraphics[width=.65\textwidth]{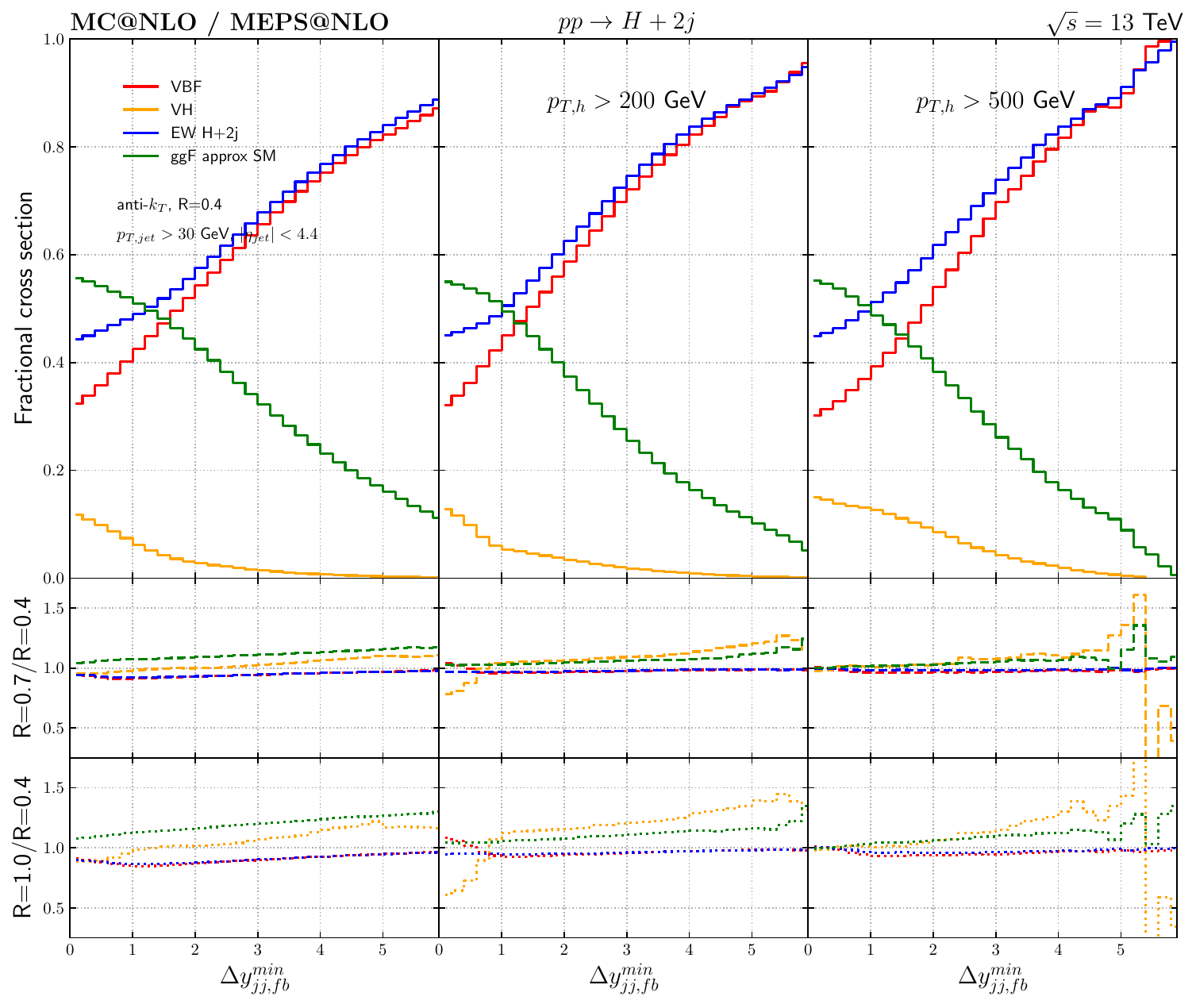}
  \includegraphics[width=.65\textwidth]{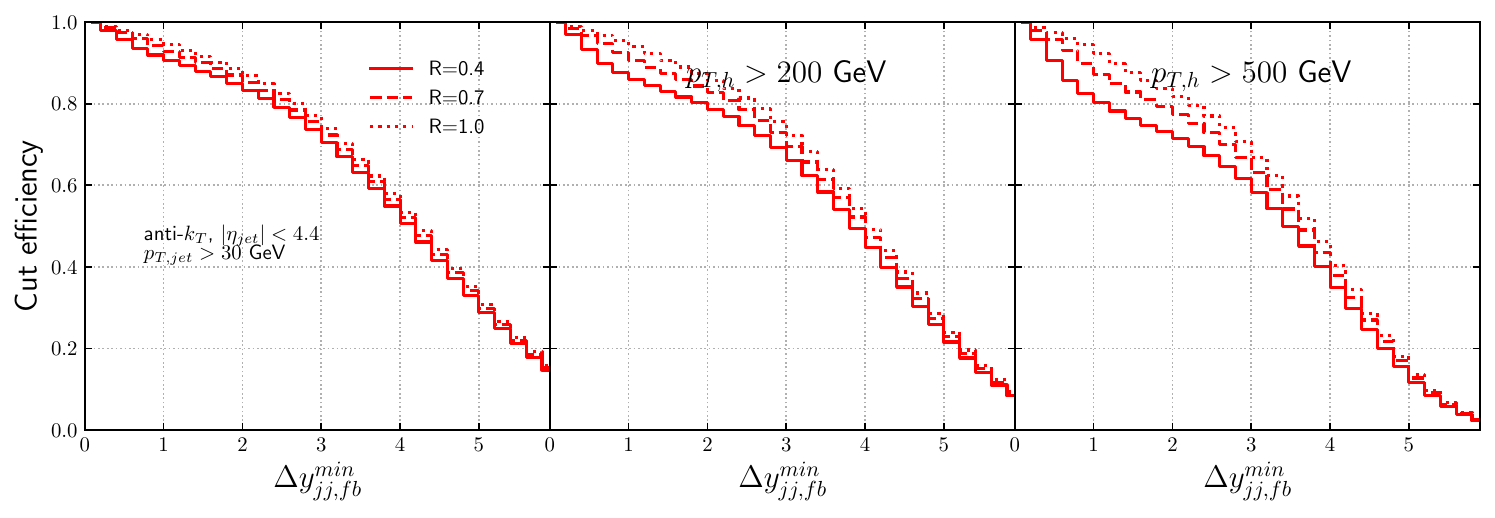}
\caption{Fractional cross sections of the various production channels as a function of the minimal dijet rapidity separation using the two most forward/backward jets ($\Delta y_{jj,fb}$).
In the bottom panel we show the acceptance probability for the \VBF{} production as a function of $\Delta y_{jj,fb}^{\min}$ for different jet radii.
}
\label{fig:incl_delta_y_jj12_fraction_fb}
\end{figure}

As discussed previously, configurations with a boosted Higgs are very likely to have, at the Born level, an {\it original} leading jet balancing almost entirely the Higgs $p_T$, with a second jet produced with smaller $p_T$. Radiative corrections can lead to the formation of two or more high-$p_T$ jets from the {\it original} leading jet, and as a result the {\it} original second leading jet will not be tagged. To avoid this,  the two jets with the largest rapidity separation can be used instead, as illustrated in the top panel of Fig.~\ref{fig:incl_delta_y_jj12_fraction_fb}, where we show the fractional cross section  as a function of $\Delta y_{jj}$ between the most forward and backward jets (fb)  for the three different Higgs boson $p_T$ cuts. In general, the distributions look  similar to those shown in Fig.~\ref{fig:incl_delta_y_jj12_fraction}. \VBF becomes the dominant process in the $\Delta y_{jj}$ range greater than 1.5 for all $p_T$ cuts, but again  the \VH process increases in importance at higher $p_T$. The fractional cross section for \VBF (and EW $Hjj$) for a $\Delta y_{jj} \approx 3$ actually decreases slightly compared to the use of the two most energetic jets.
Indeed, comparing Fig.~\ref{fig:incl_delta_y_jj12_fraction} and Fig.~\ref{fig:incl_delta_y_jj12_fraction_fb}, we notice that the yield of the gluon-gluon fusion channel for larger $\Delta y_{jj}^{\min}$ increases when using the two jets with the largest rapidity.

The $R$-dependence is shown in the middle panel of Fig.~\ref{fig:incl_delta_y_jj12_fraction_fb}, and it is milder for  \VH compared to the previous case using the two
most energetic jets to tag.
The acceptance for the \VBF process as a function of minimum $\Delta y_{jj,fb}$ using the two jets with the largest rapidity separation is shown in the bottom panel. As before, there is no $R$-dependence for inclusive production and the $R$-dependence becomes larger with higher Higgs boson transverse momentum. The acceptance using the two most forward-backward jets above the $p_T$ threshold is only very slightly improved over using the two jets with largest transverse momentum. Thus, there appears to be no significant gain with the use of the two jets with largest rapidity separation for the separation of \VBF and gluon-gluon fusion at high Higgs boson $p_T$.
Furthermore, using the most-forward rapidity jets can enhance the sensitivity to soft physics effects~\cite{Aaboud:2017pou}.

\begin{figure}[p]
  \centering
  \begin{minipage}{.28\textwidth}
    \includegraphics[width=\textwidth]{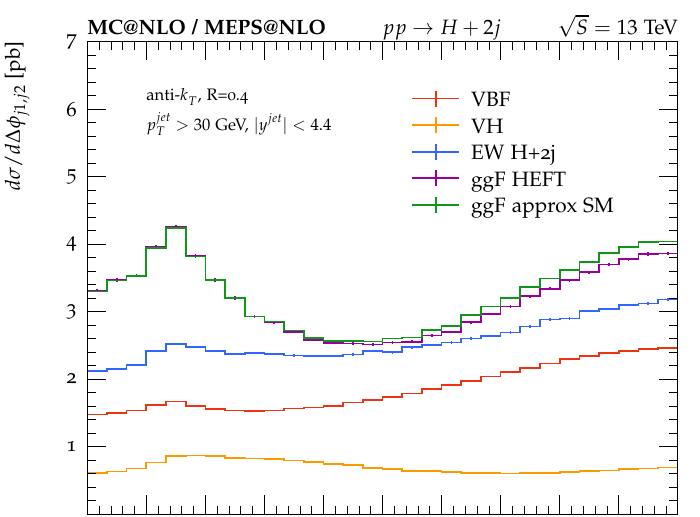}
    \includegraphics[width=\textwidth]{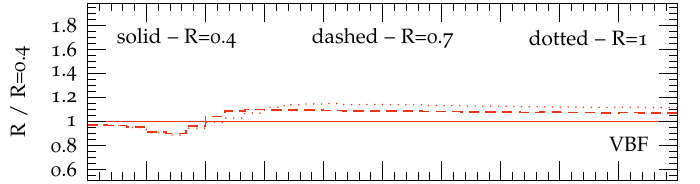}
    \includegraphics[width=\textwidth]{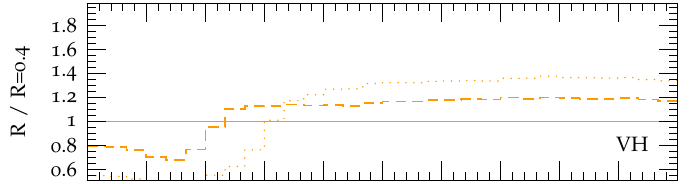}
    \includegraphics[width=\textwidth]{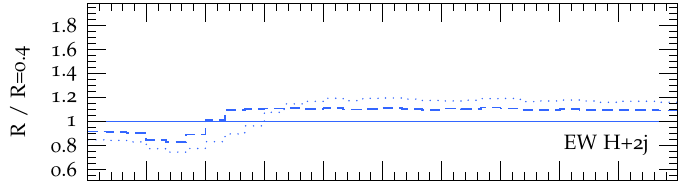}
    \includegraphics[width=\textwidth]{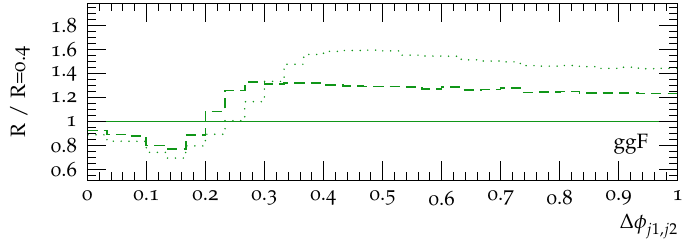}
  \end{minipage}\hfill
  \begin{minipage}{.28\textwidth}
    \includegraphics[width=\textwidth]{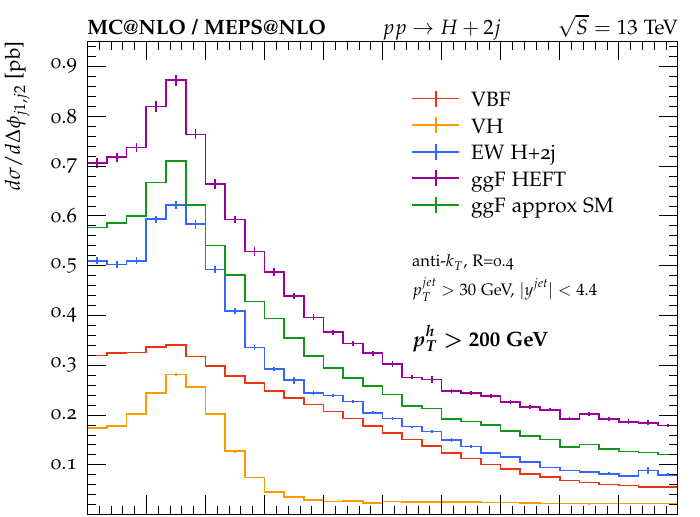}
    \includegraphics[width=\textwidth]{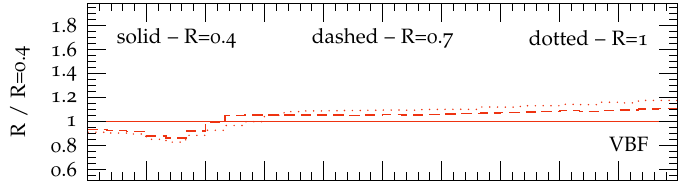}
    \includegraphics[width=\textwidth]{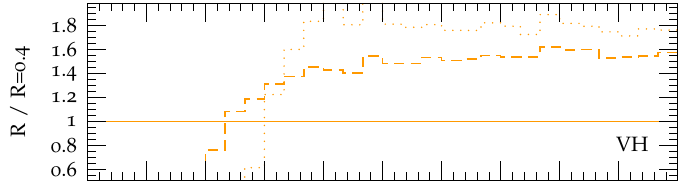}
    \includegraphics[width=\textwidth]{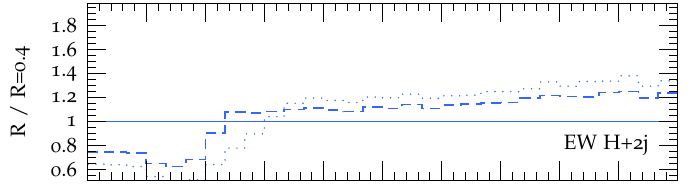}
    \includegraphics[width=\textwidth]{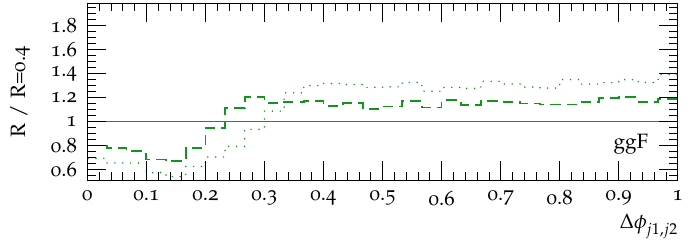}
  \end{minipage}\hfill
  \begin{minipage}{.28\textwidth}
    \includegraphics[width=\textwidth]{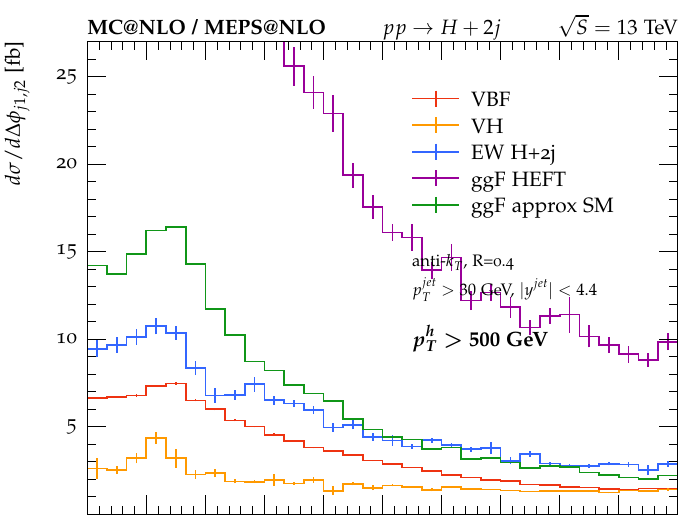}
    \includegraphics[width=\textwidth]{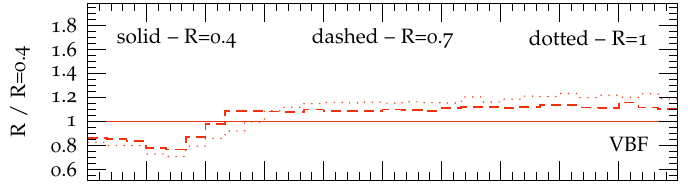}
    \includegraphics[width=\textwidth]{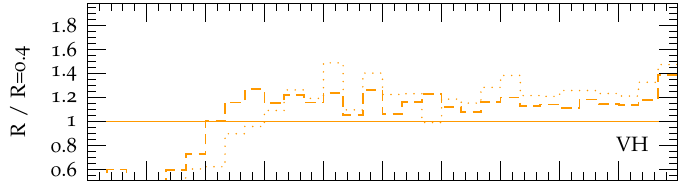}
    \includegraphics[width=\textwidth]{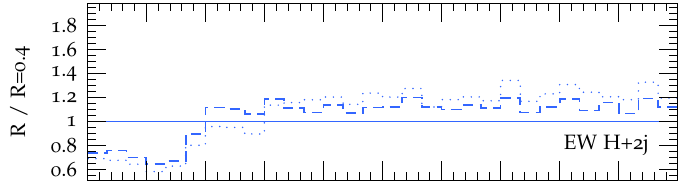}
    \includegraphics[width=\textwidth]{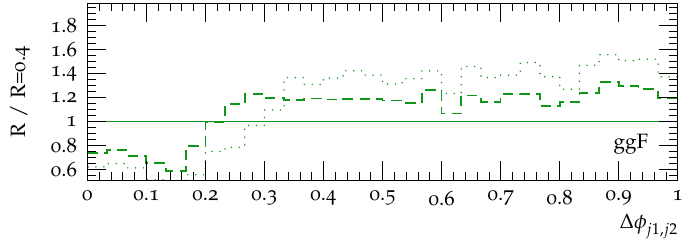}
  \end{minipage}
\caption{Inclusive $\Delta\phi_{jj}$ distribution, using the two leading jets.
The left panels show inclusive predictions, while the middle and right panels
show results for a minimum Higgs transverse momentum of 200 and 500~GeV.
See Fig.~\ref{fig:incl_njets} and the main text for details.}
\label{fig:incl_delta_phi_jj12}
\end{figure}
\begin{figure}[p]
    \centering
    \begin{minipage}{0.28\textwidth}
     \includegraphics[width=\textwidth]{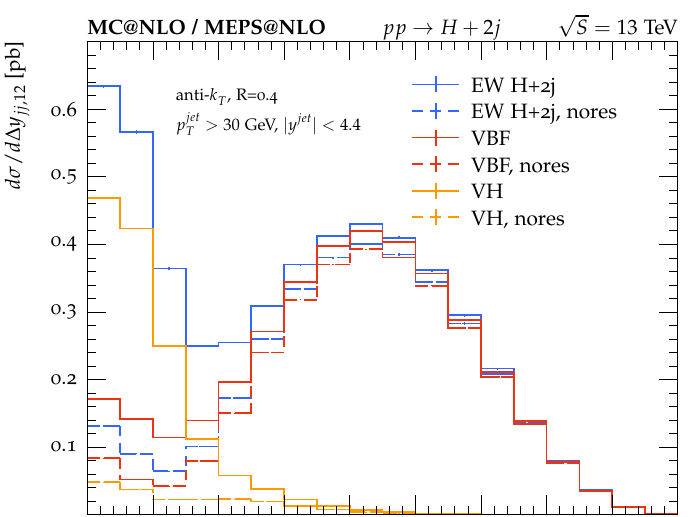}
     \includegraphics[width=\textwidth]{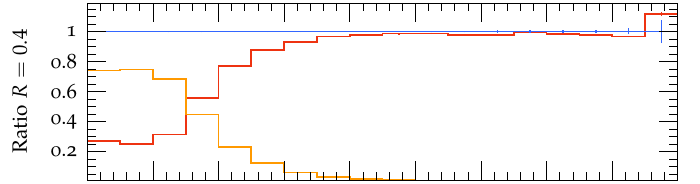}
    \includegraphics[width=\textwidth]{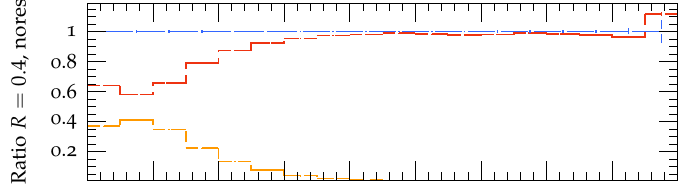}
    \includegraphics[width=\textwidth]{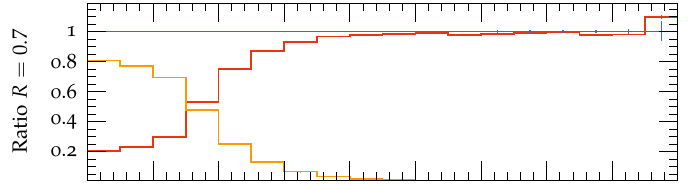}
    \includegraphics[width=\textwidth]{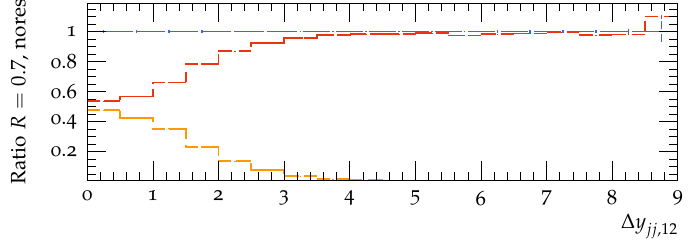}
    \end{minipage}\hfill
    \begin{minipage}{0.28\textwidth}
     \includegraphics[width=\textwidth]{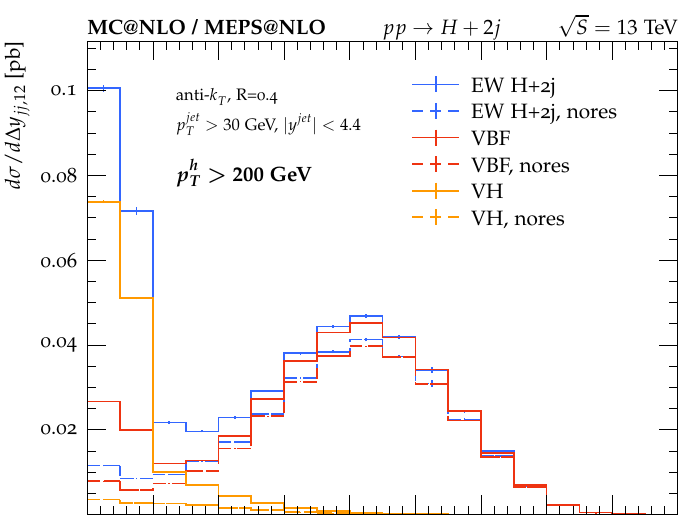}
     \includegraphics[width=\textwidth]{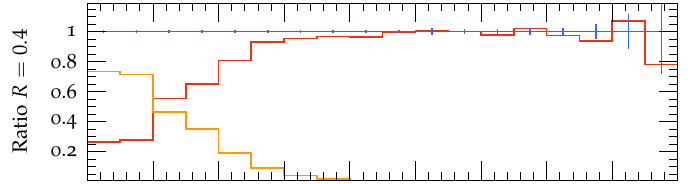}
    \includegraphics[width=\textwidth]{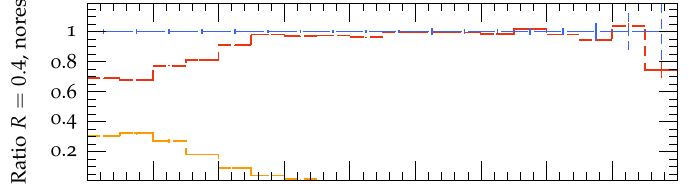}
    \includegraphics[width=\textwidth]{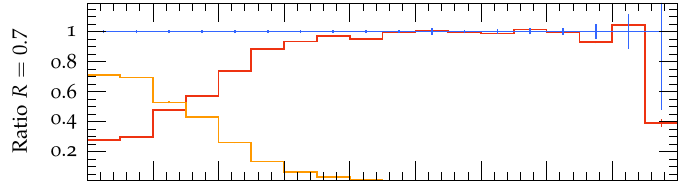}
    \includegraphics[width=\textwidth]{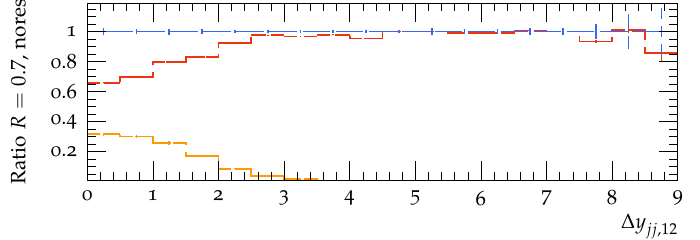}
    \end{minipage}\hfill
    \begin{minipage}{0.28\textwidth}
     \includegraphics[width=\textwidth]{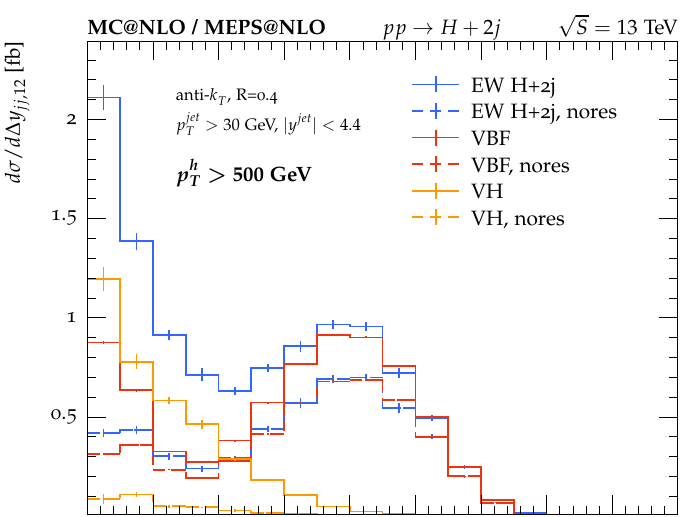}
     \includegraphics[width=\textwidth]{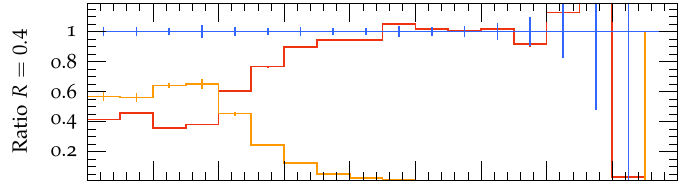}
    \includegraphics[width=\textwidth]{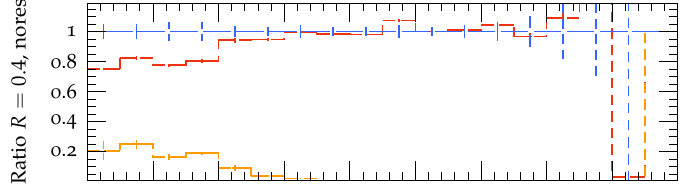}
    \includegraphics[width=\textwidth]{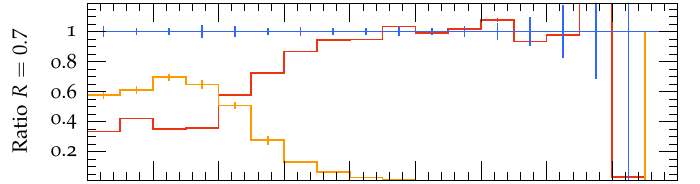}
    \includegraphics[width=\textwidth]{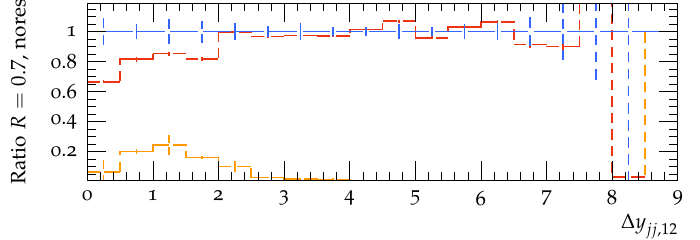}
    \end{minipage}
    \caption{$\Delta y_{jj}$ distribution, using the two leading jets for three Higgs boson transverse-momentum cuts: $p_{T,h}>0$ (left panels), $p_{T,h}>200$~GeV (middle) and $p_{T,h}>500$~GeV (right).
    Predictions with dashed lines~(nores) have been obtained imposing a jet-mass veto, while solid lines are used for more inclusive predictions.
    In the first (third) ratio plot, we show the ratio with respect to the full EW $Hjj$ contribution for $R=0.4$ ($R=0.7$) without the imposition of the resonance veto, while in the second (fourth) ratio plot we compute the ratio after having imposed this cut.}
    \label{fig:deltay_massveto}
\end{figure}

The $\Delta\phi_{jj}$ distribution is shown in Fig.~\ref{fig:incl_delta_phi_jj12}. We notice that any shape differences between \VBF and gluon-gluon fusion decrease as the Higgs boson transverse momentum increases. The $R$-dependence is relatively mild for \VBF production, but can be significantly larger for \VH and gluon-gluon fusion.

\subsection{Resonance mass vetoes}
\label{sec:resonance}

As already discussed, a significant fraction of the $Hjj$ cross section is produced through \VH production, with the vector boson decaying hadronically. This contribution is enhanced when the $W$ or $Z$ boson is on its mass-shell, as shown in Fig.~\ref{fig:incl_m_jj12} which shows that the \VH channel contribution is peaked for $m_{jj}\approx100~$GeV.
However, as we observe in the right panel of Fig.~\ref{fig:incl_m_jj12}, when the boson is highly boosted, all its decay products are clustered in a single jet.
Thus, to reduce the \VH background, we can exclude configurations where the mass of any of three hardest jets, or of the system comprising two of the three hardest jets, or all the three hardest jets, is between 50 and 150~GeV. We remark that we need to consider the three hardest jets, and not only the first two, because initial-state radiation may produce a jet with larger transverse momentum than those arising from the weak boson decay, or because three jets can arise from the $V$ decay due to final state radiation.

We can investigate if the imposition of this resonance veto can lead to a smaller $\Delta y_{jj}$ cut to isolate the \VBF{} contribution.
The $\Delta y_{jj}$ distribution is shown in Fig.~\ref{fig:deltay_massveto}, where dashed (solid) lines have been obtained with(out) this veto.
We notice that applying a jet mass cut constructed looking at any combination of the three hardest jets significantly reduces the relative size of the \VH{} contribution for very small $\Delta y_{jj}$. However, without imposing any $p_{T,h}$ cuts the minimum value $\Delta y_{jj}$ at which the \VH{} contribution becomes negligible is not appreciably decreased by  the imposition of a jet mass veto, but we notice a small improvement for boosted topologies, where the decay products of a vector boson resonance are more likely to be clustered in the same jet.
Thus we concluded that when looking at very boosted topologies, imposing a cut on the mass of the hardest jets and not only on the dijet system, would allow for a better signal-to-background ratio. Our findings also indicate that in phase space regions which cannot distinguish efficiently between the different contributions, a full calculation including all interferences and off-shell effects should be preferred over an approximate one which is appropriate for a strict VBF or VH selection.

\section{Further event topology cuts}
\label{sec:topology}

\begin{figure}[p]
  \centering
  \begin{minipage}{.2925\textwidth}
    \includegraphics[width=\textwidth]{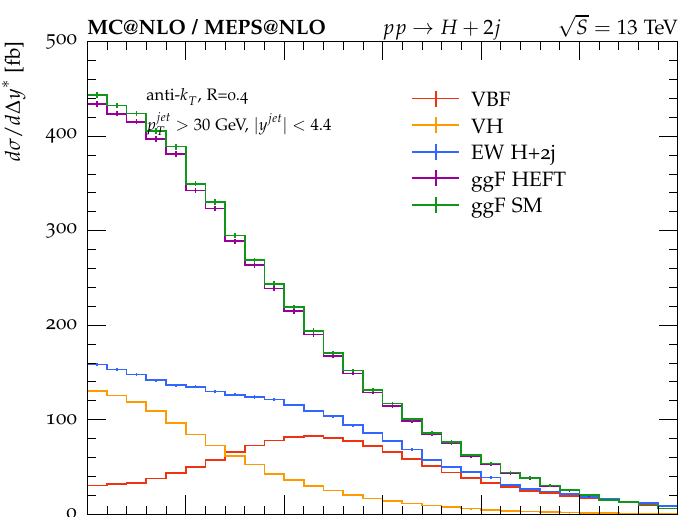}
    \includegraphics[width=\textwidth]{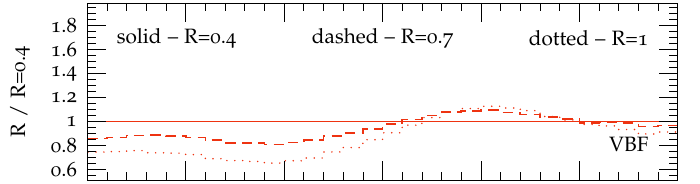}
    \includegraphics[width=\textwidth]{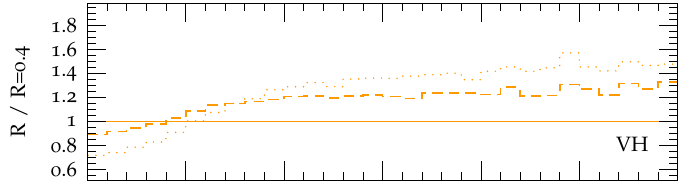}
    \includegraphics[width=\textwidth]{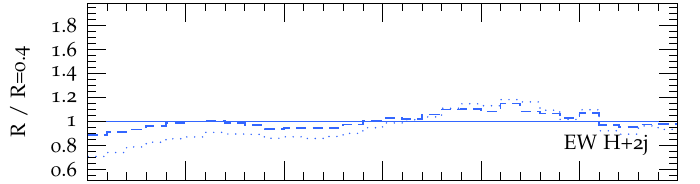}
    \includegraphics[width=\textwidth]{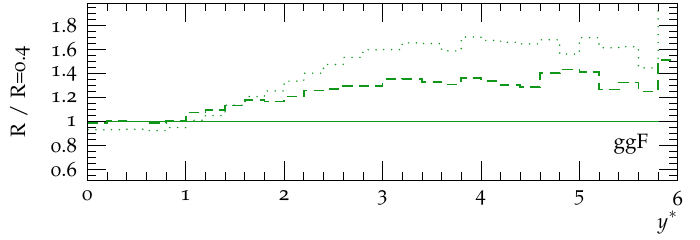}
  \end{minipage}\hfill
  \begin{minipage}{.2925\textwidth}
    \includegraphics[width=\textwidth]{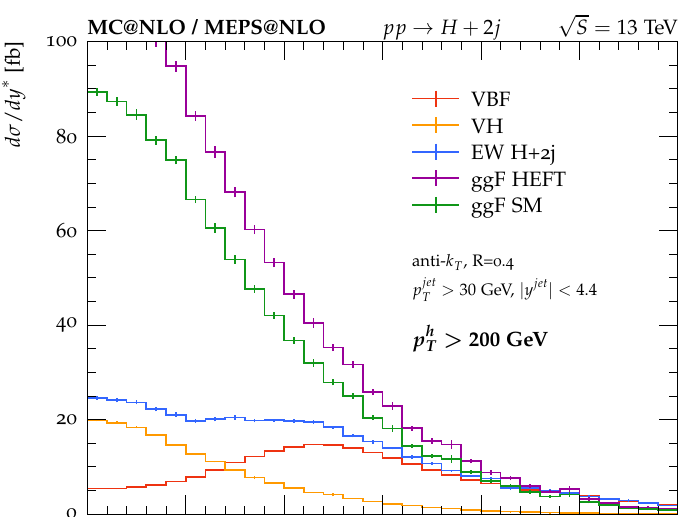}
    \includegraphics[width=\textwidth]{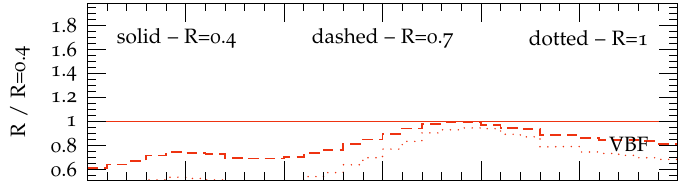}
    \includegraphics[width=\textwidth]{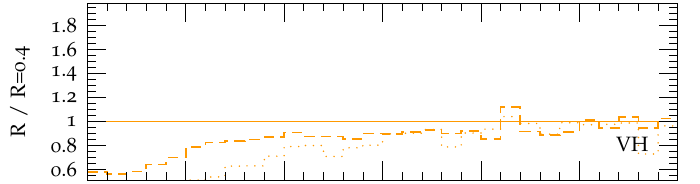}
    \includegraphics[width=\textwidth]{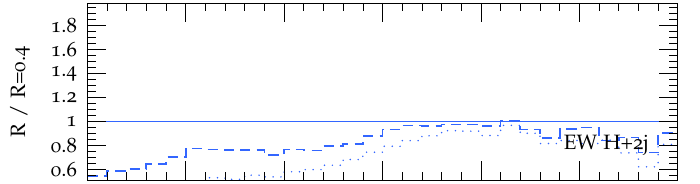}
    \includegraphics[width=\textwidth]{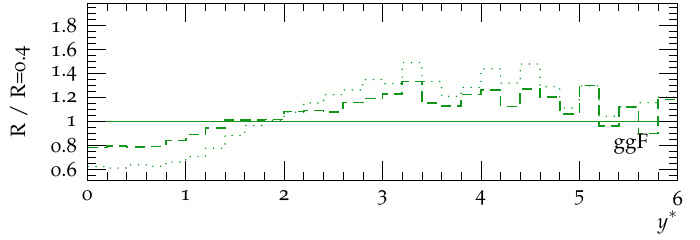}
  \end{minipage}\hfill
  \begin{minipage}{.2925\textwidth}
    \includegraphics[width=\textwidth]{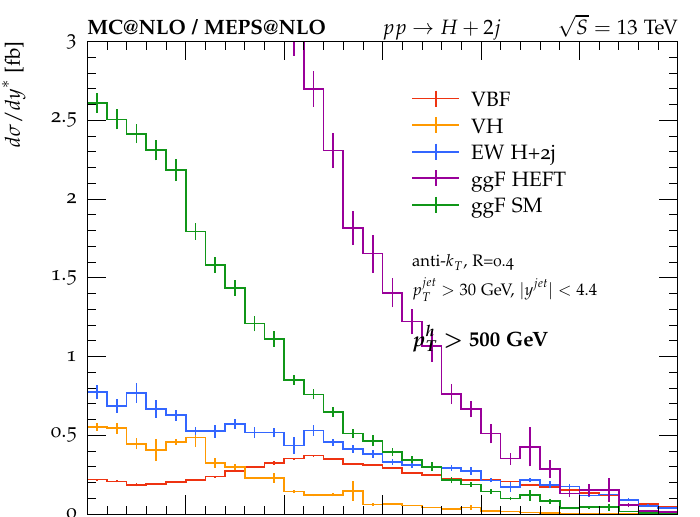}
    \includegraphics[width=\textwidth]{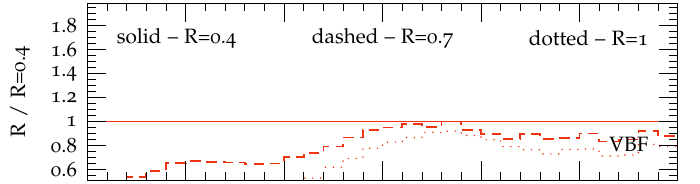}
    \includegraphics[width=\textwidth]{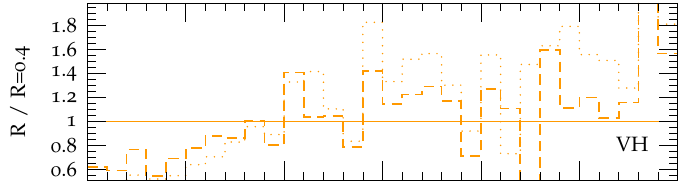}
    \includegraphics[width=\textwidth]{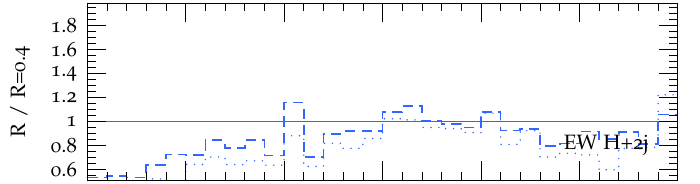}
    \includegraphics[width=\textwidth]{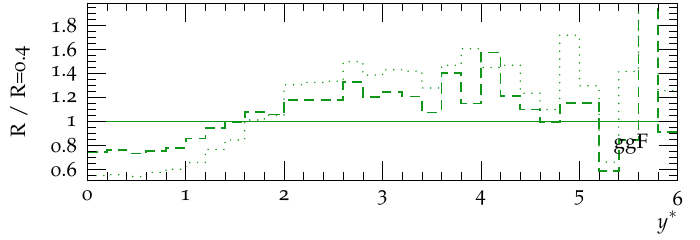}
  \end{minipage}
\caption{$y^*$ distribution.
The left panels show inclusive predictions, while the middle and right panels
show results for a minimum Higgs transverse momentum of 200 and 500~GeV.
See Fig.~\ref{fig:incl_njets} and the main text for details.}
\label{fig:incl_y_star}
\end{figure}

In Ref.~\cite{Rainwater:1996ud} it was suggested to use variables sensitive to the rapidity-gap between the two hardest jets to discriminate processes which proceed via a colorless $t$-channel signal exchange from the QCD background.
Distributions of the so-called Zeppenfeld variable $y^*$, defined as the shifted rapidity of the third leading jet, \emph{i.e.}
\begin{equation}
    y^* = \left| y_{j_3}- \frac{y_{j_1}+y_{j_2}}{2}\right|,
\end{equation}
are shown in Fig.~\ref{fig:incl_y_star} for the different Higgs boson production processes for the three different Higgs boson transverse momentum cuts.
We notice that the gluon-gluon and \VH{} contributions, which essentially are $s$-channel processes, are peaked at $y^*=0$, while \VBF is peaked around $y^*\approx 2.3$. For increasing Higgs $p_T$ cuts however the \VBF{} distribution becomes more and more flat.
\begin{figure}[p]
  \centering
  \begin{minipage}{.295\textwidth}
    \includegraphics[width=\textwidth]{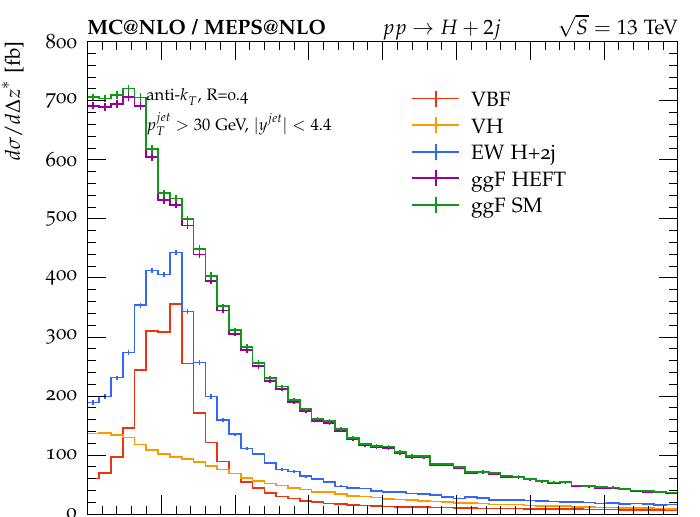}
    \includegraphics[width=\textwidth]{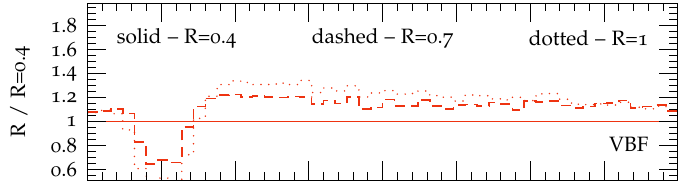}
    \includegraphics[width=\textwidth]{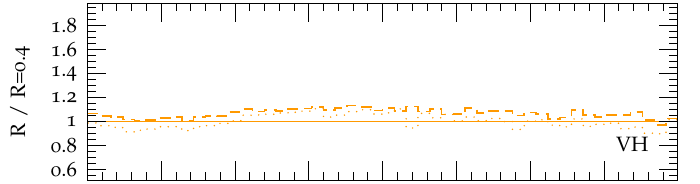}
    \includegraphics[width=\textwidth]{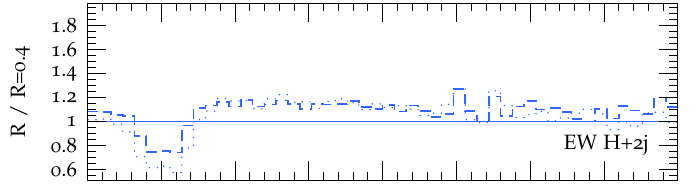}
    \includegraphics[width=\textwidth]{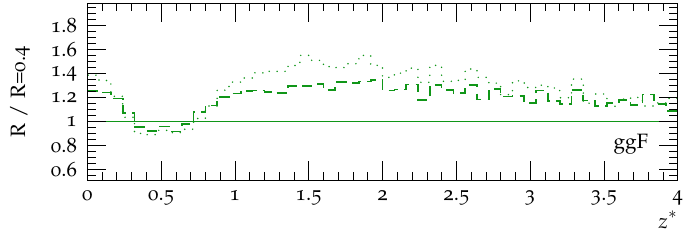}
  \end{minipage}\hfill
  \begin{minipage}{.295\textwidth}
    \includegraphics[width=\textwidth]{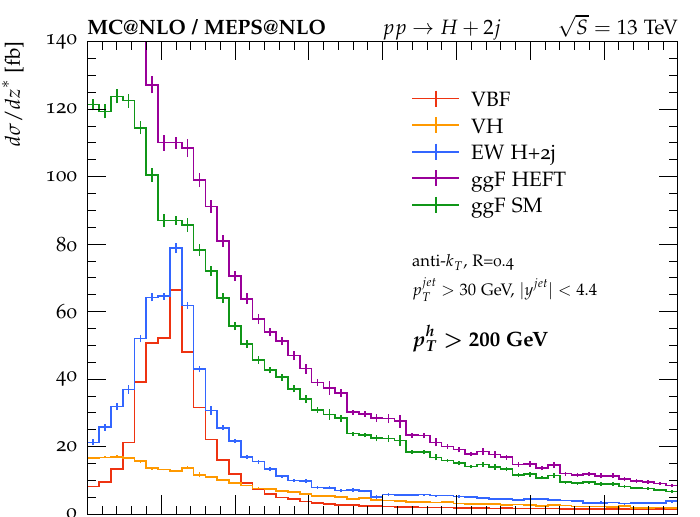}
    \includegraphics[width=\textwidth]{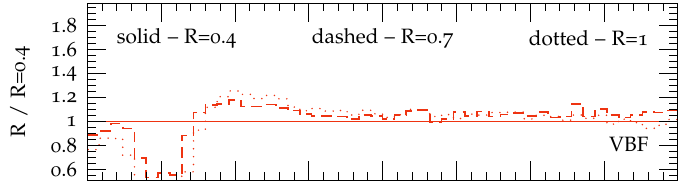}
    \includegraphics[width=\textwidth]{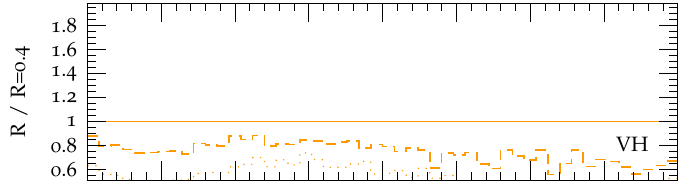}
    \includegraphics[width=\textwidth]{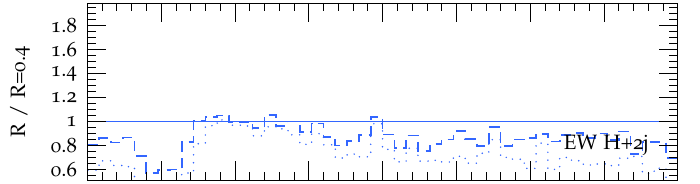}
    \includegraphics[width=\textwidth]{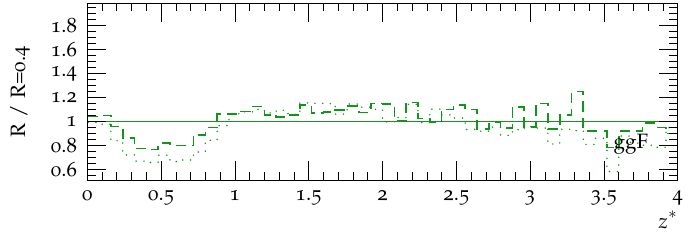}
  \end{minipage}\hfill
  \begin{minipage}{.295\textwidth}
    \includegraphics[width=\textwidth]{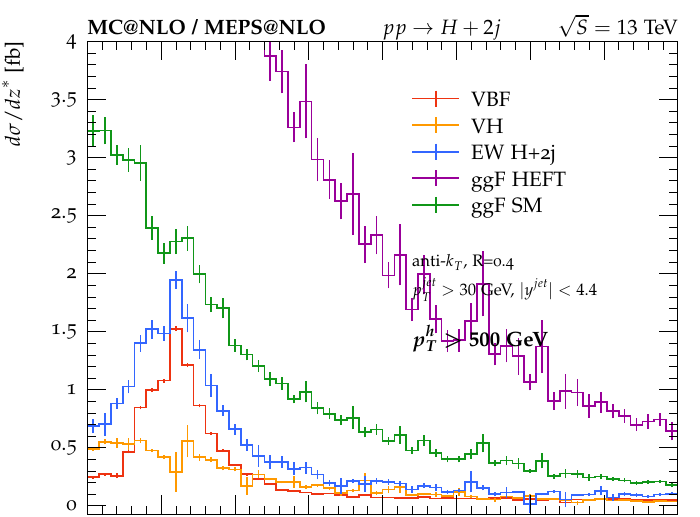}
    \includegraphics[width=\textwidth]{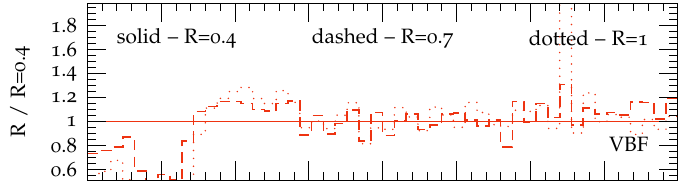}
    \includegraphics[width=\textwidth]{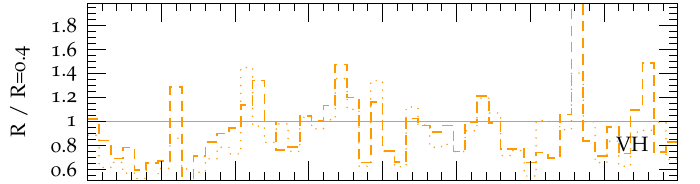}
    \includegraphics[width=\textwidth]{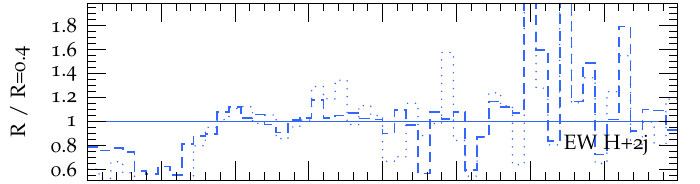}
    \includegraphics[width=\textwidth]{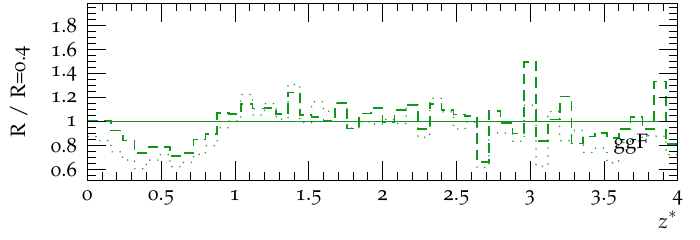}
  \end{minipage}
\caption{$z^*$ distribution.
The left panels show inclusive predictions, while the middle and right panels
show results for a minimum Higgs transverse momentum of 200 and 500~GeV.
See Fig.~\ref{fig:incl_njets} and the main text for details.}
\label{fig:incl_z_star}
\end{figure}

\begin{figure}[t]
  \centering
  \includegraphics[width=.65\textwidth]{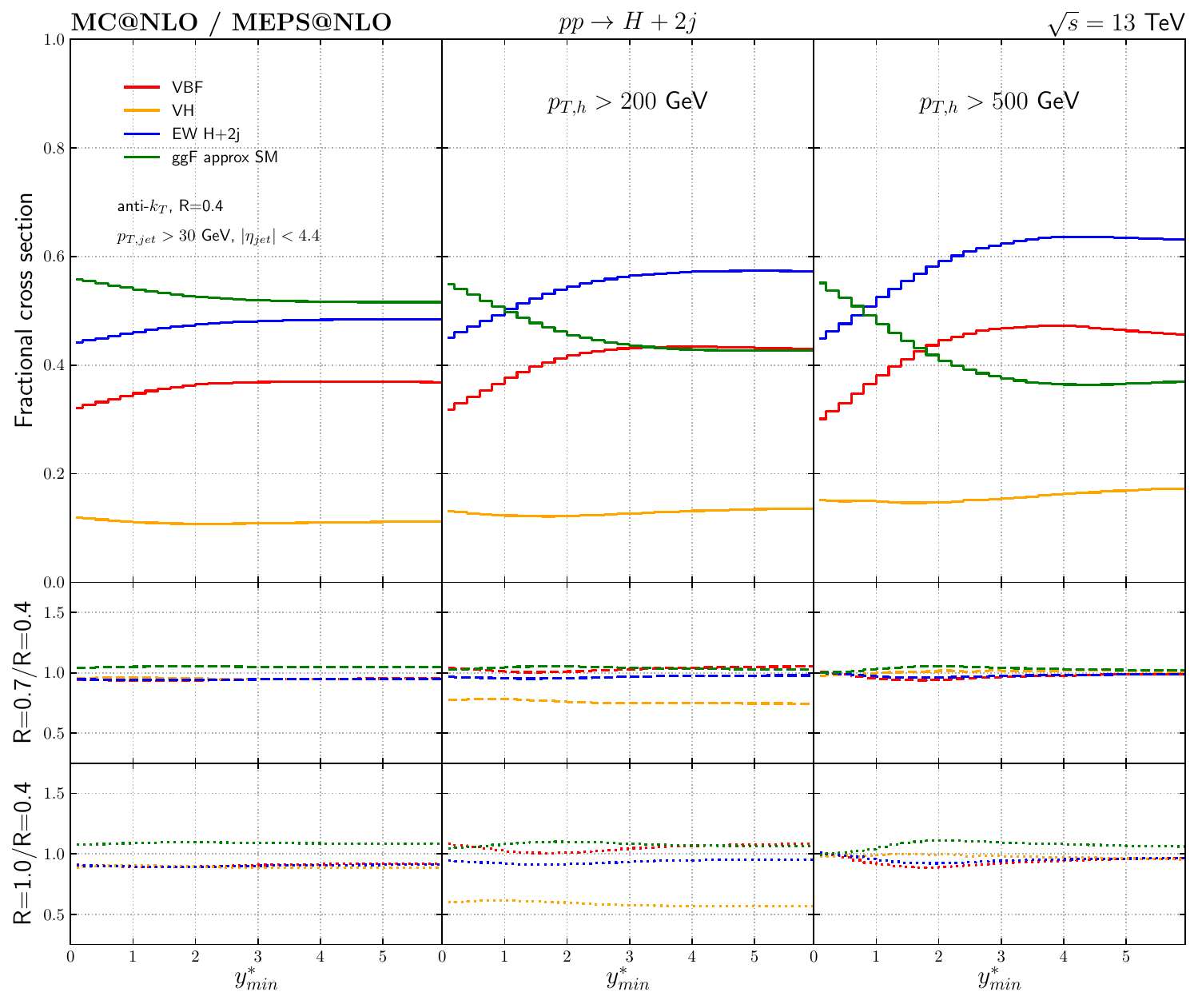}
  \includegraphics[width=.65\textwidth]{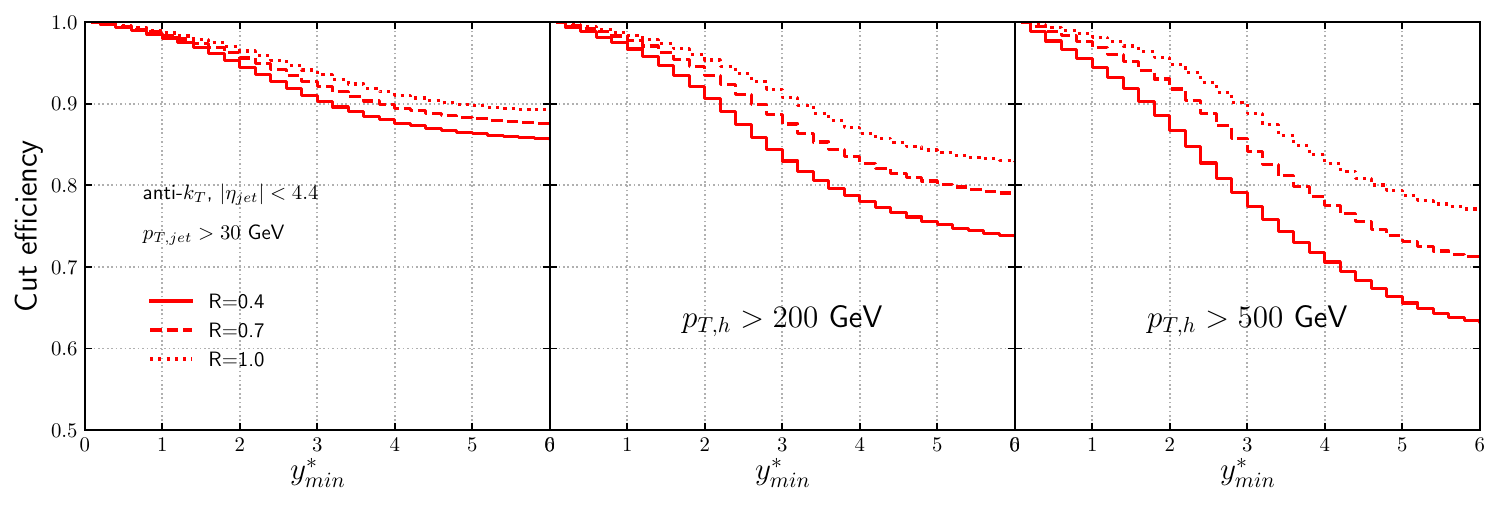}
\caption{Fractional cross sections of the various production channels as a function of the Zeppenfeld variable $y^*_{\min}$.
The offset is given by the exclusive two-jets contribution, which has been added to the three-jets inclusive predictions with $y^*>y^*_{\min}$.
In the bottom panel we show the acceptance probability for the \VBF{} contribution for three different jet radii.}
\label{fig:incl_y_star_fraction}
\end{figure}

Since requiring only two tagged jets in the final state allows for a higher signal-to-background ratio, and jets in gluon-fusion are predominantly produced in the central region, one can think about adding to this sample events with at least three jet with $y^*>y^*_{\min}$. This is commonly referred to as a central jet veto.
The corresponding fractional cross sections and cut efficiencies are shown in Fig.~\ref{fig:incl_y_star_fraction}, where the abscissa represents the minimal value of $y^*$ employed to define the integrated contribution for the events with at least three jets, such that the bin corresponding to $y^*_{\min}=0$ coincides with the two-jets inclusive results, while for $y^*_{\min}\gg0$ we recover the exclusive two-jet cross-section.
We notice that the fractional cross sections becomes essentially flat for large $y^*_{\min}$.
Marginal improvements in the \VBF (EW $Hjj$) separation can be achieved with cuts on $y^*$, more effective at high Higgs boson $p_T$. In particular, for $p_{T,h}>500$~GeV, we notice that the relative size for \VBF{} is larger than 50\% for $y_{\min}^*\ge 3.5$. Adding the vetoed 3-jets sample with $y_{\min}^*=3.5$ increases the \VH{} acceptance to 10--20\% (larger increase for smaller $R$), thus it does not lead to a significant improvement on the sample statistics.

\begin{figure}[t]
  \centering
  \includegraphics[width=.65\textwidth]{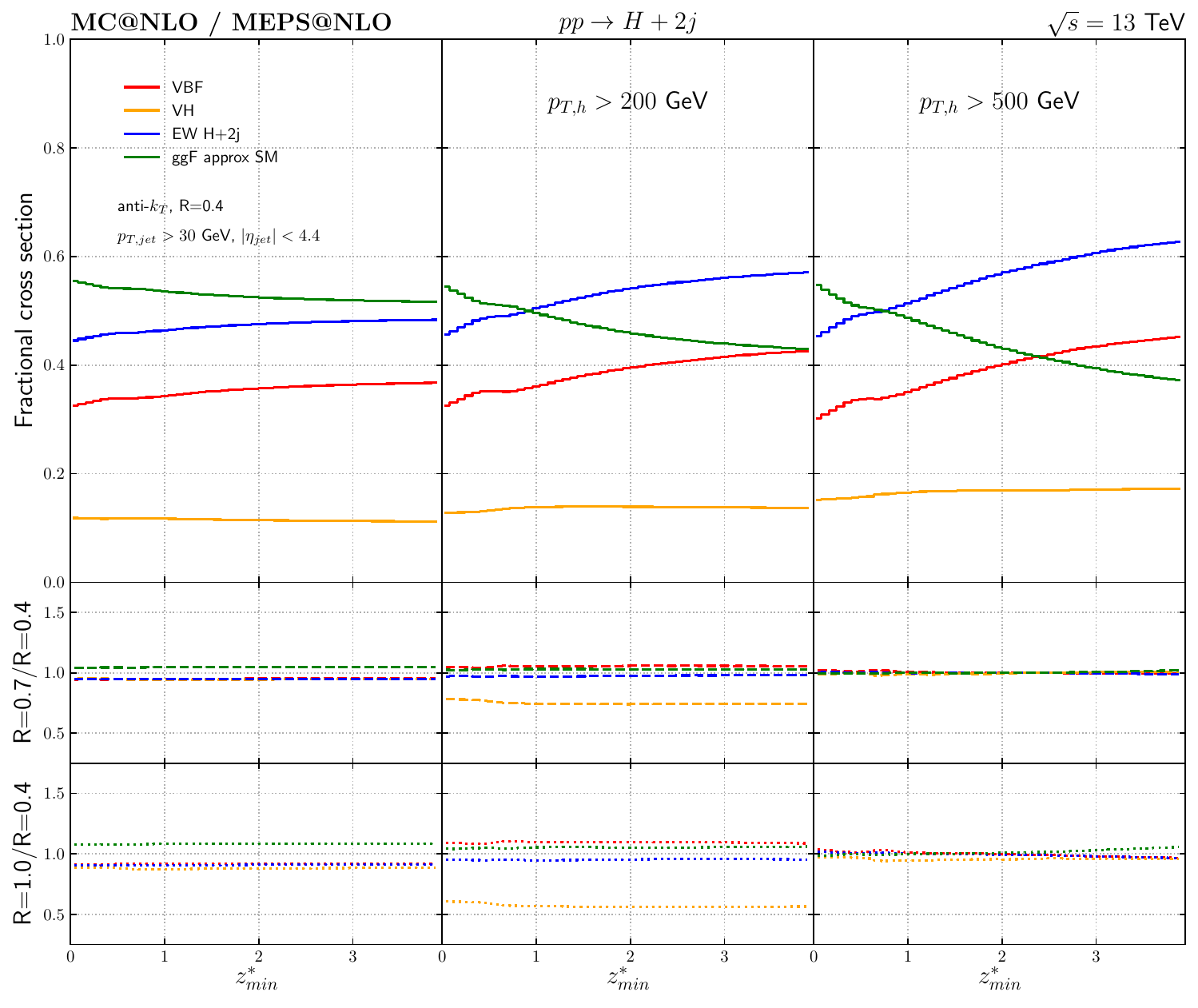}
  \includegraphics[width=.65\textwidth]{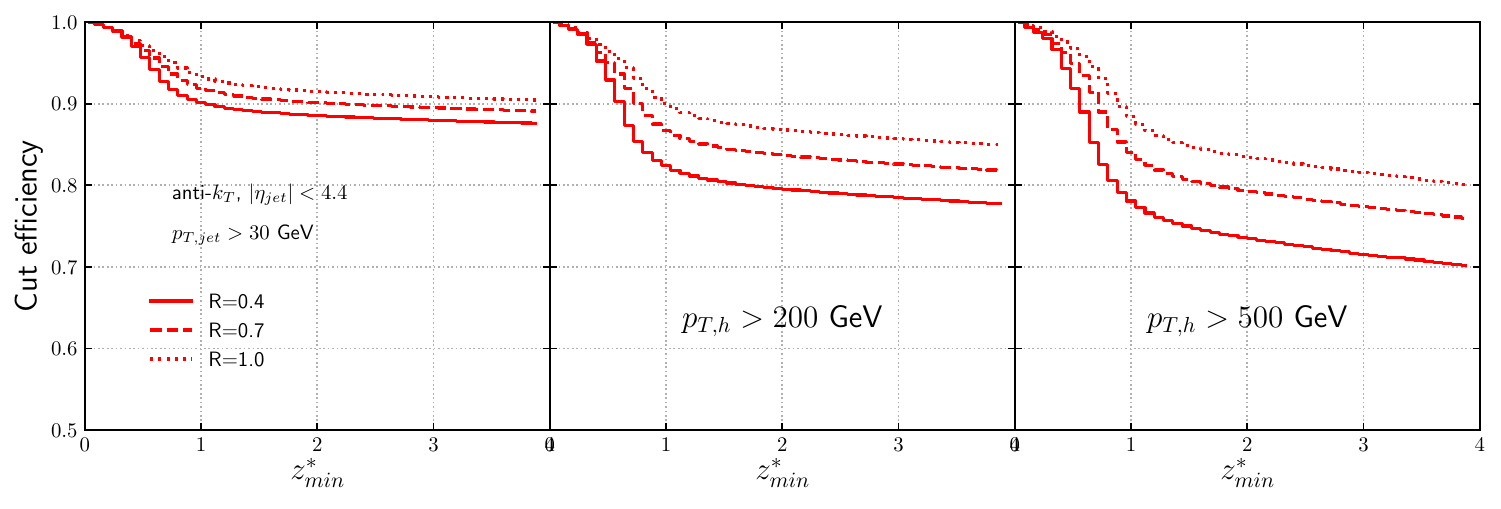}
\caption{Fractional cross sections of the various production channels as a function of the Zeppenfeld variable $z^*_{\min}$.
The offset is given by the exclusive two-jets contribution, which has been added to the three-jets inclusive predictions with $z^*>z^*_{\min}$.
In the bottom panel we show the acceptance probability for the \VBF{} contribution for three different jet radii.}
\label{fig:incl_z_star_fraction}
\end{figure}

Distributions of the normalized Zeppenfeld variable $z^*= y^*/\Delta y_{j_1 j_2}$ are shown in Fig.~\ref{fig:incl_z_star} for the different Higgs boson production processes for the three different Higgs boson transverse momentum cuts. Again we notice that ggF and \VH are peaked for $z^*=0$, while \VBF{} is peaked around $z^*=0.6$.
The fractional cross sections and cut efficiencies are shown in Fig.~\ref{fig:incl_z_star_fraction}.
We notice that, conversely to the $y^*$ case, the fractional cross section does not reach the asymptotic value given by the 2-jets exclusive sample fast enough, thus adopting a jet veto with a $z^*_{\min}$ cut that increases significantly the \VBF{} yield is less effective than using $y^*_{\min}$.

We conclude that the addition of events of large $y^*$ or $z^*$ to the exclusive 2-jet sample only yields marginal improvements in terms of signal-to-background ratio. On the other hand, it significantly complicates the event analysis due to the more intricate structure of the resummed higher-order QCD corrections, which contain non-global logarithms~\cite{Forshaw:2009fz,DuranDelgado:2011tp}, as it is not yet fully clear to which level of accuracy they are accounted for in parton showers.
We caution that the small improvement in statistics from events with $y^* >3.5$ for very large values of the Higgs transverse momentum will only be effective if it is matched by the theoretical precision, and that in absence of accurate predictions for non-global logarithms the global jet veto will be more practical. The corresponding computations can in principle be performed in an automated fashion with current technology~\cite{Gerwick:2014gya,Baberuxki:2019ifp,Baron:2020xoi,Caletti:2021oor}.

\section{The Higgs boson transverse momentum and its dependence on the jet cone size}
\label{sec:conesize}

With the comparisons of characteristic observables from gluon-gluon fusion, VBF and VH production channels of the Higgs boson, we haved revealed distinct properties of QCD activity accompanying Higgs boson production. Due to its $t$-channel topology, VBF stands out among gluon-gluon fusion and VH production channels in fiducial regions with large dijet invariant mass and rapidity separation. However, we would like to also emphasize the sizable event rates in all three production channels for boosted Higgs ($p_{T,h}>500$ GeV) with \emph{small} di-jet invariant mass (Fig.~\ref{fig:incl_m_jj12}) and rapidity separation (Fig.~\ref{fig:incl_delta_y_jj12}). Traditional fiducial cuts known as ``VBF cuts'' suggested in~\cite{Figy:2004pt,Ciccolini:2007ec,deFlorian:2016spz}, are likely to veto a significant number of events and alter acceptances of boosted Higgs boson production that might bear hints of new physics. In the following we perform a more detailed analysis of the VBF production channel in order to assess the stability and systematic uncertainty of the theoretical predictions.

\begin{figure}[p]
\centering
\includegraphics[scale=0.44]{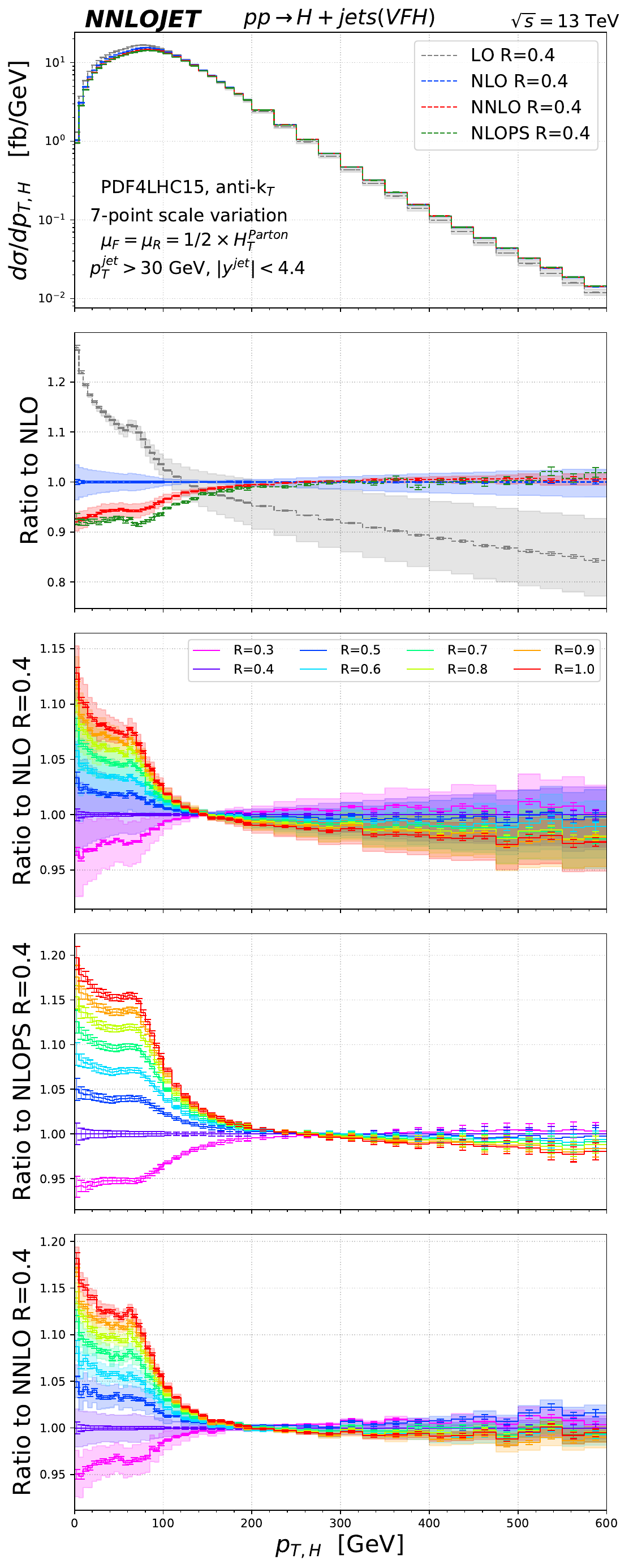}\hfill
\includegraphics[scale=0.44]{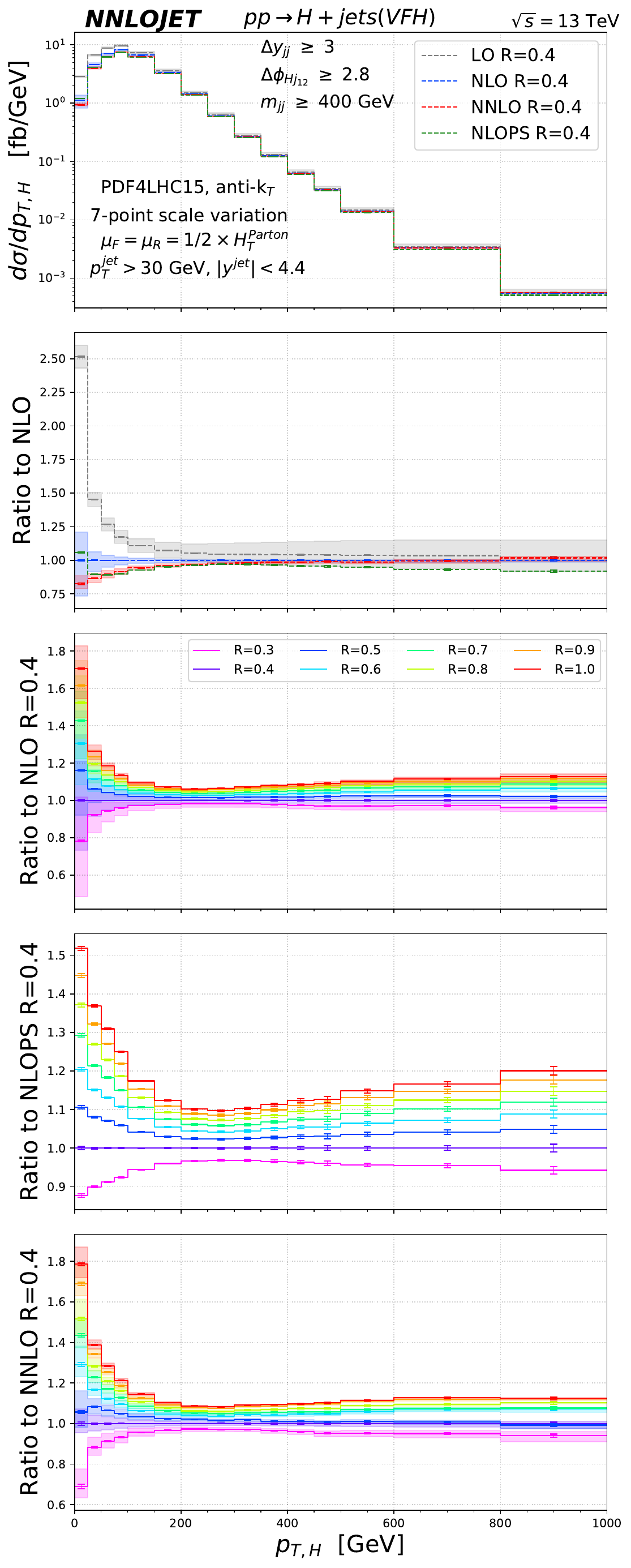}
\caption{The Higgs boson transverse momentum distribution from the \VBF production channel as a function of jet radius. The left panel is inclusive in the dijet final state phase space while the right panel is with VBF fiducial cuts used by ATLAS.}
\label{fig:fig_Higgs_pT_NNLO_incandATLAS}
\end{figure}

\begin{figure}[t]
\centering
\includegraphics[scale=0.3]{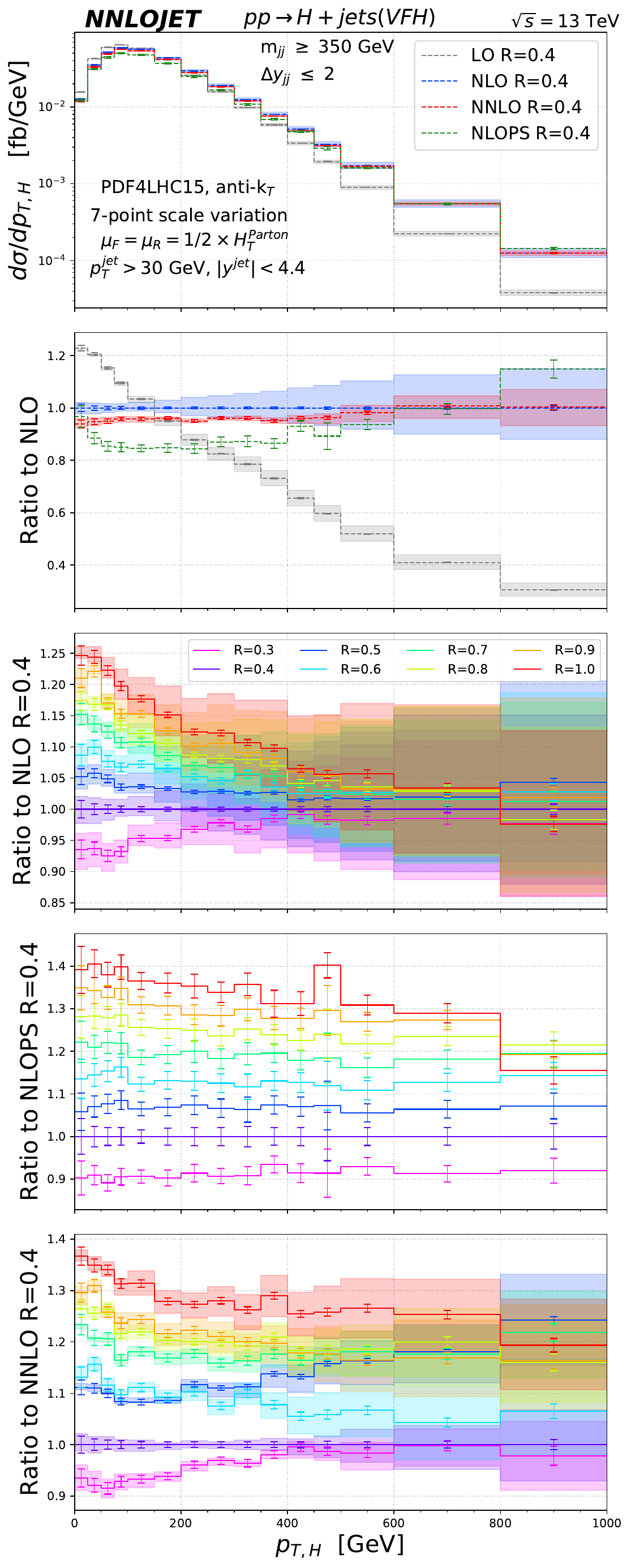}\hfill
\includegraphics[scale=0.3]{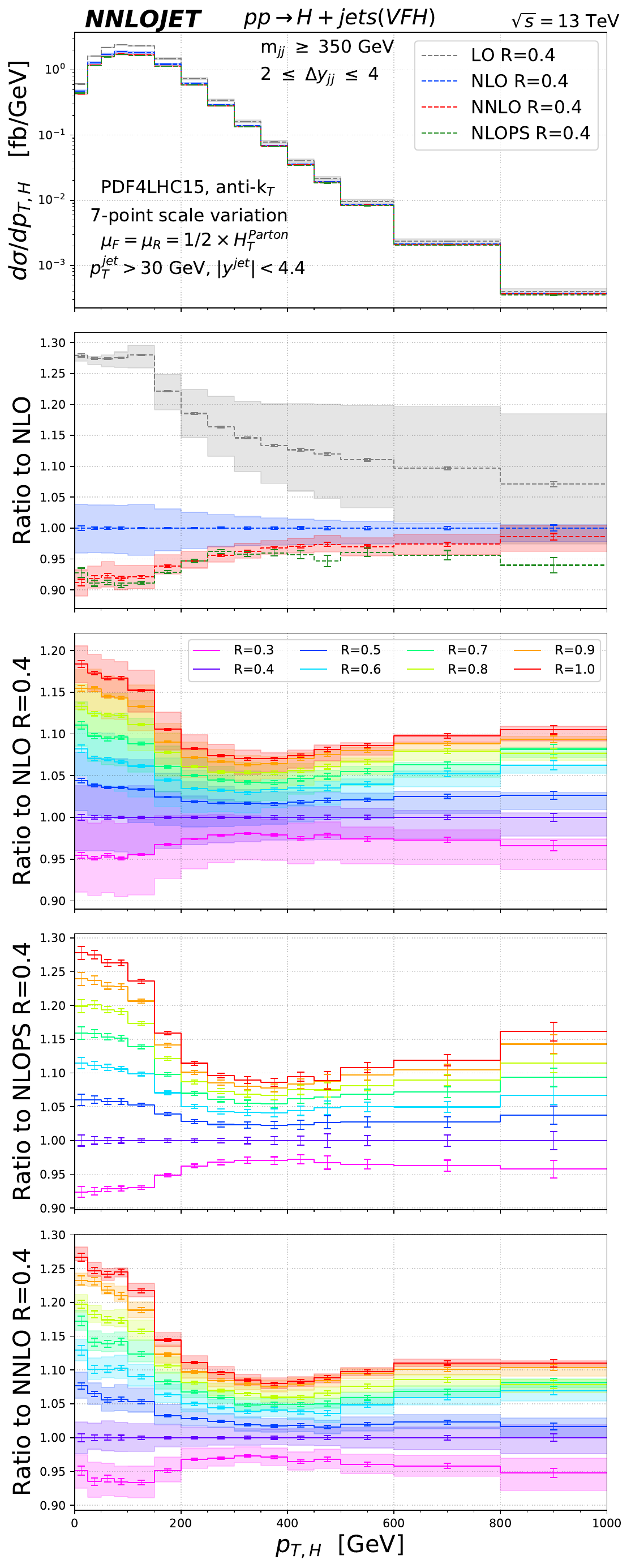}\hfill
\includegraphics[scale=0.3]{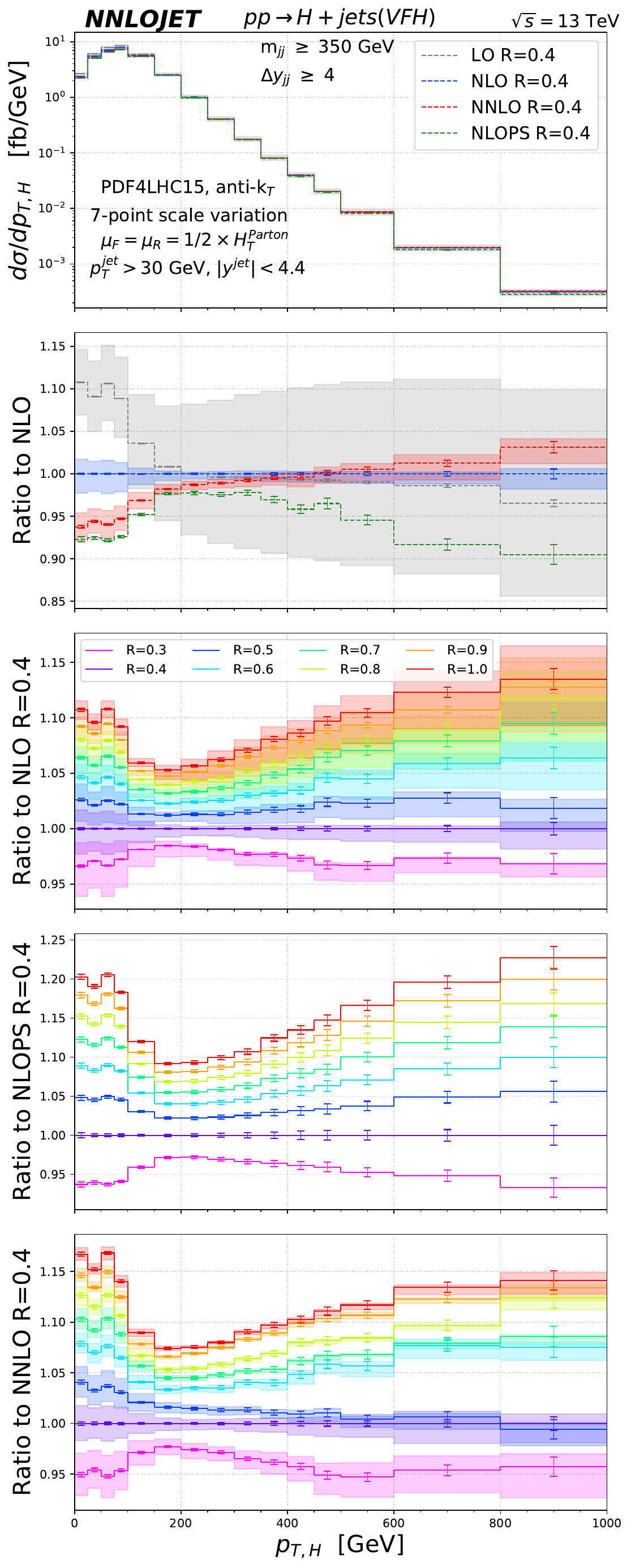}
\caption{The Higgs boson transverse momentum distribution from the \VBF sub-process as a function of jet radius using the \textit{heavy-center} (left), \textit{heavy-intermediate} (middle) and \textit{heavy-forward} (right) cuts.}
\label{fig:fig_Higgs_pT_NNLO_large_heavy-center}
\end{figure}
\begin{figure}[t]
\centering
\includegraphics[scale=0.3]{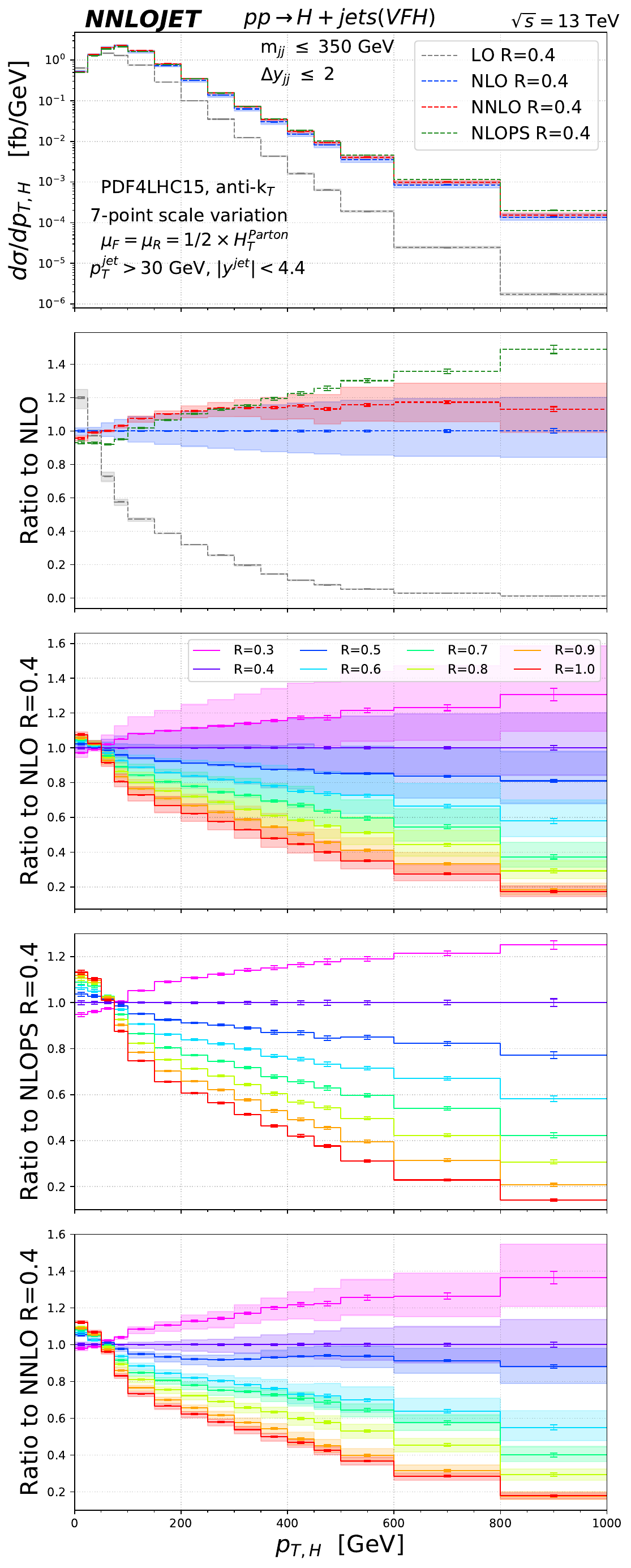}\hfill
\includegraphics[scale=0.3]{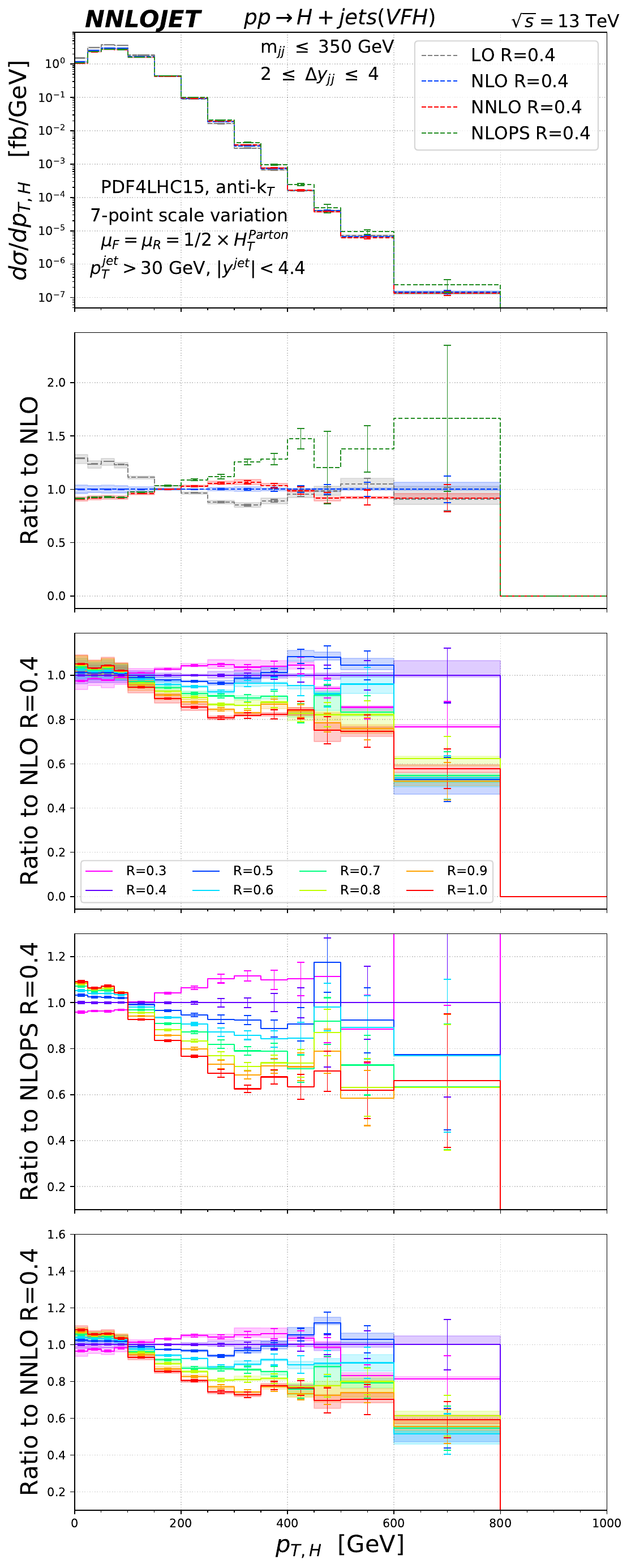}
\includegraphics[scale=0.3]{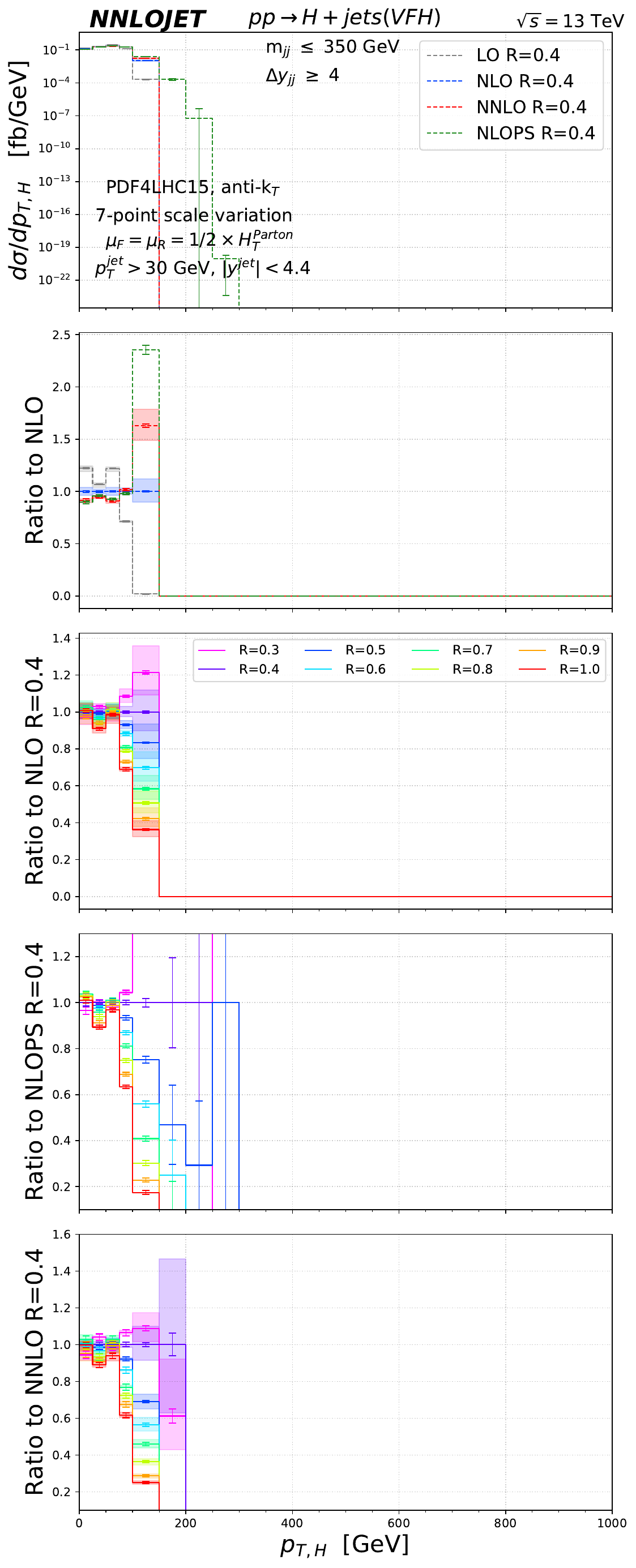}
\caption{The Higgs boson transverse momentum distribution from the \VBF sub-process as a function of jet radius using the \textit{light-center} (left), \textit{light-intermediate} (middle) and \textit{heavy-forward} (right) cuts.}
\label{fig:fig_Higgs_pT_NNLO_large_light-center}
\end{figure}

As a result of two jets being present already at the Born level, it is interesting to further investigate jet-radius-dependent effects.
In a previous study, some of us compared the effects of varying the jet radius $R$ in the anti-$k_t$ jet algorithm for two-to-two scattering processes at both fixed-order (with up to NNLO QCD corrections) and for NLOPS predictions, for dijet, Higgs-plus-jet and vector-boson-plus-jet productions~\cite{Bellm:2019yyh}. Here we present a complementary study for two-to-three processes from \VBF production. The jet-radius-dependent effects of \VBF production were first studied in~\cite{Rauch:2017cfu} with up to NNLO QCD corrections for fiducial total cross sections with ``VBF-cuts". In this work, we extend the study to differential observables. We apply different scenarios of fiducial cuts to exam their impact on the Higgs transverse momentum distribution and compare the fixed order predictions with parton shower corrections.
The fixed-order calculations include corrections up to NNLO QCD  using \NNLOJET, and NLOPS simulations are obtained from the Sherpa framework (Catani-Seymour dipoles).%
\footnote{NLOPS predictions obtained with Sherpa, Herwig, POWHEG (matched to Pythia or Herwig) are found to be in excellent agreement, thus in this section we only show Sherpa predictions.}
We observe in general good agreement between  the NNLO and NLOPS predictions,  with the exception of very large Higgs transverse momentum region with the application of VBF cuts. We will return to this mismatch later in this section.

The left panels of Fig.~\ref{fig:fig_Higgs_pT_NNLO_incandATLAS} illustrate the Higgs boson $p_T$ distribution as a function of the jet radius $R$, normalized to the results obtained with $R=0.4$, for inclusive selection cuts. The top panel shows a comparison between the central scale predictions from fixed-order calculations and Sherpa (NLOPS). The
higher order corrections and their scale uncertainties are normalized to those at NLO in the second row. Within the structure function approach, the LO results  have only a factorization scale dependence that arises from the DGLAP evolution of the quark PDFs. For a  Higgs boson produced with zero $p_T$, the center of mass energy is approximately 200~GeV (accounting for  the mass of the Higgs boson and the transverse momenta of the two (or more) jets). This mass/energy scale results in a small DGLAP evolution dependence.
For larger Higgs boson transverse momenta (200 to 600~GeV),  DGLAP evolution occurs  with a larger scale and the LO scale uncertainty increases from 5\% to 10\%.  With the addition of  higher order corrections, new initial state partonic channels  contribute and the resulting scale uncertainties depend on their interplay with the Born level subprocesses,  as well as on virtual corrections and the running of the strong coupling $\alpha_s$. We observe significant corrections from LO to NLO, with the NLO scale variation band outside the range of LO uncertainties, throughout the full transverse momentum region. Perturbative corrections start to converge at $p_{T,h}>200$ GeV from NLO with scale uncertainty about $\pm2\sim 3\%$. The NNLO corrections further reduce the scale uncertainty, to the level of $\pm 1$\%. The NNLO corrections also modify the central predictions, for $p_{T,h}\le 200$~GeV, so that the cross sections lie outside the NLO scale variation band. For $p_{T,h}\ge 200$~GeV, the NNLO scale variation band overlaps with the NLO uncertainty band. NLOPS predictions in this range  agree well with the NNLO fixed-order results. The effects of multiple gluon radiation from the parton shower seem to agree well with  the modifications of the NLO distribution resulting from the NNLO corrections.

In the third to the fifth row, we present  normalized distributions (with respect to the cross sections for $R=0.4$) for eight different $R$ values from 0.3 to 1.0.  Jet-radius-dependent effects are shown for NLO (third row), NLO plus parton shower (fourth row) and NNLO (fifth row). For $p_{T,h}\ge 150-200$~GeV, there is a relatively small dependence of the cross section on the jet size. However, there is a sizable variation of the cross section in the small transverse momentum region. This is due to the requirement that there be at least two jets  in the event, each with $p_T \ge 30$~GeV.
For small $m_{jj}$ and  $\Delta  y_{j1,j2}$, the two leading jets and the accompanying real radiation tend to be combined into a single jet as the jet radius increases, and thus the event fails the minimum two jet requirement (this can be seen also in Figs.~\ref{fig:incl_delta_y_jj12_mc} and~\ref{fig:incl_m_jj12_mc}). In the large $m_{jj}$ and the large $\Delta y_{j1,j2}$ regions, the final state partons are well separated and more likely to be identified as separate jets.
With increasing jet radius, more radiation is added to the tagging jets, allowing the event to more easily pass the threshold $p_T$ requirement for the two final-state jets.
For $p_{T,h}\le 200$~GeV, the two leading jets tend to have
a moderate value of  $m_{jj}$ (peaked at 300 GeV in Fig.~\ref{fig:incl_m_jj12_mc}) and large $\Delta  y_{j1,j2}$ (peaked at 4.2 in Fig.~\ref{fig:incl_delta_y_jj12_mc}) resulting in the enhancements (or reduction for R=0.3) for the various cone sizes shown in the third to fifth row of Fig.~\ref{fig:fig_Higgs_pT_NNLO_incandATLAS}.
There is a shoulder for $p_{T,h}$ on the order of 70~GeV due to the minimum two jet requirement, after which all ratios converge to unity.

For $p_{T,h}\ge 200$~GeV, according to Figs.~\ref{fig:incl_delta_y_jj12_mc} and~\ref{fig:incl_m_jj12_mc}, the two leading  transverse momenta jets often originate from the same parent parton (due to hard gluon radiation off of the lead quark jet, and the relative softness of the 2nd quark tagging jet at high Higgs boson $p_T$).  As a result, the enhancement in the small $m_{jj}$ (peaked at 100 GeV) and small $\Delta y_{j1,j2}$ (peaked at 0.5) region results in the  differential $p_{T,H}$ cross section  decreasing mildly with increasing jet radius, due to the merging of the two jets arising from the same quark line.

Compared to the inclusive cross section with R$=0.4$, we observe a ``pinch point'', for NLO, NLOPS and NNLO, where the \VBF cross sections are insensitive to the  jet cone size. At this pinch point,  the predictions also have the minimum scale variation. In the vicinity of the pinch point, most of the events have two well-separated jets, for which a  variation of jet radius or a variation of the evolution scale would not affect the cuts applied to the inclusive Higgs $p_T$ cross section.
As a result, the scale uncertainty is also reduced.

The pinch point is at 150~GeV for NLO compared to 200~GeV for NNLO. One possible explanation for the difference is that the additional gluon radiation present at NNLO provides extra recoil for the Higgs boson.
The scale uncertainties around the pinch point are reduced from $\pm$0.2\% at NLO to $\pm$0.1\% at NNLO, with the numerical integration error at the same level.

In general for the Higgs $p_T$ spectrum, the scale variations are smaller for larger values of  jet R.  This supports the argument that there is a better cancellation of the real and virtual corrections as the \VBF phase space becomes more inclusive (more radiation absorbed in larger R jets).
This is in contrast to the gluon-gluon fusion channel of Higgs-plus-jet, vector-boson-plus-jet or dijet production in the study of~\cite{Bellm:2019yyh}, where, instead, scale variation uncertainties mildly increase with a larger choice of jet radius. We attribute the phenomenology to the accidental cancellation among the virtual and real radiation at smaller values of jet radius. In general, the impact of such behavior needs to be studied on a process-by-process basis.

The $R$-dependence for the NNLO \VBF cross section using the ATLAS fiducial selection criteria for \VBF production is shown in the right panel of Fig.~\ref{fig:fig_Higgs_pT_NNLO_incandATLAS}. In comparison to the inclusive case (in the left panel), the requirement that the two leading jets be well separated ($\Delta y_{jj} \ge 3$, $\Delta \phi_{Hj_{12}} \geq 2.8 $ and $m_{jj} \geq 400$~GeV) results in increasingly positive corrections as the  jet radius increases.
The enhancement observed  for Higgs transverse momenta near zero, for the largest cone sizes, has increased from 10\% (15\%) to 70\% (80\%) at NLO (NNLO). The  pinch point disappears at both NLO and NNLO.
The NLOPS predictions agree with the NNLO fixed-order results for $p_{T,h}\le 400$~GeV, and differ  from NNLO by 10\% as the Higgs boson $p_T$ approaches 1~TeV. Both the $R$-dependence of the jet cross sections and the size of the NLO and NNLO corrections depend on the amount of real radiation. Thus, it is no surprise that there are similarities between the NLO and NNLO K-factors, and the observed normalized $R$ variations for the cross sections.

In order to investigate further the dynamics of the \VBF production mode for different kinematics, we now analyze the differential distributions of the Higgs boson transverse momenta using the following fiducial selections for $m_{j1j2}$ and $\Delta y_{jj}$:
\begin{itemize}
    \item light: $m_{j1j2} \le 350$~GeV
    \item heavy: $m_{j1j2} \ge 350$~GeV
    \item center: $\Delta y_{jj} \le 2$
    \item intermediate: $2 \le \Delta y_{jj} \le 4$
    \item forward: $\Delta y_{jj} \ge 4$.
\end{itemize}
In Fig.~\ref{fig:fig_Higgs_pT_NNLO_large_heavy-center} and~\ref{fig:fig_Higgs_pT_NNLO_large_light-center}, the double differential $p_{T,h}$ distributions are shown in the six fiducial regions that arise from combining the  cuts described above: \textit{heavy-center}, \textit{heavy-intermediate}, \textit{heavy-forward}, \textit{light-center}, \textit{light-intermediate} and \textit{light-forward}. Typical experimental \VBF cuts correspond to the \textit{heavy-forward} case. From the distributions, it can be observed that the application of the  \textit{heavy-forward} cuts captures the bulk of the \VBF cross section, while the \textit{heavy-intermediate}, \textit{light-center} and \textit{light-intermediate} fiducial regions contain smaller, but still significant, fractions.

Figure~\ref{fig:fig_Higgs_pT_NNLO_large_heavy-center} reveals similar higher order corrections as observed in  Fig.~\ref{fig:fig_Higgs_pT_NNLO_incandATLAS}, as well as a similar jet radius dependence in the \textit{heavy-intermediate} and the \textit{heavy-forward} fiducial regions. However, the agreement between fixed order and NLOPS predictions deteriorates in the \textit{heavy-forward} case at large transverse momenta. This effect is identical to the one observed in Fig.~\ref{fig:fig_Higgs_pT_NNLO_incandATLAS} (right) and seems worrisome because it occurs in a region that should nominally not be affected by resummation effects. We have analyzed it in detail and found it to be related to an inappropriate choice of the factorization scales.

We remind the reader that the fixed-order calculations are performed in the structure function approach, which is an excellent approximation to the full Standard Model result for VBF-type processes in the phase-space regions singled out by the \textit{heavy-forward} jet requirements. It implies in particular that there are two entirely independent sets of higher-order QCD corrections to the perturbatively computed hadronic tensor for each incoming beam. In order to minimize these corrections, the perturbative calculation should be performed using the characteristic energy scale at which the hadronic tensor is probed by the reaction. This energy scale should be set by the $t$-channel virtualities of the exchanged electroweak bosons which fuse into the Higgs boson. In the case of large Higgs $p_T$ and \textit{heavy-forward} jet cuts, these $t$-channel virtualities will exhibit a large hierarchy, because the Higgs boson tends to recoil against a single, high-$p_T$ leading jet, with the second jet having a small transverse momentum. This effect is not captured by the global factorization scale in Eq.~\eqref{eq:murf}. Similar effects also appear in the higher order corrections to single top quark production in the $t$-channel~\cite{Campbell:2020fhf}. In order to reflect the true dynamics of the reaction, one would need to define two scales, $H_T^{\text{parton}}$, related to the two independent beams. Any longitudinally boost invariant global scale will inevitably induce potentially large logarithmic corrections that are due to the DGLAP evolution of the PDFs. These are corrected for at fixed order, leading to very stable results when going from NLO to NNLO. However, the agreement between LO and NLO could be improved substantially by choosing appropriate independent factorization scales. The same is true for the agreement between NNLO and NLOPS predictions, because the latter suffer from the fact that the resummed higher-order corrections implemented by the NLOPS are assumed to be associated with a single, global factorization scale of the form of Eq.~\eqref{eq:murf}. We have verified this hypothesis using a leading-order parton shower simulation that was equipped with a true structure function approximation in the sense that we chose $\mu_F$ to be the individual $t$-channel virtualities of the electroweak bosons.

For the \textit{heavy-center} fiducial region, at small $p_{T,h}$, the jet cone size dependence and the higher order corrections are similar to those observed in the  \textit{heavy-intermediate} and \textit{heavy-forward} region, while being substantially different at large $p_{T,h}$. The rapidly falling LO contribution at large $p_{T,h}$ indicates that events are being discarded for not satisfying the requirement of reconstructing two jets from two partons.
With the presence of multiple real emissions from higher order corrections and parton showers, we observe a sensitivity to the choice of jet radius  in the NLOPS and NNLO predictions. An enhancement of the cross section with increasing jet cone size at  high $p_{T,h}$ is observed, which is due to the inclusion of more energy  in the two leading jets, leading to a greater potential for passing the requirement of $m_{j1j2} \ge 350$~GeV.

For the light sector shown in Fig.~\ref{fig:fig_Higgs_pT_NNLO_large_light-center}, the steeply falling  distributions at large $\Delta y_{jj}$  indicate that the two leading jets rapidly run out of available phase space  to recoil from a Higgs boson with large $p_{T,h}$. Higher order, as well as parton shower corrections, are essential to provide more phase space volume.
Increasing the jet radius results in more partonic emissions being clustered into one jet, which  tends to remove the event from the defined  light phase space.
We observe that the effects described above result in  a decrease of the cross section in the \textit{light-center} region  of more than 600\% when the jet radius changes from 0.3 to 1.0 (for $p_{T,h}$ at 1 TeV).

With a comprehensive study on the Higgs boson transverse momentum distributions in this section, we have quantified the phenomenological impact of different choices of jet cone size, as well as the properties of $p_{T,h}$ distributions outside the fiducial region defined by the ``VBF-cuts". With the observation of excellent agreement between NLOPS and NNLO in general, we also discover mild deviations in the boosted $p_{T,h}$ region when requiring large $m_{jj}$ and $\Delta y_{jj}$. To further quantify the consistency between various NLOPS tools and fixed-order calculations, we present comparisons of characteristic differential observables from the VBF production channel in the following section.

\section{Comparison of theoretical approaches}
\label{sec:tools}
In this section we compare the different computational approaches in some more detail, and in particular we examine
fixed-order and NLOPS predictions for the specific case of \VBF production for a jet size of $R=0.4$,
also quantifying the differences between different NLOPS simulations. Table~\ref{tab:xs_nlops} lists the 
fiducial cross sections in the various approaches.
\begin{table}[t]
    \centering
    \begin{tabular}{cc|clclccclcl}
         Code &&& \multicolumn{3}{c}{$\sigma$ [fb] (incl)}
         &&& \multicolumn{3}{c}{$\sigma$ [fb] (excl 2j)} \\\hline
         \NNLOJET (NLO) &&&  $1963.9_{-17.9}^{+15.4}$ & $\pm$ & 0.1
         &&&  $1645.2_{-67.7}^{+55.6}$ & $\pm$ & 0.1\\
         \NNLOJET (NNLO) &&& $1891.9_{-18.4}^{+18.5}$ & $\pm$ & 2.9
         &&& $1550.6_{-20.0}^{+27.5}$ & $\pm$ & 2.8\\
         Herwig Dipole &&& 1851.1 & $\pm$ & 3.2
         &&& 1561.1 & $\pm$ & 3.2\\
         Herwig $\tilde{q}$ &&& 1858.7 & $\pm$ & 8.9
         &&& 1555.2 & $\pm$ & 8.8\\
    \end{tabular}\hspace*{5mm}
    \begin{tabular}{cc|clclccclcl}
         Code &&& \multicolumn{3}{c}{$\sigma$ [fb] (incl)}
         &&& \multicolumn{3}{c}{$\sigma$ [fb] (excl 2j)} \\\hline
         POWHEG+Herwig &&& 1814.6 & $\pm$ & 0.3
         &&& 1534.9 & $\pm$ & 0.3\\
         POWHEG+Pythia &&& 1821.9 & $\pm$ & 0.3
         &&& 1555.1 & $\pm$ & 0.3\\
         Sherpa CSS &&& 1860.1 & $\pm$ & 0.7
         &&& 1586.5 & $\pm$ & 0.6\\
         Sherpa Dire &&& 1860.1 & $\pm$ & 2.7
         &&& 1594.0 & $\pm$ & 2.6\\
    \end{tabular}
    \caption{Cross sections for $R=0.4$ at NLO+PS accuracy, in comparison to fixed-order results.}
    \label{tab:xs_nlops}
\end{table}

\begin{figure}[tp]
  \centering
  \begin{minipage}{.295\textwidth}
    \includegraphics[width=\textwidth]{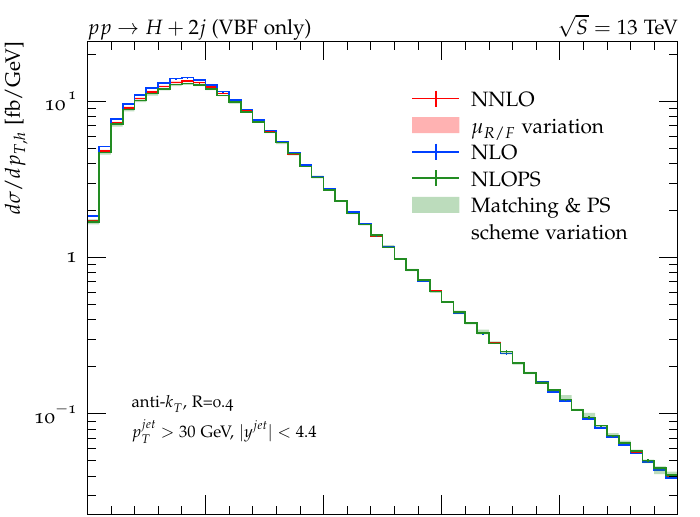}
    \includegraphics[width=\textwidth]{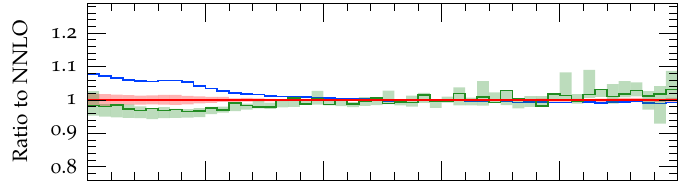}
    \includegraphics[width=\textwidth]{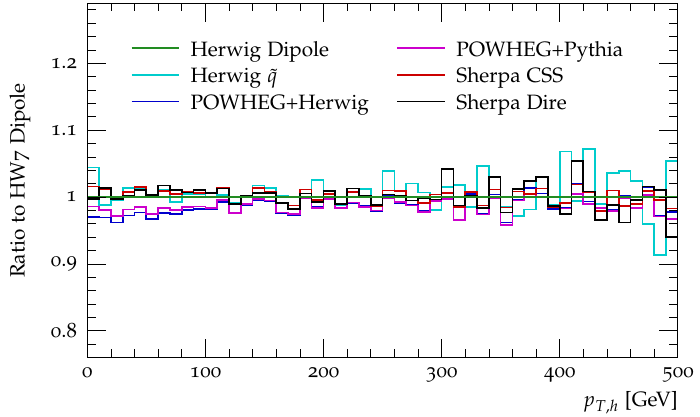}
  \end{minipage}\hskip 1cm
  \begin{minipage}{.295\textwidth}
    \includegraphics[width=\textwidth]{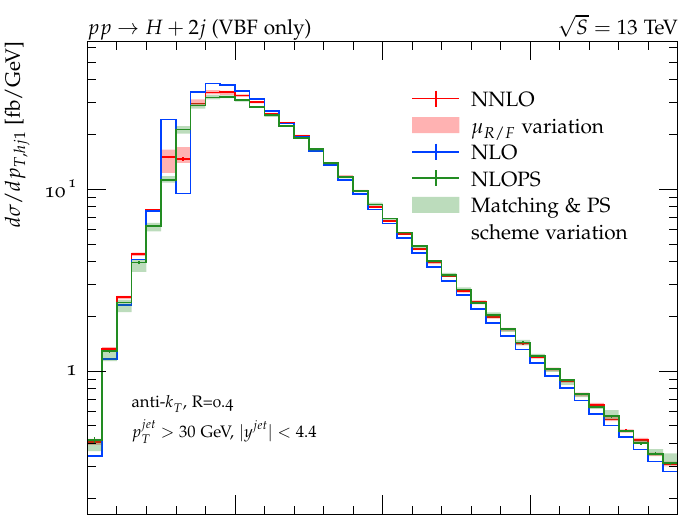}
    \includegraphics[width=\textwidth]{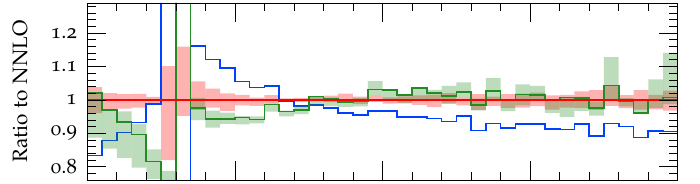}
    \includegraphics[width=\textwidth]{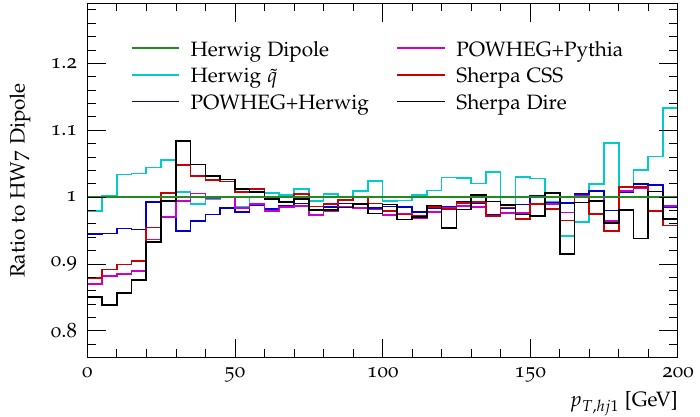}
  \end{minipage}\hfill
\caption{Higgs boson transverse momentum (left) and Higgs plus leading jet transverse momentum (right) distribution. We compare NLO and NNLO fixed-order predictions to NLOPS calculations from various event generators. The reference NLOPS generator is taken to be \HW{} standalone (MC@NLO matching, dipole shower).
    See the main text for details.}
\label{fig:incl_pth_mc}
\end{figure}

\begin{figure}[tp]
  \centering
  \begin{minipage}{.32\textwidth}
    \includegraphics[width=\textwidth]{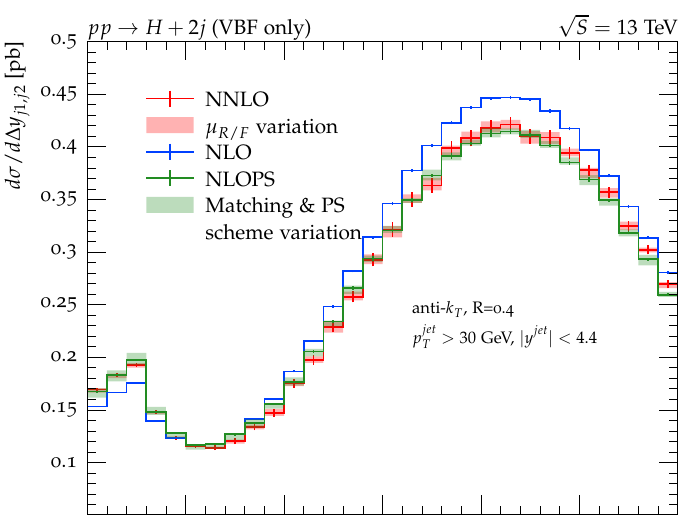}
    \includegraphics[width=\textwidth]{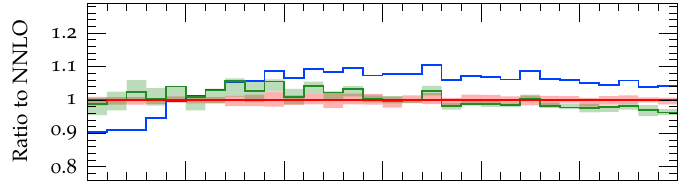}
    \includegraphics[width=\textwidth]{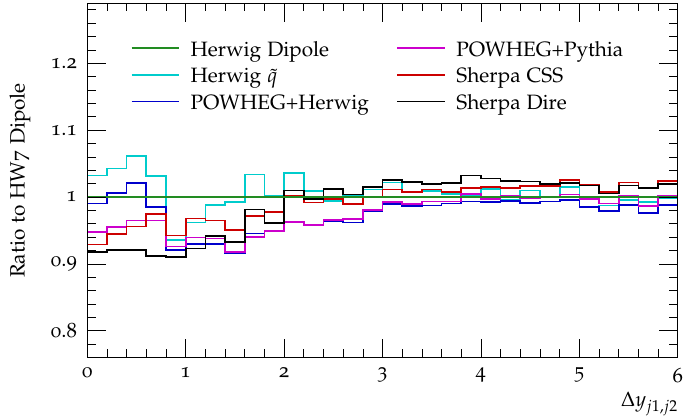}
  \end{minipage}\hfill
  \begin{minipage}{.32\textwidth}
    \includegraphics[width=\textwidth]{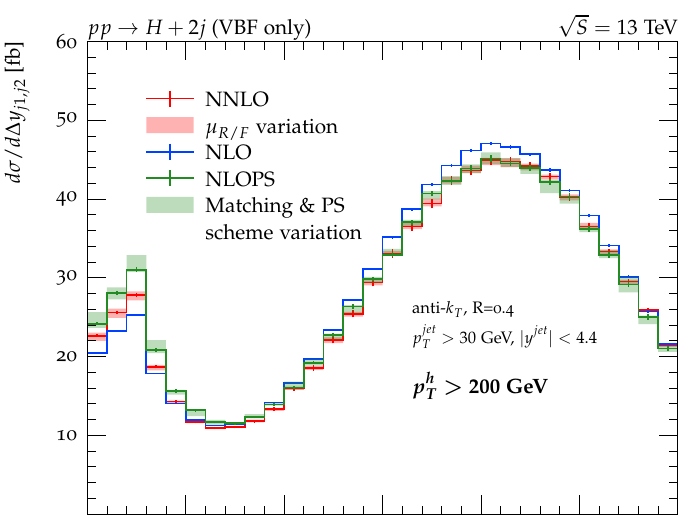}
    \includegraphics[width=\textwidth]{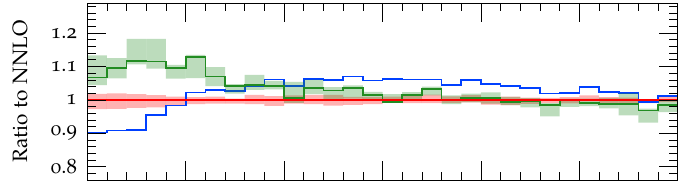}
    \includegraphics[width=\textwidth]{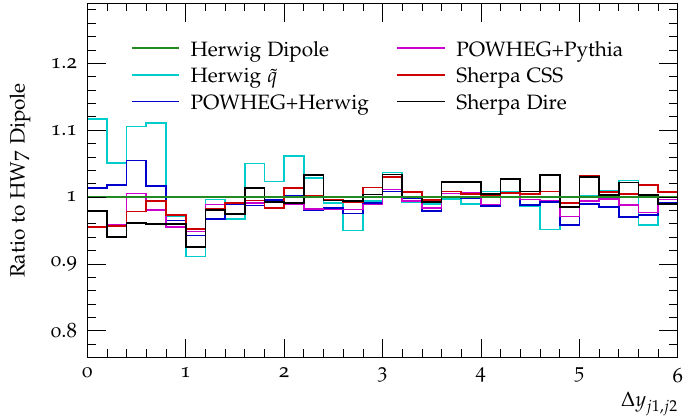}
  \end{minipage}\hfill
  \begin{minipage}{.32\textwidth}
    \includegraphics[width=\textwidth]{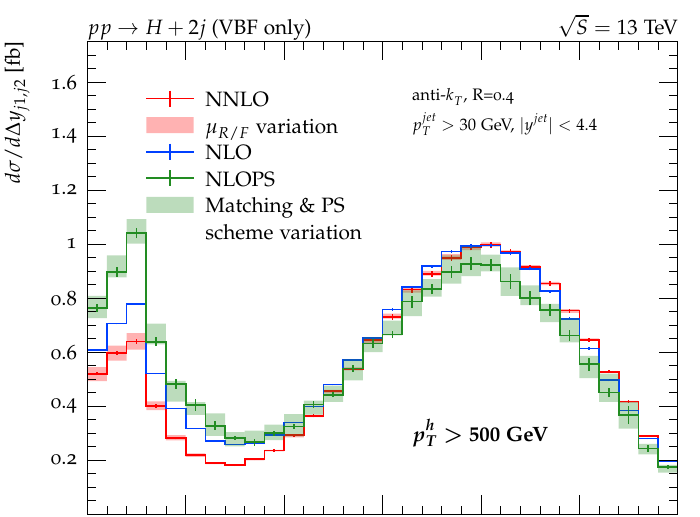}
    \includegraphics[width=\textwidth]{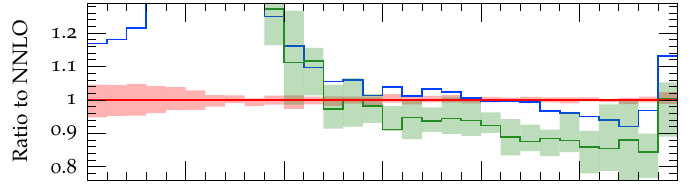}
    \includegraphics[width=\textwidth]{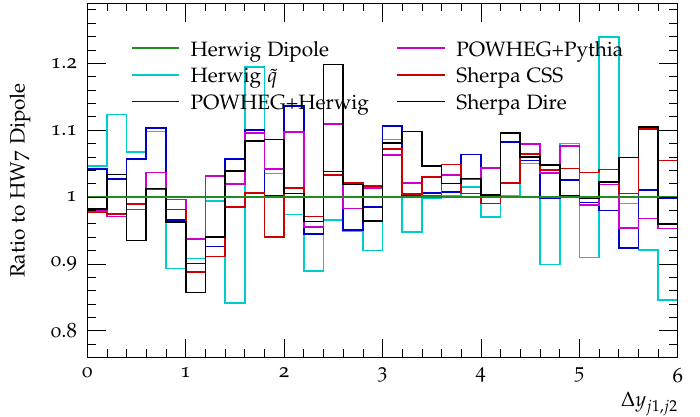}
  \end{minipage}
\caption{$\Delta y_{jj}$ distribution, using the two leading jets.
The left panels show inclusive predictions, while the middle and right panels
show results for a minimum Higgs transverse momentum of 200 and 500~GeV.
See Fig.~\ref{fig:incl_pth_mc} and the main text for details.}
\label{fig:incl_delta_y_jj12_mc}
\end{figure}

\begin{figure}[tp]
  \centering
  \begin{minipage}{.32\textwidth}
    \includegraphics[width=\textwidth]{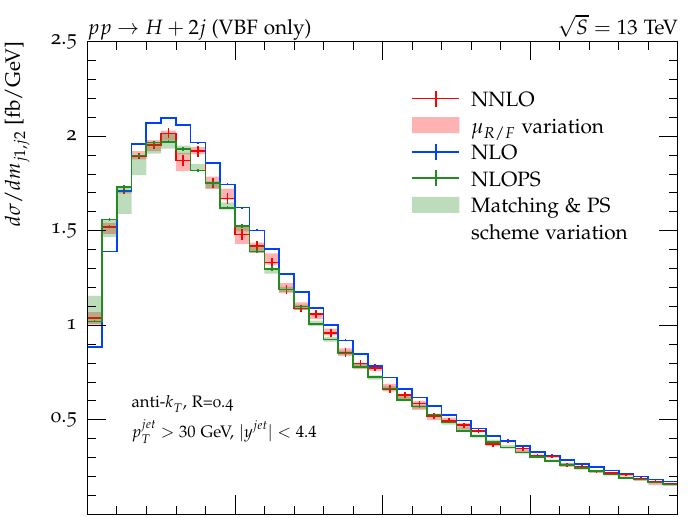}
    \includegraphics[width=\textwidth]{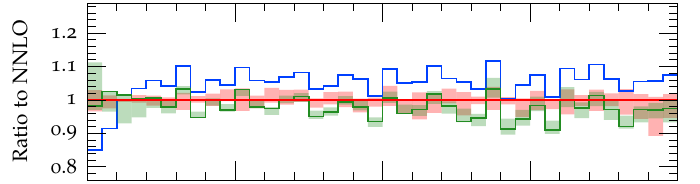}
    \includegraphics[width=\textwidth]{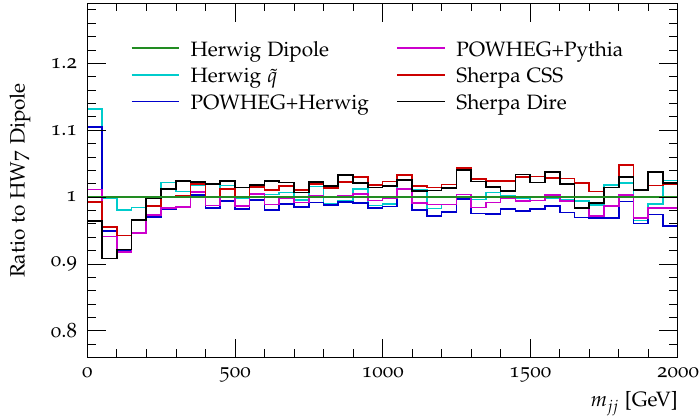}
  \end{minipage}\hfill
  \begin{minipage}{.32\textwidth}
    \includegraphics[width=\textwidth]{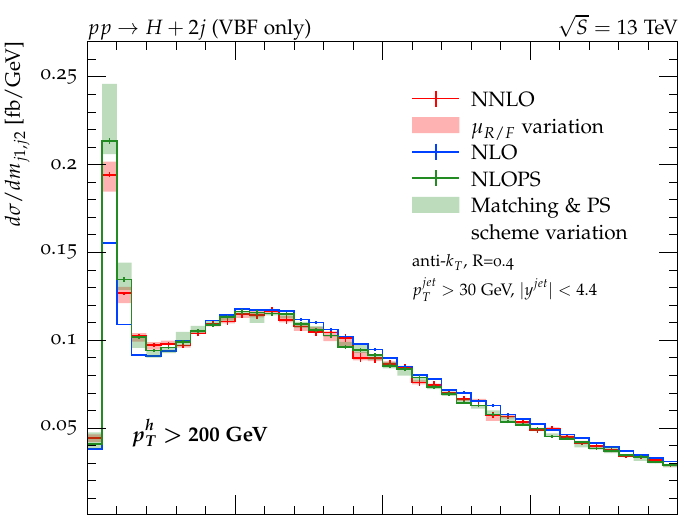}
    \includegraphics[width=\textwidth]{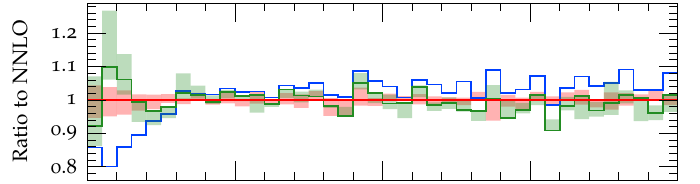}
    \includegraphics[width=\textwidth]{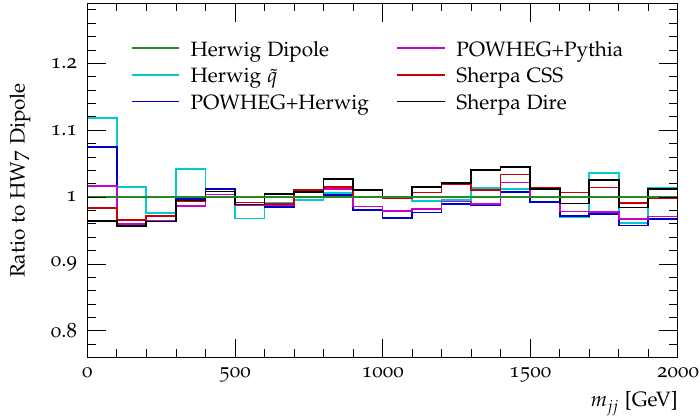}
  \end{minipage}\hfill
  \begin{minipage}{.32\textwidth}
    \includegraphics[width=\textwidth]{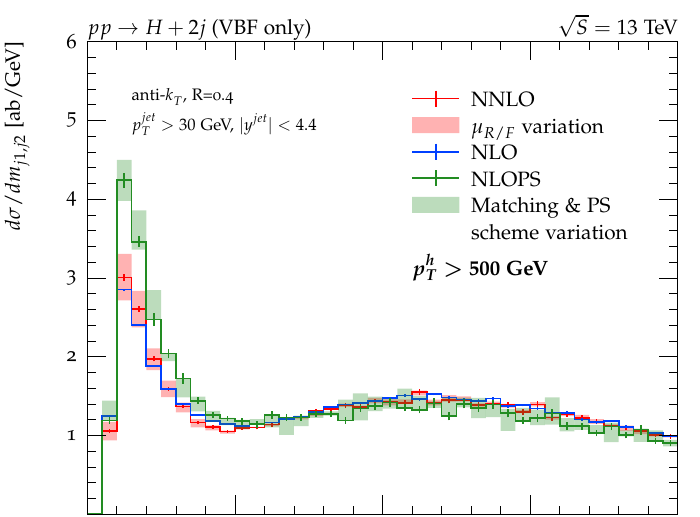}
    \includegraphics[width=\textwidth]{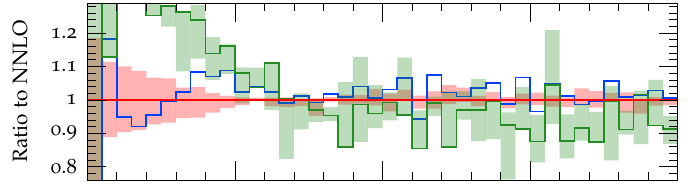}
    \includegraphics[width=\textwidth]{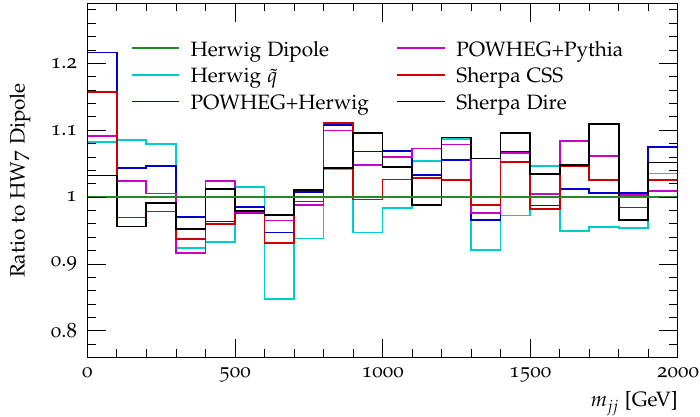}
  \end{minipage}
\caption{Dijet invariant mass distribution. The left panels show inclusive predictions,
while the middle and right panels show results for a minimum Higgs transverse momentum of 200 and 500~GeV.
See Fig.~\ref{fig:incl_pth_mc} and the main text for details.}
\label{fig:incl_m_jj12_mc}
\end{figure}

\begin{figure}[tp]
  \centering
  \begin{minipage}{.32\textwidth}
    \includegraphics[width=\textwidth]{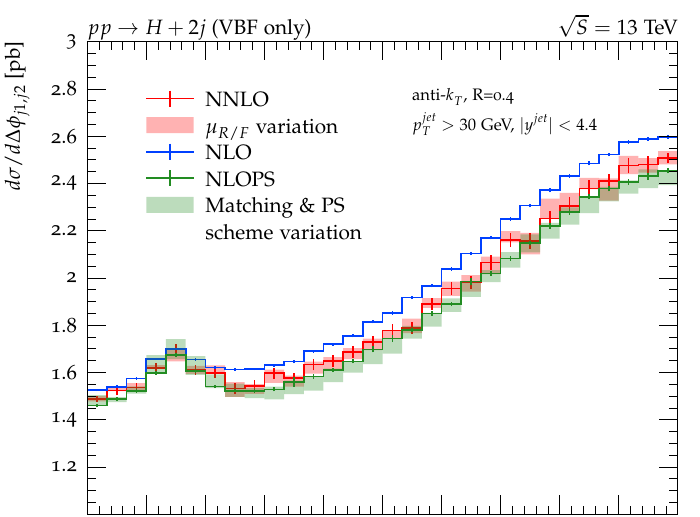}
    \includegraphics[width=\textwidth]{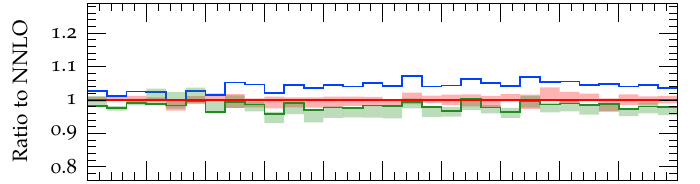}
    \includegraphics[width=\textwidth]{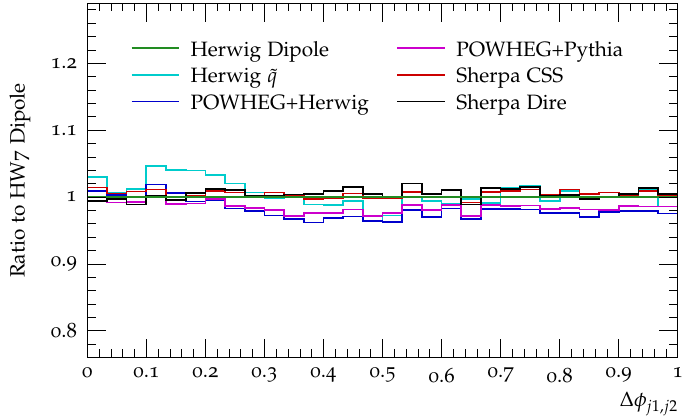}
  \end{minipage}\hfill
  \begin{minipage}{.32\textwidth}
    \includegraphics[width=\textwidth]{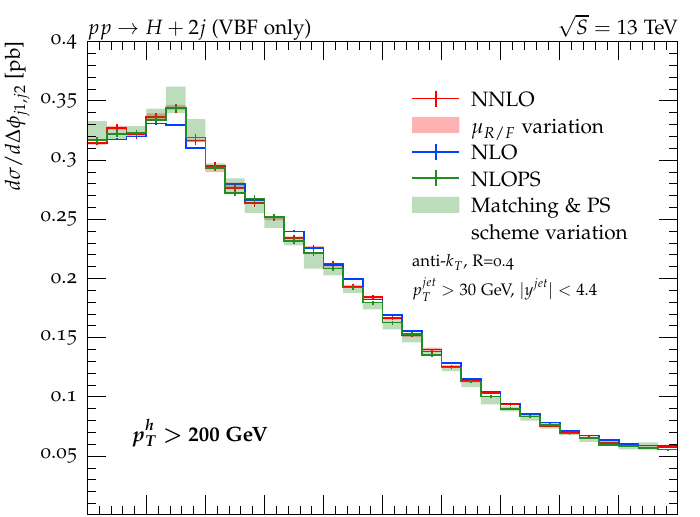}
    \includegraphics[width=\textwidth]{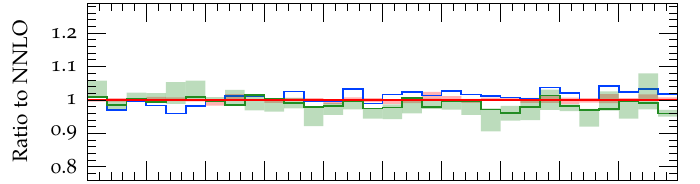}
    \includegraphics[width=\textwidth]{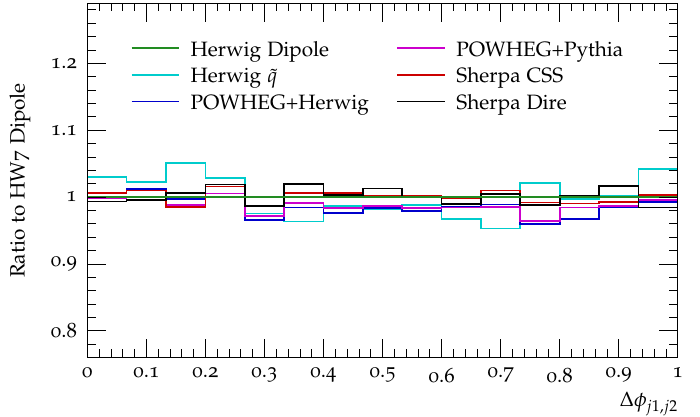}
  \end{minipage}\hfill
  \begin{minipage}{.32\textwidth}
    \includegraphics[width=\textwidth]{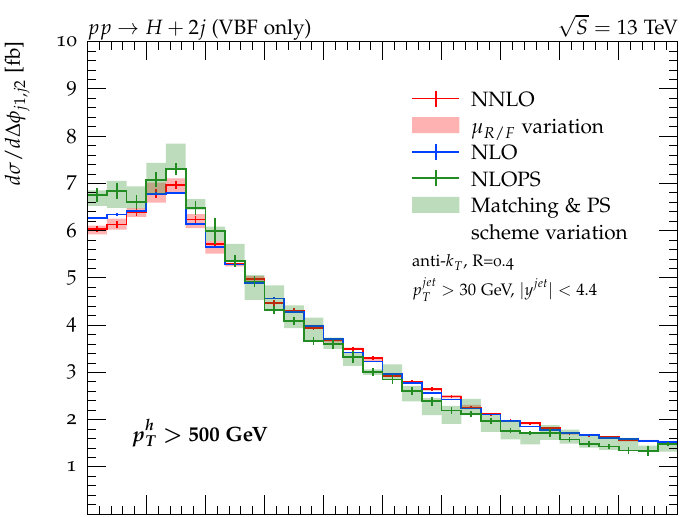}
    \includegraphics[width=\textwidth]{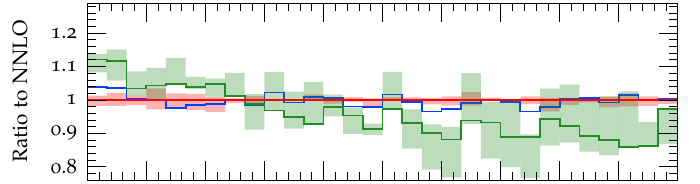}
    \includegraphics[width=\textwidth]{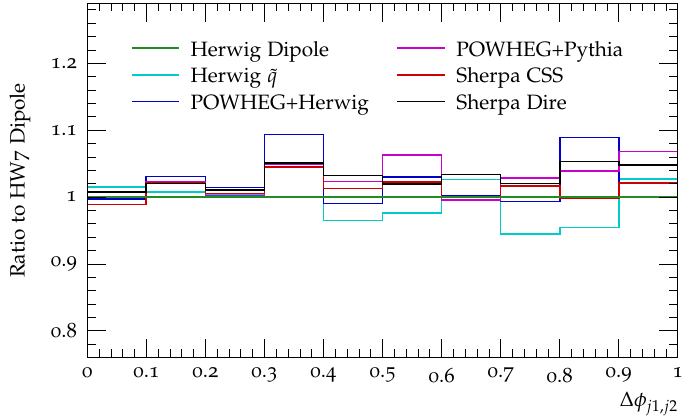}
  \end{minipage}
\caption{$\Delta \phi_{jj}$ distribution, using the two leading jets.
The left panels show inclusive predictions, while the middle and right panels
show results for a minimum Higgs transverse momentum of 200 and 500~GeV.
See Fig.~\ref{fig:incl_pth_mc} and the main text for details.}
\label{fig:incl_delta_phi_jj12_mc}
\end{figure}

\label{sec:resum}
\begin{figure}[tp]
  \centering
  \includegraphics[width=.475\textwidth]{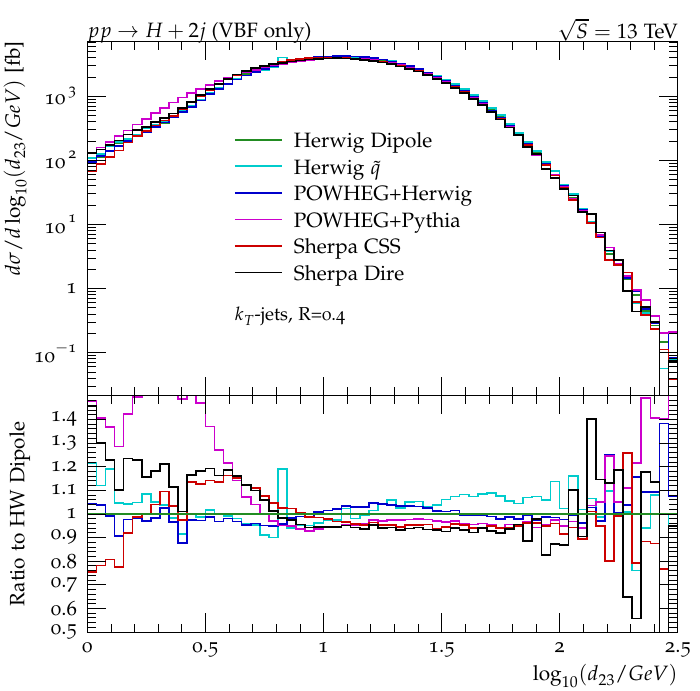}\hfill
  \includegraphics[width=.475\textwidth]{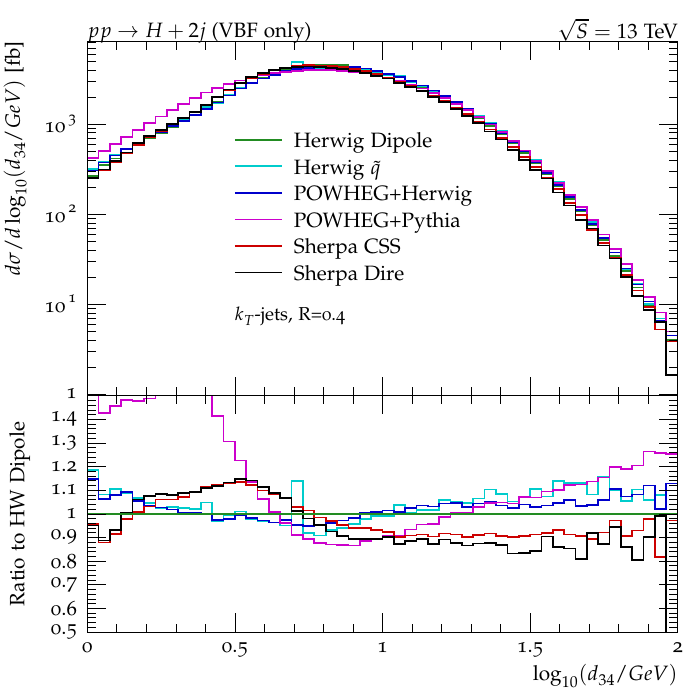}
\caption{$k_T$ jet rates at $R=0.4$}
\label{fig:kt_jet_rates}
\end{figure}

The top panels of Fig.~\ref{fig:incl_pth_mc} show a comparison between fixed-order calculations from \NNLOJET  for the Higgs boson and Higgs plus jet transverse momentum to \HW{} standalone (MC@NLO matching, dipole shower) results.
The green uncertainty band corresponds to the envelope of all the NLOPS predictions considered in this study. The second panels display the ratio of the fixed-order and the reference NLOPS predictions to the NNLO.
The bottom panels display the ratio of several NLOPS distributions to the reference one.
In the left panels, \emph{i.e.} for the Higgs transverse-momentum spectrum, we notice a good agreement between the NLOPS and NNLO results.
If instead we look at the transverse momentum of the system comprising the Higgs boson and the hardest jet, that corresponds to the transverse momentum of the second leading jet at LO, (right panels), we notice a good agreement only for $p_{T,hj_1}>50$~GeV. This is expected as the small transverse-momentum region is sensitive to the resummation of soft and collinear radiation. We also notice that for $p_{T,hj_1}<15$~GeV the difference between the several NLOPS predictions is smaller than 20\%, indicating a relatively small residual uncertainty.

In Fig.~\ref{fig:incl_delta_y_jj12_mc} we display the jet rapidity separation at the inclusive level, and with cuts on the Higgs $p_T$ of 200 and 500~GeV. Again, the NLOPS predictions agree to the percent level, except at small $\Delta y_{jj}$, where differences can grow up to 10\%. In comparison to NNLO fixed-order results, we find larger differences at high Higgs $p_T$. All the NLOPS simulations predict a larger peak at low $\Delta y_{jj}$ than the fixed-order predictions, as perhaps expected due to the radiative nature of that peak. It is surprising that the NLOPS predictions agree better with NLO than with NNLO for the highest Higgs boson $p_T$ cut. Note that matching scheme and parton-shower scheme variations are larger in relative size at large Higgs $p_T$. This is expected based on the additional scale hierarchies present in these configurations.

Figure~\ref{fig:incl_m_jj12_mc} shows similar comparisons between fixed-order calculations from \NNLOJET and the various NLOPS simulations for the dijet invariant mass spectrum at the inclusive level, and with cuts on the Higgs $p_T$ of 200 and 500~GeV. The NLOPS predictions agree to the percent level, except in the region around $m_{jj}\approx0$, where differences can grow up to 10\%. In comparison to NNLO fixed-order results, we find differences of up to 20\% in the low-$m_{jj}$ region, depending on the Higgs transverse momentum cut. The differences arise from the same source as for Fig.~\ref{fig:incl_delta_y_jj12_mc}.

Figure~\ref{fig:incl_delta_phi_jj12_mc} shows a comparison between fixed-order calculations from \NNLOJET and the various MC event generators for the jet azimuthal angle separation at the inclusive level, and with cuts on the Higgs $p_T$ of 200 and 500~GeV. The MC predictions agree to the percent level throughout, and the agreement with fixed-order NNLO calculations is at the same level, except for Higgs $p_T$ larger than 500~GeV, where differences can reach 10\%.

Most observables shown in this section are sensitive to physics simulations
at large scales, and are dominated by radiative effects that are best described by
fixed-order calculations at higher jet multiplicity. In order to demonstrate the
behavior of the simulations at low to intermediate momentum scales, where Sudakov
effects become important, we investigate the $2\to3$ and $3\to4$ jet resolution scales
in the longitudinally invariant $k_T$ algorithm~\cite{Catani:1993hr} with $R=0.4$.
The resolution variables $d_{nn-1}$ are similar to the Durham jet resolution scales
$y_{nn-1}$ in $e^+e^-$ collisions and are defined as the scale at which $n$ subjets
merge into $n-1$. The predictions from the various MC event generators
are shown in Fig.~\ref{fig:kt_jet_rates}. These observables are highly sensitive
to the QCD radiation pattern and show differences at the 10-20\% level between most
NLOPS simulation, except {\tt POWHEG+Pythia} which predicts more radiation at smaller scales.
It will be interesting to investigate these differences in more detail and compare
the NLOPS results with NLL resummed predictions, once these become available.

We conclude this section by noting that the differences among the various parton-shower matched predictions, 
and also the differences between the fixed-order predictions and matched results, rarely exceed the few percent level.
Significant deviations are observed only in the regions of large Higgs bosons $p_T$, where the scale set 
by the transverse momentum induces an additional hierarchy. 
This situation is very different from the observations in Ref.~\cite{Ballestrero:2018anz}, where a large deviation between
the different parton-shower matched predictions was observed in the third jet rapidity, for example (see Fig.~19 ibidem).
The deviations observed in Ref.~\cite{Ballestrero:2018anz} can be attributed to an inconsistent setting 
of parton shower parameters, in particular a recoil scheme in \PY{} which does not maintain the virtuality
of $t$-channel electroweak propagators. We refer the reader to Ref.~\cite{Hoche:2021mkv} for a more detailed 
discussion of this effect, and for a more comprehensive analysis of matching uncertainties in the \PY{} event generator.

\section{Conclusion}
\label{sec:conclusions}
Electroweak Higgs boson plus two-jet production is one of the prime channels to study the structure of the electroweak sector of the Standard Model at the Large Hadron Collider. Future running at the  LHC will allow the accumulation of a very large data sample, pushing the Higgs boson measurements out to high transverse momentum, where the possibility of the discovery of new physics exists. The vector-boson fusion production mode in particular offers the opportunity for a detailed measurement of Higgs boson interactions involving the massive gauge bosons. Precise theoretical predictions for both signal and background processes are of utmost importance in order to harness the full statistical power of LHC event samples. Experimental analyses do however depend on a more detailed modeling of events, including in particular resummation effects, underlying event and hadronization. The proper description of these aspects of QCD hinges on Monte Carlo event generators, which describe the hard scattering process only at NLO accuracy. In this manuscript we have presented a comparison of the two types of calculations, \emph{i.e.} NLOPS and NNLO, and we have compared a wide range of parton showers, including in particular Herwig, Pythia and Sherpa,
and both  additive and multiplicative matching algorithms (MC@NLO and POWHEG). We have studied the jet-radius dependence of the cross sections as a means of better understanding the higher order corrections and the scale uncertainties at each order. We have investigated the behavior of the cross sections at high Higgs boson transverse momentum, a region not fully explored from the phenomenological perspective, but one accessible with the data to be taken at the high-luminosity LHC.

We have found generally good agreement between NLOPS and NNLO predictions, apart from some intricate effects related to choices of the factorization scale, which can induce large logarithmic corrections.   We have studied the \VBF production mode and associated backgrounds in detail over the full Higgs boson transverse momentum range. We have also found that there is a large $R-$dependence on the \VBF cross sections  that can be understood in terms of the kinematic  cuts applied to the event. The $R-$dependent effects are largest at low to moderate Higgs boson $pT$, and with the application of typical \VBF selection cuts.  We have examined the impact of a resonance dijet mass veto to reduce the \VH contribution, and have shown that at high $p_T$, a broader definition of the veto must be considered to achieve the best separation between the two processes. We have examined the effectiveness of a $\Delta y_{jj}$ cut, as a function of jet $R$, Higgs boson $p_T$ and definition of the dijet system and have shown that the effectiveness of such cuts decreases as the Higgs boson $p_T$ increases. At high $p_T$, the Higgs boson recoils primarily against the lead tagging jet, with the second (tagging) jet being relatively soft. The emission of an additional jet from the lead jet grows with the lead jet transverse momentum. Given the softness of the second tagging jet, the emitted jet often becomes the second highest $p_T$ jet in the event, often leading to low values of $\Delta y_{jj}$, if the two highest $p_T$ jets are chosen to define the \VBF kinematics.  The further improvement of the \VBF Higgs boson signal to background is possible through application of cuts on additional jets present in the event. We  have proposed the usage of a global rather than a central jet veto to enhance the signal-to-background ratio, both because of the effectiveness of such a cut and the relative ease of understanding its impact on the accuracy of the theoretical predictions. This will provide the opportunity to further increase the  precision of perturbative calculations by combining existing NNLO predictions with NLL resummation, also opening the possibility for further validation of Monte Carlo event generators in the Sudakov region.

A better understanding of processes involving jets at the LHC can be gained by a comprehensive study of the processes using both fixed order (NLO and NNLO) and NLOPS calculations, all conducted within a common framework. The variation of the jet radius allows a more detailed understanding of the higher order corrections and on the remaining scale dependence. This paper follows a study~\cite{Bellm:2019yyh} with a similar spirit carried out for dijet, Z-boson + jet and Higgs boson + jet production at the LHC.

\section{Acknowledgments}
\label{sec:acknowledgments}
We are grateful to Carlo Oleari and Alexander Karlberg for useful discussions and comments on the manuscript.
This research was supported by the Fermi National Accelerator Laboratory (Fermilab), a U.S. Department of Energy, Office of Science, HEP User Facility. Fermilab is managed by Fermi Research Alliance, LLC (FRA), acting under Contract No. DE--AC02--07CH11359. T.G. and X.C. are supported by the Swiss National Science Foundation (SNF) under contract 200020-175595.
T.G. and X.C. would like to thank the University of Zurich S3IT (http://www.s3it.uzh.ch) and Swiss National Supercomputing Centre (CSCS) with project ID UZH10 for providing support and computational resources.
J.C.-M.\ is supported by the European Research Council (ERC) under the European Union’s Horizon 2020 research and innovation programme (grant agreement n.740006, N3PDF).
S.F.R.’s work was supported by the
European Research Council (ERC) under the European Union’s Horizon 2020 research and innovation program (grant agreement No. 788223, PanScales) and by the UK Science and Technology Facilities Council (grant number ST/P001246/1).
E.W.N.G.\ and M.S.\ are supported by the UK Science and Technology Facilities Council
(STFC) under grant \linebreak ST/P001246/1.
S.P.\ and M.S.\ are acknowledging support from the European Union's Horizon 2020
research and innovation programme as part of the Marie Sklodowska-Curie Innovative
Training Network MCnetITN3 (grant agreement no. 722104).
M.S.\ is funded by the Royal Society through a University Research Fellowship
(URF\textbackslash{}R1\textbackslash{}180549).
J.L. is supported by the Science and Technology Research Council (STFC) under the Consolidated Grant ST/T00102X/1 and the STFC
Ernest Rutherford Fellowship ST/S005048/1.

\end{document}